\documentclass[10pt]{aa}

\pdfoutput = 1

\usepackage[]{graphicx}
\usepackage[]{exscale}

\def\aj#1#2#3{#1,      {AJ, }{\bf#2}, #3}
\def\aa#1#2#3{#1,      {A\&A, }{\bf#2}, #3}

\def\apj#1#2#3{#1,     {ApJ, }{\bf#2}, #3}

\def\apjsupp#1#2#3{#1, {ApJS, }{\bf#2}, #3}

\def\mn#1#2#3{#1,      {MNRAS, }{\bf#2}, #3}
\def\Nature#1#2#3{#1,  {Nature, }{\bf#2}, #3}

\newfont{\msbm}{msbm8 scaled 1600} 
\DeclareTextFontCommand{\Bbb}{\msbm}

\newcommand{\ud}{\mathrm{d}}

\begin{document}

\title{Turbulent transport and its effect on the dead zone in
protoplanetary discs}

\author{Martin~Ilgner \& Richard~P.~Nelson}


\institute{Astronomy Unit, Queen Mary, University of London, 
Mile End Road, London E1 4NS, U.K. \\
\email{M.Ilgner@qmul.ac.uk, R.P.Nelson@qmul.ac.uk}}

\date{Received 21 December 2007 / Accepted 13 February 2008}

\titlerunning{Turbulent transport and the dead zone}

\authorrunning{M.~Ilgner, R.P.~Nelson}

\abstract
{Protostellar accretion discs have cool, dense midplanes where externally originating
ionisation sources such as X--rays or cosmic rays are unable to penetrate. This suggests 
that for a wide range of radii, MHD turbulence can only be sustained in the surface layers 
where the ionisation fraction is sufficiently high. A dead zone is expected to exist near the 
midplane, such that active accretion only occurs near the upper and lower disc surfaces.
Recent work, however, suggests that under suitable conditions the dead zone may be 
enlivened by turbulent transport of ions from the surface layers into the dense interior. }
{In this paper we present a suite of simulations that examine where, and under which 
conditions, a dead zone can be enlivened by turbulent mixing.}
{We use three--dimensional, multifluid shearing box MHD simulations, which include 
vertical stratification, ionisation chemistry, ohmic resistivity, and ionisation due to X--rays 
from the central protostar. We compare the results of the MHD simulations with a simple 
reaction--diffusion model.}
{The simulations show that in the absence of gas-phase heavy metals, such as magnesium,
turbulent mixing has essentially no effect on the dead zone. The addition of a relatively low
abundance of magnesium, however, increases the recombination time and allows 
turbulent mixing of ions to enliven the dead zone completely beyond a distance of 5 AU 
from the central star, for our particular disc model.}
{During the late stages of protoplanetary disc evolution, when small grains have been
depleted and the disc surface density has decreased below its high initial value, the 
structure of the dead zone may be significantly altered by the action of turbulent
transport. This may have important consequences for ongoing planet formation
in these discs. }


\maketitle

\section{Introduction}
Observations of young stars in a broad variety of star forming environments
show that they are surrounded by
gaseous and dusty circumstellar discs (e.g. Beckwith \& Sargeant 1996;
O'Dell et al. 1993; Prosser et al 1994; Furlan et al. 2006).
These systems usually show signatures of active gas accretion onto the
central star with a range of mass flow rates, but with the typical
value being $\simeq 10^{-8}$ M$_{\odot}$ yr$^{-1}$ (e.g. Sicilia--Aguilar et al. 2004).
The mechanism by which the disc transports angular momentum internally to cause
accretion is
not yet fully understood, but is likely to have its origin in disc turbulence.
So far only one mechanism has been shown to be robust in generating
turbulence in Keplerian discs, 
namely the magnetorotation instability (MRI) (Balbus \& Hawley 1991; Hawley \& Balbus 1991).
In addition to providing the internal stress required for accretion, turbulence may also
have important consequences for planet formation and evolution in protoplanetary discs.
It will lead to mixing of the dust, and may prevent settling toward the midplane
(Carballido et al. 2005; Johansen \& Klahr 2005; Fromang \& Papaloizou 2006).
It will also cause an increase in planetesimal velocity dispersion, and may
prevent runaway growth of planetesimals into planetary embryos (Nelson 2005). It will
modify the migration of low mass protoplanets (Nelson \& Papaloizou 2004; Nelson 2005),
and will provide the effective viscous stress in the disc needed to drive the so--called
type II migration of giant planets (Nelson \& Papaloizou 2003). It has recently
been suggested that it may lead to planetesimal formation through
gravitational instability (Johansen et al. 2007).

There are continuing questions, however, about the applicability of the MRI
to cool, dense protostellar discs, as the ionisation fraction is expected 
to be low (Blaes \& Balbus 1994). A protostellar disc model has been
presented by Gammie (1996) in which the main source of ionisation
is Galactic cosmic rays. Such a disc is predicted
to have magnetically ``active zones" near the disc surface in which turbulence is
sustained by the MRI due to cosmic ray ionisation, but with a ``dead zone" near the
disc midplane where cosmic rays are unable to penetrate. Sano et al. (2000) have
examined the effects of a more complex chemical reaction network
and the influence of small dust grains.
Glassgold et al. (1997) and Igea \& Glassgold (1999) examined X--rays emitted by the
protostellar corona as a possible source of disc ionisation, since it is doubtful
that cosmic rays can penetrate the inner disc regions because of the attenuating
effect of the T Tauri wind.
Fromang et al. (2002) demonstrated the potential importance of gas phase
heavy metals such as magnesium, whose presence in trace quantities
can significantly increase the recombination
time and decrease the size of the dead zone (at least in a dust free disc).
Semenov et al. (2004) examined the disc chemistry and ionisation fraction
using a reaction network drawn from the UMIST database, and analysed 
results using a reduced reaction network.

Nonlinear numerical studies of the effects of ohmic resistivity on
MRI--driven MHD turbulence have also been presented. Fleming et al. (2000)
performed MHD simulations of resistive discs, and showed that
turbulence is not sustained in discs with a magnetic Reynolds number
$\rm Re_m$ which is below a critical value $\rm Re_m^{crit}$.
Fleming \& Stone (2003) performed simulations where resistivity decreased
as a function of height in the disc, as predicted by the Gammie (1996) model,
and showed that active zones could indeed coexist with dead zones in the disc.
They also showed that a low Reynolds stress can be maintained in the dead zone,
such that low levels of accretion are sustained there. A recent study of
resistive discs has also been presented by Turner et al. (2007), who presented
stratified shearing box simulations of discs in which resistivity varied with height,
and a multifluid simulation in which disc chemistry was evolved along with the
dynamics. This latter run showed that turbulent mixing can potentially have
an important effect in generating stresses in the dead zone. 
Fromang \& Papaloizou (2007) have recently performed a resolution study
of shearing box simulations, and showed that the level of turbulent
activity reduces as the resolution increases. An analysis of
exiting shearing box simulations in the literature by Pessah et al. (2007)
led to a similar conclusion. In a companion paper to 
Fromang \& Papaloizou (2007),
Fromang et al. (2007) also examined how turbulent activity scales with
magnetic Prandtl number (defined by $P_m = {\rm Re_m/Re}$ where $\rm Re$ is the Reynolds
number and $\rm Re_m$ is the magnetic Reynolds number). They showed that
both the Reynolds number and magnetic Prandtl number are the parameters that
control the level of turbulent activity in a disc, with $P_m \le 1$ flows
showing no sustained turbulent activity for zero net flux magnetic fields.
Clearly there is much work to be done in understanding the nature of MHD turbulence
in discs.

In a recent set of publications, we have undertaken an extensive study of 
the chemistry and ionisation
fraction in protoplanetary discs. In Ilgner \& Nelson (2006a) (hereafter paper I) we examined the
dead zone structure in standard $\alpha$--disc models as predicted by a
number of different chemical reaction networks, and showed that the
simple model of Oppenheimer \& Dalgarno (1974) gives good agreement with
more complex models based on the UMIST database (Le Teuff et al. 2000).
We also demonstrated that grain depletion by factors even lower than 
$10^{-6}$ are required to reduce significantly the sizes of dead zones.
In Ilgner \& Nelson (2006b) (hereafter paper II) we constructed
a reaction--diffusion model to examine the role of turbulent mixing on
dead zone structure. The results showed that turbulent mixing has
essentially no effect throughout the disc in the absence of gas phase
heavy metals such as magnesium.  In the presence of trace quantities
of magnesium, however, turbulent mixing was able to enliven the
dead zone out beyond a few AU, where the mixing time becomes shorter
than the recombination time. In a third paper of the series (Ilgner \& Nelson 2006c),
we examined the effect of X--ray flares on dead zone structure.

In this paper we present a suite of shearing box simulations
of stratified local disc models in which we evolve the
disc chemistry along with the magnetohydrodynamic equations. The
primary aim is to re--examine the results of paper II using multifluid
MHD simulations, and 
determine whether, and under what conditions, turbulent mixing can enliven a dead zone
by mixing ions from the surface layers down toward the midplane.
We use the simple reaction scheme of Oppenheimer \& Dalgarno (1974),
which we incorporate into a multifluid MHD code, and assume that
dust grains are absent and ionisation is caused primarily by X--rays from the central star. 
We examine the effects of mixing as a function of gas phase magnesium abundance
and distance from the central star. Our results are in very good agreement with
the predictions of paper II. Disc models which contain no gas phase magnesium
show that the dead zone structure is essentially unmodified by turbulent mixing.
In the presence of magnesium, however, our simulations show that the dead zone can be enlivened
completely beyond a distance of 5 AU from the central star.

The paper is organised as follows. In Sect.~\ref{basic-eqns} we present the basic equations and
the chemical reaction network that we solve. In Sect.~\ref{reaction-diffusion} 
we discuss the reaction--diffusion
model, which we use to compare with the MHD simulations. In Sect.~\ref{ionisation} we discuss the 
method used for calculating the X--ray ionisation rate, and in Sect.~\ref{previous} we discuss previous
simulations that have examined dead zone structure. In Sect.~\ref{results} we present our simulation
results, and finally in Sect.~\ref{conclusions} we draw our conclusions.

\section{The dynamical and chemical model}
\label{basic-eqns}
In this section we give a detailed description of the chemical model
used in our simulations, and present the multifluid MHD equations
that we solve.
\subsection{Chemical model}
For the purposes of simplicity and computational tractability, we have applied 
the simple kinetic model of Oppenheimer \& Dalgarno (1974) to evolve the 
gas-phase chemistry within the simulations. This reaction network has been 
described in Ilgner \& Nelson (2006a), where it was compared with more complex 
reaction networks and found to predict electron fractions that were slightly higher
on average, due to the lower number of molecular ion species present in the 
simpler model. The Oppenheimer \& Dalgarno reaction network may be written: 
\begin{eqnarray}
&&{\rm H}_2 + h \nu  \rightarrow  {\rm H}_2^+ + {\rm e}^{-}  \label{chem1} \\
&&{\rm H}_2^+ + {\rm e}^- \rightarrow {\rm H}_2 \label{chem2} \\
&&{\rm H}_2^+ + {\rm Mg} \rightarrow  {\rm H}_2 + {\rm Mg}^+ \label{chem3}  \\
&&{\rm Mg}^+ + {\rm e}^- \rightarrow {\rm Mg} + h \nu \label{chem4} \\
\nonumber
\end{eqnarray}
This reaction scheme involves five species: molecular hydrogen and its ionised
counterpart (which act as a representative molecule and molecular ion),
atomic magnesium and its singly ionised counterpart (which act as representative
heavy metal and metal ion), and free electrons. The free electrons are treated
as a dependent species and their local fractional abundance is calculated 
assuming local charge neutrality: $x[{\rm e}^- ]= x[{\rm H}_2^+] + x[{\rm Mg}^+]$.
In our model, Eqns.~(\ref{chem1}) -- (\ref{chem4}) represent ionisation of 
${\rm H}_2$ by X--rays, recombination of ${\rm H}_2^+$, charge transfer
between ${\rm H}_2^+$ and ${\rm Mg}$, and recombination of ${\rm Mg}^+$, 
respectively. The associated reaction rates are given in table~\ref{rates:oppenheimer},
with the rate coefficients being denoted by
${\tilde \zeta}$, ${\tilde \alpha}$, ${\tilde \beta}$, ${\tilde \gamma}$, 
respectively. The recombination rate of ${\rm Mg}^+$ is five
orders of magnitude lower than of ${\rm H_2}^+$, indicating that
a magnesium--abundant gas will sustain a higher ionisation fraction 
than a metal--poor one because of charge transfer reactions, a point 
already noted by Fromang et al. (2002) in the context of protostellar discs.

Eqns.~(\ref{chem1}) -- (\ref{chem4}) form a set of stiff coupled ordinary 
differential equations, and we use the Gear method for their solution at 
each point in the simulation domain and for each simulation time step. 
The major source of ionisation that we consider is X--rays from the central 
protostar, and our approach to calculating the ionisation rate is described 
below in Sect.~\ref{ionisation}.
 
\begin{table}[t]
\caption{Rate coefficients for the Oppenheimer \&
Dalgarno model.}
\begin{center}
\begin{tabular}{lll}\hline\hline
$\tilde{\zeta}$ & $\zeta_{\mathrm{eff}}^{}$ & $\rm s^{-1}_{}$\\
$\tilde{\alpha}$ & $3 \times 10^{-6}_{}/ \sqrt{T}$ & $\rm cm^3_{} \ s^{-1}_{}$  \\
$\tilde{\beta}$  & $3 \times 10^{-9}_{}$ & $\rm cm^3_{} \ s^{-1}_{}$\\
$\tilde{\gamma}$ & $3 \times 10^{-11}_{}/ \sqrt{T}$ &$ \rm cm^3_{} \ s^{-1}_{}$  \\ 
\hline\hline
\end{tabular}
\end{center}
\label{rates:oppenheimer}
\end{table}

\subsection{Multifluid MHD equations}
\noindent
As the mean--free path for collisions, and the ion gyro--radii, are very much
smaller than the length scales we consider in our calculations, we adopt
a multifluid MHD approach to incorporating chemical evolution of the gas
during the dynamical evolution of our disc models. We use the shearing
box representation of a local patch of the protostellar disc
(Goldreich \& Lynden-Bell 1965),
in which the fluid is described using Cartesian coordinates ($x$, $y$, $z$).
The origin of this coordinate system rotates with the local Keplerian
angular velocity, $\Omega$, and the $x$ coordinate represents the radial
direction, the $y$ coordinate the azimuthal direction, and $z$ the vertical
direction.
The standard shearing box equations for MHD, including ohmic
resistivity, are:
\begin{eqnarray*}
\lefteqn{ \frac{\partial \varrho}{\partial t}  + 
{\bf \nabla \cdot } \left( \varrho {\bf v} \right) = 0.} & & 
\end{eqnarray*}
%
\begin{eqnarray}
\lefteqn{
\frac{\partial {\bf v}}{\partial t} + {\bf v} \cdot {\bf \nabla} {\bf v} =
- 2 \Omega {\bf \hat{z} \times v} + 3 \Omega_{}^2 x {\bf \hat{x}}
 } & & \label{eq5} \\ 
\lefteqn{
- \frac{1}{\varrho} {\bf \nabla} P -
  \frac{1}{\varrho c} {\bf J \times B}  - \Omega^2  z {\hat {\bf z}}} & & \nonumber
\nonumber
 \end{eqnarray}
%
\begin{eqnarray}
\lefteqn{ \frac{\partial {\bf B}}{\partial t} =
 {\bf \nabla \times} \left( {\bf v} {\bf \times B}  -
 \eta {\bf \nabla \times B} \right) } \label{eq8}
\end{eqnarray}
%
The expressions above represent the continuity, momentum and induction
equation, respectively. $\varrho$ represents the gas density, ${\bf v}$ the
velocity, $P$ the pressure, ${\bf B}$ the magnetic field, ${\bf J}$ the current 
density, and $\Omega$ is the local Keplerian angular velocity. The resistivity 
is denoted by $\eta$.

In our scheme the five species are treated as indivdual but tightly coupled 
fluids which move with the bulk velocity ${\bf v}$, and so we must solve
a continuity equation for each of them. When combined with the possibility 
that the local abundance of species may change because of chemical 
evolution as well as advection, then the continuity equation for each species 
$i$ may be written:
\begin{equation}
\frac{\partial \varrho_i^{}}{\partial t}  + 
{\bf \nabla \cdot } \left( \varrho_i^{} {\bf v} \right) = 
m_i \sum_{j = 1}^{r} \nu_{ij}^{}J_{j}^{} \ 
\;\;\;\; {\rm with \hspace*{.5cm}} i = 1,\dots,n 
\label{eq2}
\end{equation}
%
where $m_i$ is the particle mass for species $i$, $J_j^{}$ denotes 
the chemical  reaction rate associated with the $j$th chemical reaction, 
while $\nu_{ij}^{}J_{j}^{}$ is the formation/destruction rate of the $i$th 
fluid component due to the $j$th chemical reaction. We assume an 
isothermal equation of state such that 
\begin{equation}
P = c_s^2 \varrho
\label{estate}
\end{equation}
and calculate the resistivity according to (Blaes \& Balbus 1994)
\begin{equation}
\eta = \frac{234}{x[{\rm e_{}^-}]} T_{}^{1/2}.
\label{eta}
\end{equation}
The thermal structure of the disc model we use is described in the 
next section.

\subsection{Disc model}
\label{discmodel}
The MHD simulations that we performed were calculated using a system 
of dimensionless variables, as is convenient when performing shearing 
box simulations. In order to evolve the chemistry, however, whose reaction 
rates depend on local temperature and density, we need to ascribe physical 
units to these simulated quantities. We assume that the central star is of 
solar mass, and we adopt a disc model for which the surface density varies 
according to $\Sigma = 1000 \ [R/1 \ \rm AU]^{-3/2} \ gcm^{-2}$, and where 
the volume density varies with disc height as a Gaussian:
\begin{equation}
\varrho(z) \propto \exp \left\{ - \frac{1}{2} 
\left( \frac{z}{H} \right)_{}^2 \right\}.
\label{eq9}
\end{equation}
%
We assume that the disc aspect ratio has a constant value $H/r=0.05$,
and that the sound speed $c_s = H \Omega$. The temperature is then 
related to the local sound speed according to $c_s^2 = {\cal R} T/ \mu$, 
where the mean molecular weight is assumed to be $\mu =2.33$. The 
local physical parameters for each shearing box simulation are then fully 
specified once the orbital radius, magnetic field, chemical abundances 
and the local ionisation rate are defined.

\subsection{Numerical method}
We use the ZEUS finite difference MHD code (Stone \& Norman 1992), modified to
allow for the treatment of an arbitrary number of coupled fluids, to perform all our 
simulations. Time step control was achieved using the standard Courant condition,
with the Courant number set to 0.5. As the simulations use explicit time stepping,
the time step size is also determined by the ohmic diffusion rate of the magnetic 
field. The chemical kinetic equations were evolved using the Gear method.

\subsection{Initial and boundary conditions}
The basic disc model used is described in Sect.~\ref{discmodel}. The initial 
velocities of the gas in the simulations were taken to be the shearing box 
equilibrium values ${\bf v} = (0,-3/2 \cdot \Omega x,0)$, but with random 
fluctuations imposed with maximum amplitude equal to $10^{-3}$ of the sound 
speed. The initial magnetic field is a zero net flux vertical field given by
$B_z = B_0 \sin(2\pi x/H)$, where $B_0$ is defined by the requirement for 
the volume averaged plasma parameter $\beta = 100$.

At the beginning of each shearing box simulation, the equilibrium particle 
concentrations $x_{\infty}^{} [Y]$ of species $\rm Y \in \{ Mg ,Mg^+, H_2, H_2^+\}$ 
are taken as initial abundances. Note that we use different concentrations
of magnesium for our models, and we simulate local patches
of the disc at different radii from the central star. In each case
we calculate the local equilibrium chemistry prior to initiating
the MHD simulations.
\noindent

\indent
We use the same computational set-up as Fromang \& Papaloizou (2006). 
The computational domain is given by $[-H/2, H/2]$, $[0, 2\pi H]$, and 
$[-3H, 3H]$ in $x$, $y$, and $z$. We use a grid resolution of 
$32 \times 100 \times 192$. Standard periodic boundary conditions apply 
in $y$ and $z$, while periodic boundary conditions in shearing coordinates 
are used for $x$. For a detailed description of the shearing box set up and
boundary conditions see Hawley, Gammie \& Balbus (1995). Following 
Fromang  \& Papaloizou (2006) we introduced a vertical length scale 
$H_0 = 2.4 \ H$ which is used to prevent unphysical fluctuations due to 
the non vanishing vertical component of the gravitational force at the $z$ 
boundary. By applying Eq.~(30) of Fromang \& Papaloizou (2006) we 
ensure that the vertical gravity acts on vertical length scales $L < H_0$ only. 

\section{Reaction--diffusion model}
\label{reaction-diffusion}
In paper II we calculated the ionisation fraction for conventional 
$\alpha$--disc models and examined the effect of turbulent mixing 
by modelling the diffusion of chemical species vertically through 
the disc. To recap: applying a one dimensional reaction--diffusion 
model, we assumed that vertical mixing arises because of turbulent 
diffusion, and adopted the approximation ${\cal D} = \nu_{\rm t}^{}$ 
using the $\alpha$ prescription to calculate $\nu_t$. Here 
$\nu_{\rm t}^{}$ is the (turbulent) kinematic viscosity that drives the 
radial diffusion of mass through the protostellar disc, and $\cal D$ 
denotes the vertical diffusion coefficient.\\
\indent
Instead of just mimicking the effects of turbulent mixing in this way, 
we now model the turbulent transport of chemical species by solving 
the corresponding non-ideal MHD equations in a three dimensional 
shearing box as discussed above. The mixing now is a direct outcome 
of the MHD turbulence.\\
\indent
One purpose of this paper, however, is to examine whether or not the effects 
of turbulent mixing on the ionisation fraction described in paper II, can also 
be observed in shearing box simulations when MHD turbulence operates. 
Hence, we calculated the ionistion fraction obtained for the kinetic model 
of Oppenheimer \& Dalgarno by applying the reaction--diffusion model at 
the corresponding cylindrical radius $R$, in order to aid a direct comparison 
between the results obtained for the reaction--diffusion and the shearing 
box model. We have good reason to make this comparison because
Balbus \& Papaloizou (1999) have shown that the mean flow dynamics of 
the MHD turbulence follows the $\alpha$ prescription. \\
\indent
For the reaction-diffusion model we assume the same vertical density 
profile we use for the shearing box simulation at $t = 0$. The same 
applies for the gas temperature. We further adopt the approximation 
${\cal D} = \nu_{\rm m}^{}$ with
\begin{equation}
\nu_{\rm m}^{} = \alpha_m c_{\rm s}^2 / \Omega \label{eq10} .
\end{equation}
where $\alpha_m$ is a dimensionless number used to quantify the
efficiency of vertical mixing of chemical species by the turbulence.
A range of $\alpha_m$ values are used to examine how the 
reaction--diffusion model matches the simulations. Since we expect 
the largest gradients in the electron fraction to be in vertical ($z$) direction, 
we consider only vertical diffusion in the reaction--diffusion model. The 
rate of change of the molar density of the $i$th component of the fluid within 
a given volume due to chemical reactions and diffusion caused by 
concentration gradients is 
\begin{equation}
\frac{\partial n_i^{} }{\partial t}  = 
\frac{\ \partial}{\partial z} \left( n {\cal D} 
\frac{\ \partial}{\partial z} x_i^{} \right) + 
\sum_{j=1}^r \nu_{ij}^{} J_j^{}, \ (i=1,\dots, s)
\label{eq11}
\end{equation}
where $n_i^{}$ denotes the molar density of the $i$th component, $\cal D$ 
the diffusion coefficient, $x_i^{}=n_i^{}/n$ the fractional abundance of species 
$i$, and $n = \sum_i n_i$. $J_j^{}$ denotes the chemical reaction rate associated 
with the $j$th chemical reaction, while $\nu_{ij}^{}J_{j}^{}$ is the formation/destruction 
rate of the $i$th component due to the $j$th chemical reaction. A detailed discussion 
of the derivation of Eq.~(\ref{eq11}) is presented in paper II.\\
\indent
The numerical method applied to solve the reaction--diffusion model 
is described in paper II, and the same boundary conditions apply. The 
boundaries are at $[0,+3H]$ such that the computational domain has 
size $L_z^{} = 3 H$; the number of grid cells is $L_z^{} / \Delta z = 60$, 
ensuring that the elements and charge are conserved to high accuracy.\\
\indent
For a given metal elemental abundance, the reaction--diffusion models 
are initiated with the equilibrium composition obtained for 
${\cal D} = 0$, exactly as they are for the shearing box simulations.

\section{X-ray ionisation rate}
\label{ionisation}
\begin{figure}[t]
\includegraphics[width=.45\textwidth]{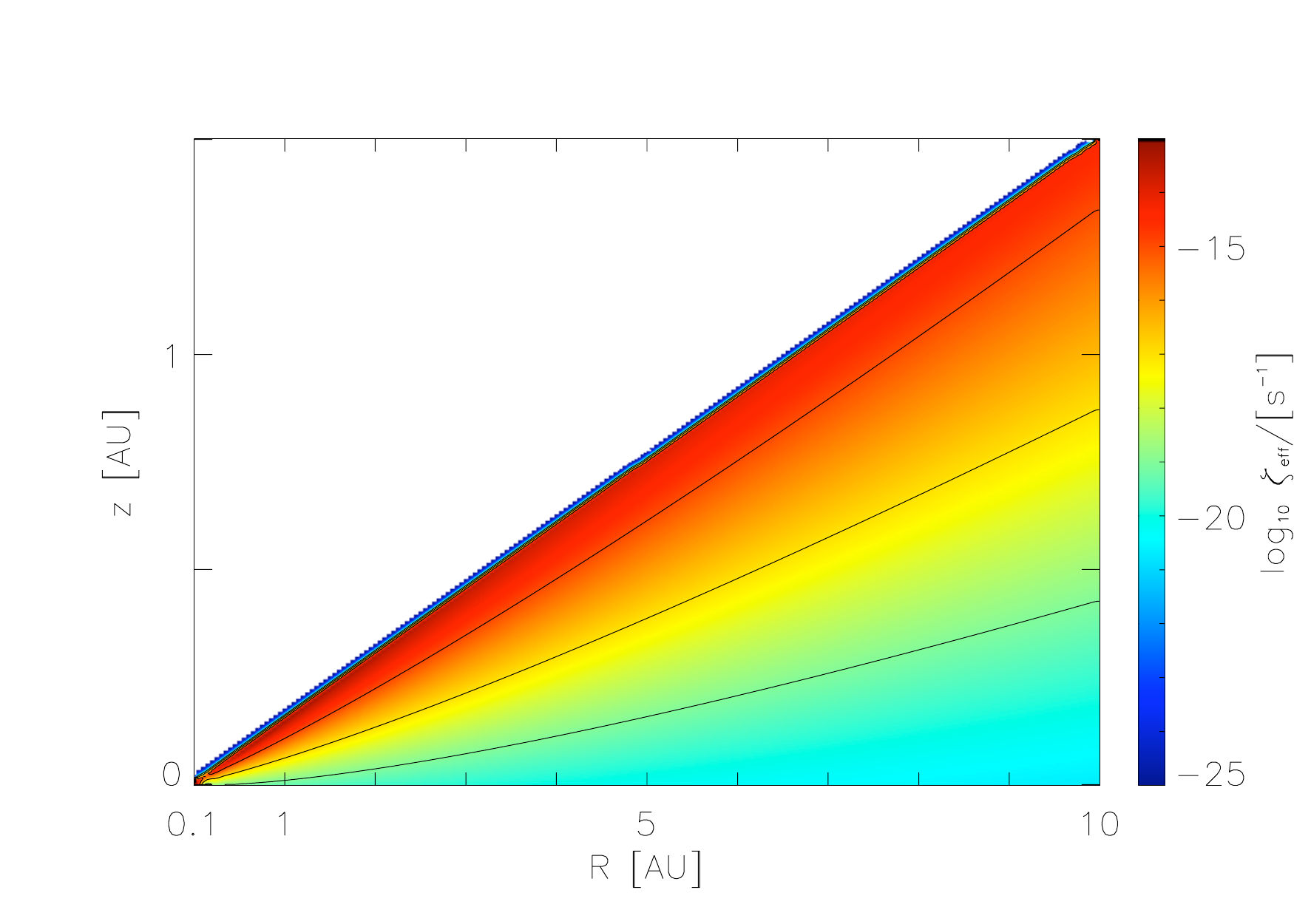}
\caption[]{The effective X--ray ionisation rate $\zeta_{\rm eff}^{}$ per hydrogen 
nucleus. The contour lines refer to values of $\zeta_{\rm eff}^{}$: $10_{}^{-15}$, 
$10_{}^{-17}$, and $10_{}^{-19} \ \rm s_{}^{-1}$.\label{figure:1}}
\end{figure}

As in paper I, we assumed that the ionisation of the disc material arises 
because of incident X-rays that originate in the corona of the central T 
Tauri star. We neglect contributions from Galactic cosmic rays as it remains 
uncertain whether they can penetrate into the inner disc regions we 
consider due to the stellar wind. The details for determining the effective 
ionisation rate $\zeta_{\rm eff}^{}$ have been described in paper I. However, 
here we do not consider standard $\alpha$-disc models as we did in that 
paper. Instead, we adopted a locally isothermal disc with the same Gaussian 
vertical profile for the mass density used for the local shearing box model at 
$t = 0$. When calculating the X-ray ionisation rate we assumed that this 
density profile did not vary with time, an assumption that is confirmed by the 
shearing box simulations which show that the density profile closely follows 
its initial Gaussian profile throughout the nonlinear evolution of the MRI, 
(see top right panel of Fig.~\ref{figure:4}).\\
\indent
We adopted values $L_{\rm X}^{} = 10^{31}_{} \ \rm erg \ s^{-1}$ and 
$k_{\rm B}^{}T_{\rm X}^{} = 5 \ \rm keV$ for the total X-ray luminosity 
$L_{\rm X}^{}$ and the plasma temperature $T_{\rm X}^{}$, respectively. 
Compared with the values applied in Ilgner \& Nelson (2006a,b), 
(i.e. $L_{\rm X}^{} = 10^{30}_{} \ \rm erg \ s^{-1}$ and 
$k_{\rm B}^{}T_{\rm X}^{} = 3 \ \rm keV$), the X-ray source considered 
here is more both harder and more luminous in order to increase the 
ionisation fraction above a theshold which makes the shearing box 
simulations feasible (a low ionisation fraction leads to a small time 
step size). However, the new values are still consistent with the 
observational constraints (e.g. Favata et al. 2005; Wolk et al. 2005).\\
\indent
The effective ionisation rate (per hydrogen nucleus) $\zeta_{\rm eff}^{}$ 
is approximated by Eq.~(3) in paper I. Calculating the X-ray optical depth 
$\tau_{\rm X}^{}$ along the line of sight between the X-ray source and the 
point in question, we derived the effective ionisation rate shown in 
Fig.~\ref{figure:1}. In particular, we applied the same data range in order 
to aid direct comparison with the ionisation rates of our previous 
$\alpha$-disc models (e.g. compare with Fig.~6 of paper I). The effective 
ionisation rate shown in Fig. ~\ref{figure:1} is higher because of the brighter 
(by one order of magnitude) and harder (more penetrating) X-ray source applied. 

\begin{figure}[t]
\includegraphics[width=.45\textwidth]{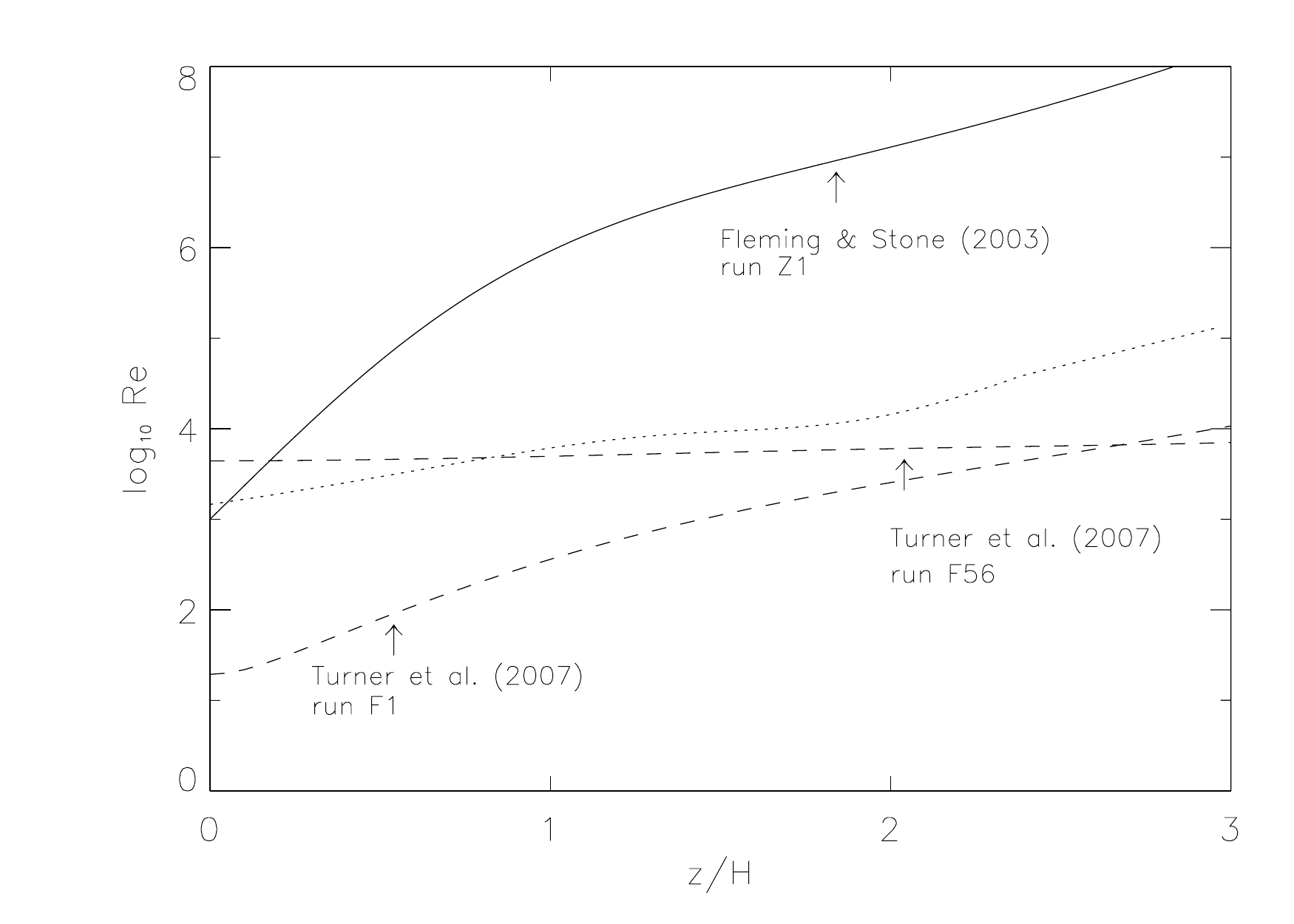}
\caption[]{Vertical profiles of the magnetic Reynolds numbers $\rm Re_m$.
The profile drawn with the solid line refers to the profile used by
Fleming \& Stone (2003) in their model with a small dead zone. 
The two dashed lines correspond to profiles used by Turner et al. 
(2007) in their models F1 and F56. The profile drawn with the dotted 
line refers to our \texttt{model1}, calculated at $R = 10 \ \rm AU$ and a 
metal abundance of $x_{\rm Mg}^{} = 5 \times 10_{}^{-11}$. Details are 
discussed in the text.}
\label{figure:2}
\end{figure}

\section{Previous simulations of dead zones}\label{previous}
As has been well documented in the literature, there remain questions about the 
applicability of the MRI to protostellar discs because of their high densities and low 
temperatures, which lead to low levels of ionisation (e.g. Blaes \& Balbus 1994; 
Gammie 1996). The first fully non linear study of the MRI including the effects
of resistivity were performed by Fleming et al. (2000), who employed shearing
box simulations to examine the conditions under which fully developed turbulence 
could be sustained. They showed that the important quantity that determines the 
outcome is the magnetic Reynolds number, $\rm Re_m$, defined by: 
\begin{equation}
{\rm Re}_{\rm m}^{} = \frac{H c_s^{}}{\eta}
\label{eq12}
\end{equation}
where the resistivity, $\eta$, is defined by Eq.~(\ref{eta}).
Simulations showed that turbulence could only be sustained if the Reynolds number 
was greater than a critical value, $\rm Re_m^{crit}$, where this critical value depends 
on the magnetic field topology. For zero net flux fields $\rm Re_m^{crit} \simeq 10^4$,
whereas for net vertical fields $\rm Re_m^{crit} \simeq 100$. The implication of
this study is that protostellar discs, in which the magnetic field is an internally
generated zero net flux field, will sustain turbulence in their surface regions where
the ionisation degree causes $\rm Re_m \ga 10^4$, but will remain in a near--laminar 
state in regions near the midplane, as envisaged by the layered disc model of Gammie 
(1996).

Shearing box simulations of stratified protostellar discs, with resistivity
varying with height, were presented by Fleming \& Stone (2003). Calculations
were presented with different vertical resistivity and magnetic Reynolds number profiles,
and it was shown that layered accretion resulted when the magnetic Reynolds number
satisfied $\rm Re_m > Re_m^{crit}$ in the surface layers, with $\rm Re_m < Re_m^{crit}$
in the midplane regions. Their results also showed that a low Reynolds stress could be sustained
in the dead zones due to the penetration of sound waves excited in the overlying active regions.
The magnetic Reynolds number profile assumed in the simulation for a ``small dead zone'' from
Fleming \& Stone (2003) is shown by the solid line in Fig.~\ref{figure:2}. 
The magnetic Reynolds number varies from 1000 at the midplane, to 
$9.23 \times 10^5$, $1.31 \times 10^7$, and $1.62 \times 10^8$ at $z/H = 1$, 
$z/H = 2$, and $z/H = 3$, respectively, resulting in a simulated disc with a dead zone 
whose vertical height is $\ll H$.

In a recent paper, Turner et al. (2007) have presented a study of dead 
zones which included shearing box simulations of vertically stratified 
discs with resistivity varying as a function of height, and also a multifluid 
simulation in which the resistivity was able to change locally because of 
chemical evolution of the gas. In this study, Turner et al. (2007) employed 
the reaction network given by Eqns.~(\ref{chem1}) -- (\ref{chem4}) in 
order to calculate the ionisation fraction and resistivity. For those runs in 
which the resistivity was kept constant in time, the initial resistivity profile 
was obtained from the equilibrium solutions to Eqns.~(\ref{chem1}) -- 
(\ref{chem4}). The run in which resistivity  varied in time and space 
employed a multifluid approach, similar to that described in Sect.~\ref{basic-eqns}. 
Ionisation was assumed to be due to cosmic rays, and the underlying disc 
model was the minimum mass solar nebula model of Hayashi (1981). In 
Fig.~\ref{figure:2} we present two of the resistivity profiles employed by 
Turner et al. (2007) corresponding to their runs F1 and F56. In run F1 the 
resistivity was a fixed function of height and corresponded to a radial location 
$R=1$ AU with the gas--phase abundance of magnesium equal to the solar 
abundance ($3.39 \times 10^{-5}$ magnesium atoms per hydrogen nucleus, 
corresponding to $x[\rm Mg] \sim 6.8 \times 10^{-5}$ in our units). This run led 
to a layered accretion flow with active surface layers and midplane dead zone, 
as expected from the steep resistivity profile. It was shown that the boundary 
between dead and active zones is well described by the criterion that MHD 
turbulence is sustained by the MRI if the Lundquist number 
$\rm Lu \equiv v_A^2/(\Omega \eta) > 1$, where $v_A$ is the Alfv\'en speed.
Run F56 had a resistivity profile corresponding to a disc at 5 AU with gas 
phase magnesium abundance equal to $10^{-6}$ below solar abundance 
(corresponding closely to our model with $x[\rm Mg] = 5 \times 10^{-11}$).
This model led to a fully turbulent disc, as expected from the flat resistivity
profile obtained because the disc provides less shielding of cosmic rays
at 5 AU.

The multifluid model presented by Turner et al. (2007), in which
the chemistry was evolved simultaneously with the dynamics,
led to an interesting and somewhat unexpected result. This disc
model corresponded to the radial position $R=1$ AU in the disc,
and assumed a gas--phase magnesium abundance equal to the 
solar value. During the early phase of the model, the disc showed 
the expected layered structure with dead zone near the midplane, 
and active zone near the disc surface. After about 60 orbits the 
situation changed after a period of more intense mixing caused by 
enhanced turbulent activity. The recombination time then exceeded 
the mixing time, allowing free electrons to mix toward the midplane, 
where net radial and azimuthal fields had built up due to field of the 
opposite polarity advecting through the vertical boundaries. The 
presence of these net magnetic fields leads to an enhance magnetic
stress that partially enlivens the dead zone during periods when the 
ionisation fraction has been increased.

In this paper, we assume that the ionisation of the disc material arises
because of X-rays that originate in the corona of the central T Tauri star. 
We neglect contributions from Galactic cosmic rays, as it remains an 
open question whether or not they can to penetrate into disc regions 
we consider. The X--ray ionisation rate decreases with cylindrical radius 
$R$, because the X-ray optical depth along the line of sight increases 
as one moves out into the disc.\\
\indent
Our use of strictly periodic boundary conditions in the vertical direction,
along with an initial magnetic field that has zero net flux, means that the 
net flux remains zero throughout the simulations. The expectation then 
is that turbulence will be sustained only in those regions where the 
magnetic Reynolds number $\rm Re_m \ga 10^4$, and we do not expect 
that large scale net--flux magnetic fields will be able to accumulate in the 
dead zones of our simulations. As such we do not expect to observe the 
behaviour shown by simulation V1 of Turner et al. (2007). We now present 
the results from a series of systematic experiments which examine the 
effects of chemical evolution and turbulent mixing on the evolution of the MRI.  

\section{Simulation results}
\label{results}
In this section we present the results of our multifluid MHD simulations
which examine the role of turbulent mixing on the structure of dead zones.
The primary aim of these simulations is to demonstrate that there exists
a region of parameter space in which turbulent mixing can enliven a 
dead zone which is otherwise predicted to exist in models that neglect
turbulent transport. A further aim is to demonstrate good agreement 
between MHD simulations and the predictions of a simple reaction--diffusion 
scheme. This latter issue is addressed in Sect.~\ref{r-d-results}.

When discussing the results of our simulations we will often refer to
certain averaged quantities. We use the same averaging procedures 
presented in recent publications studying vertically stratified disc models 
(e.g. Stone et al. 1996; Flemming \& Stone 2003; Fromang \& Papaloizou 
2006). The azimuthal and radial average of quantity $f({\bf r}, t)$ at a given 
time $t$ is defined by
\begin{eqnarray*}
F_{}^{\ast}(z,t) =  \left( f \right)_{}^{\ast } & = & 
\frac{ \int \! \! \! \int f(x_{}^{\prime},y_{}^{\prime},z,t) \ \ud x_{}^{\prime} \ 
\ud y_{}^{\prime}}{\int \! \! \! \int \ud x_{}^{\prime} \ \ud y_{}^{\prime}} 
\end{eqnarray*}
and the volume average is given by
\begin{eqnarray*}
F_{}^{\ast \ast}(t) =  \left( f \right)_{}^{\ast\ast} & = & 
\frac{\int \! \! \! \int \! \! \! \int f(x_{}^{\prime},y_{}^{\prime},z_{}^{\prime},t) \ \ud x_{}^{\prime} 
\ \ud y_{}^{\prime} \ \ud z_{}^{\prime}}{\int \! \! \! \int \! \! \! \int \ud x_{}^{\prime} \ \ud y_{}^{\prime} 
\ \ud z_{}^{\prime}} \hspace*{1.0cm} .
\end{eqnarray*}
where the symbols $\left( f \right)_{}^{\ast}$ and $\left( f \right)_{}^{\ast \ast}$,
denote the corresponding averaging procedures. The time--averaged
values are denoted by $\overline{\left( f \right)_{}^{\ast}}$ and
$\overline{\left( f \right)_{}^{\ast \ast}}$. 

A measure of the effective shear stress generated by the turbulence
is given by the parameter $\alpha$, which has contributions from
both the Reynolds and Maxwell stresses:
\begin{equation}
\alpha = \alpha_{\rm Rey}^{} + \alpha_{\rm Max}^{}
\label{eq16}
\end{equation}
where
\begin{eqnarray}
\alpha_{\rm Rey}(x,z) & = & \frac{T_{r\Phi}^{\rm Rey}}{P_0} = \frac{1}{P_0}
\varrho \left( v_x^{} - \langle v_x \rangle \right) 
\left( v_y - \langle v_y \rangle \right) \label{eq17}\\
\alpha_{\rm Max}(x,z)& = & \frac{T_{r\Phi}^{\rm Max}}{P} = 
\frac{-\langle B_x^{}B_y^{} \rangle }{4\pi P_0} \  \label{eq18},
\end{eqnarray}
and the azimuthal average (over $y$) is denoted by angled brackets, 
and $P_0$ is the initial midplane pressure. We regularly use the volume 
and time averaged values of $\alpha$ when discussing the simulation 
results below.
%
\begin{table}[t]
\caption{List of models considered. Note that the column ``recombination process 
included" specifies whether or not recombination of free electrons occurs along
with dynamical evolution.} 
\begin{tabular}{clc} \hline \hline \\[-.5em]
model & resistivity & recombination process included\\
\texttt{model1} & $\partial_t^{} \eta(t,z) = 0$ & no \\
\texttt{model2} & $\partial_t^{} \eta(t,z) \ne 0$ & yes \\
\texttt{model3} & $\partial_t^{} \eta(t,z) \ne 0$ & no \\[.1em]
\hline \hline \label{table:2}
\end{tabular}
\end{table}
In the following simulations we consider three different treatments of the resistivity 
and chemistry, and we refer to these models as \texttt{model1}, \texttt{model2} and 
\texttt{model3}. We describe each of these below, and each are summarised in 
table~\ref{table:2}.

\noindent
1) \underline{\texttt{model1}}: This model assumes a static resistivity 
profile which varies with height. The resistivity profile at $t = 0$ is obtained 
using the equilibrium solution of the kinetic model presented in 
Eqns.~(\ref{chem1}) -- (\ref{chem4}). During the simulations the local 
resistivity values are kept fixed. This model corresponds to the single fluid 
models of Fleming \& Stone (2003), and the runs F1, F52, F56, F58 of 
Turner et al. (2007).\\
2) \underline{\texttt{model2}}: This is a multifluid model in which the
resistivity varies in both time and space. The chemical reaction network 
given by Eqns.~(\ref{chem1}) -- (\ref{chem4}) is solved simultaneously 
with the dynamical evolution. The resistivity profile at $t = 0$ is obtained 
from the equilibrium solution of the kinetic model. \\
3) \underline{\texttt{model3}}: This is a multifluid model which has a 
resistivity profile which varies in time and space. The resistivity profile 
at $t = 0$ is obtained using the equilibrium solution of the kinetic model 
presented in Eqns.~(\ref{chem1}) -- (\ref{chem4}). For $t > 0$, however, 
the recombination of free electrons with ions is switched off, as is further 
ionisation of neutral species by X--rays. Local changes in the resistivity 
are due only to the turbulent mixing of ions. This model is equivalent to 
one in which the initial values of resistivity are conserved on fluid elements 
by being advected with the flow. 

We now present our simulation results in detail. We begin by highlighting
simulations which show that turbulent mixing can remove the dead zone,
before examining disc evolution at different radii and with different
magnesium abundances. 

\subsection{Dead--zone removal through turbulent mixing}
\label{mixing}
\begin{figure*}[ht]
\centering
\includegraphics[width=.45\textwidth]{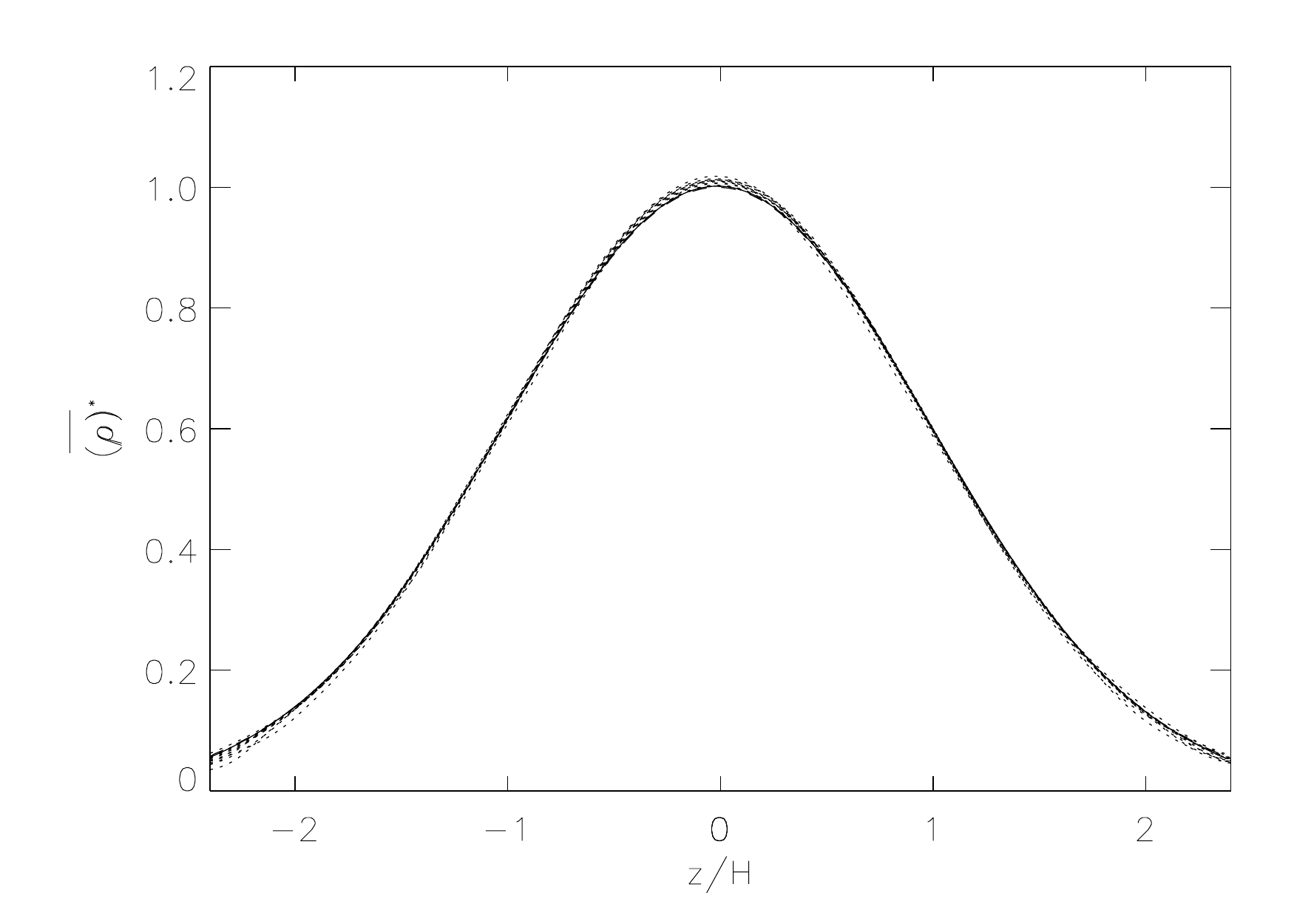}
\includegraphics[width=.45\textwidth]{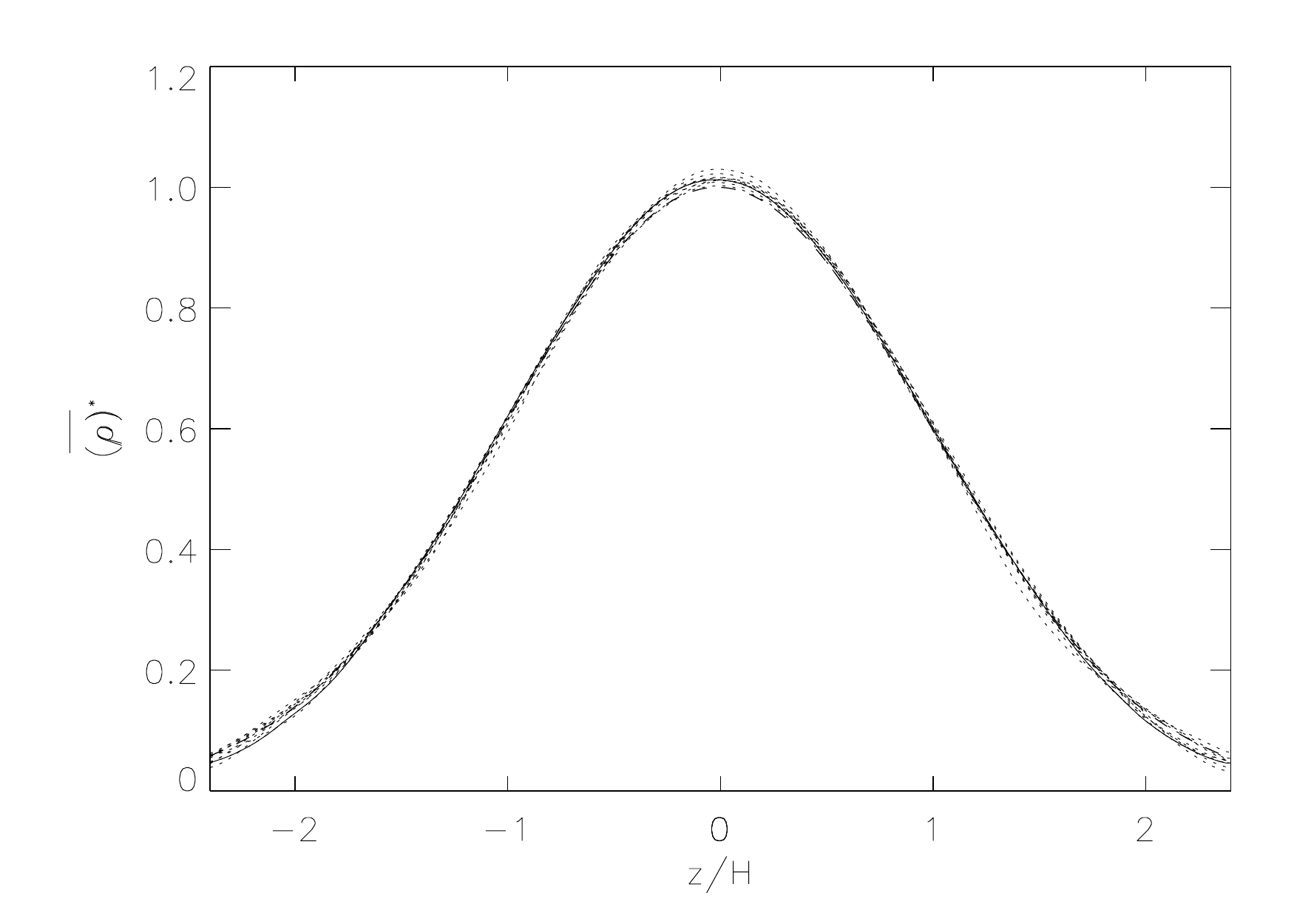}
\includegraphics[width=.45\textwidth]{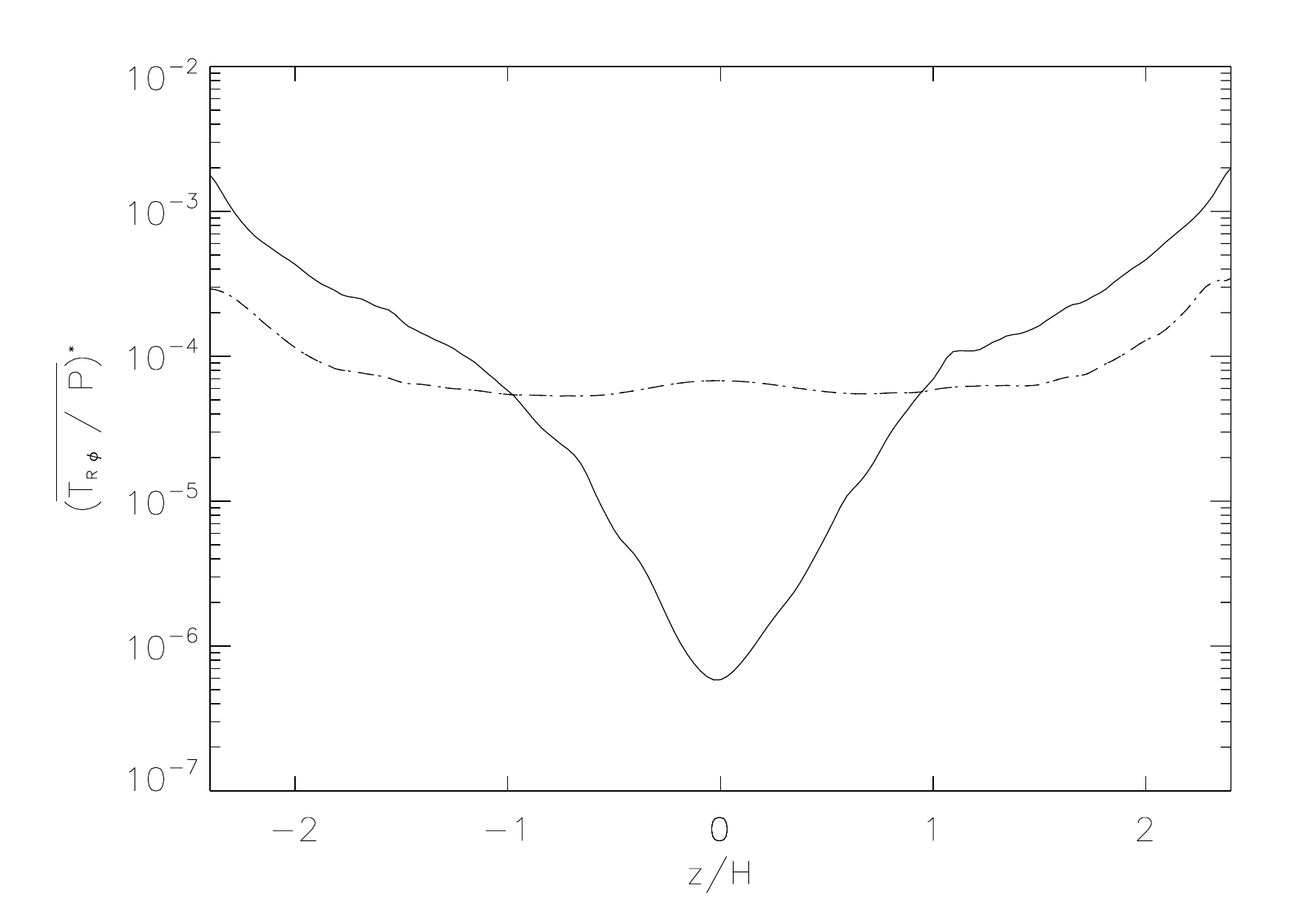}
\includegraphics[width=.45\textwidth]{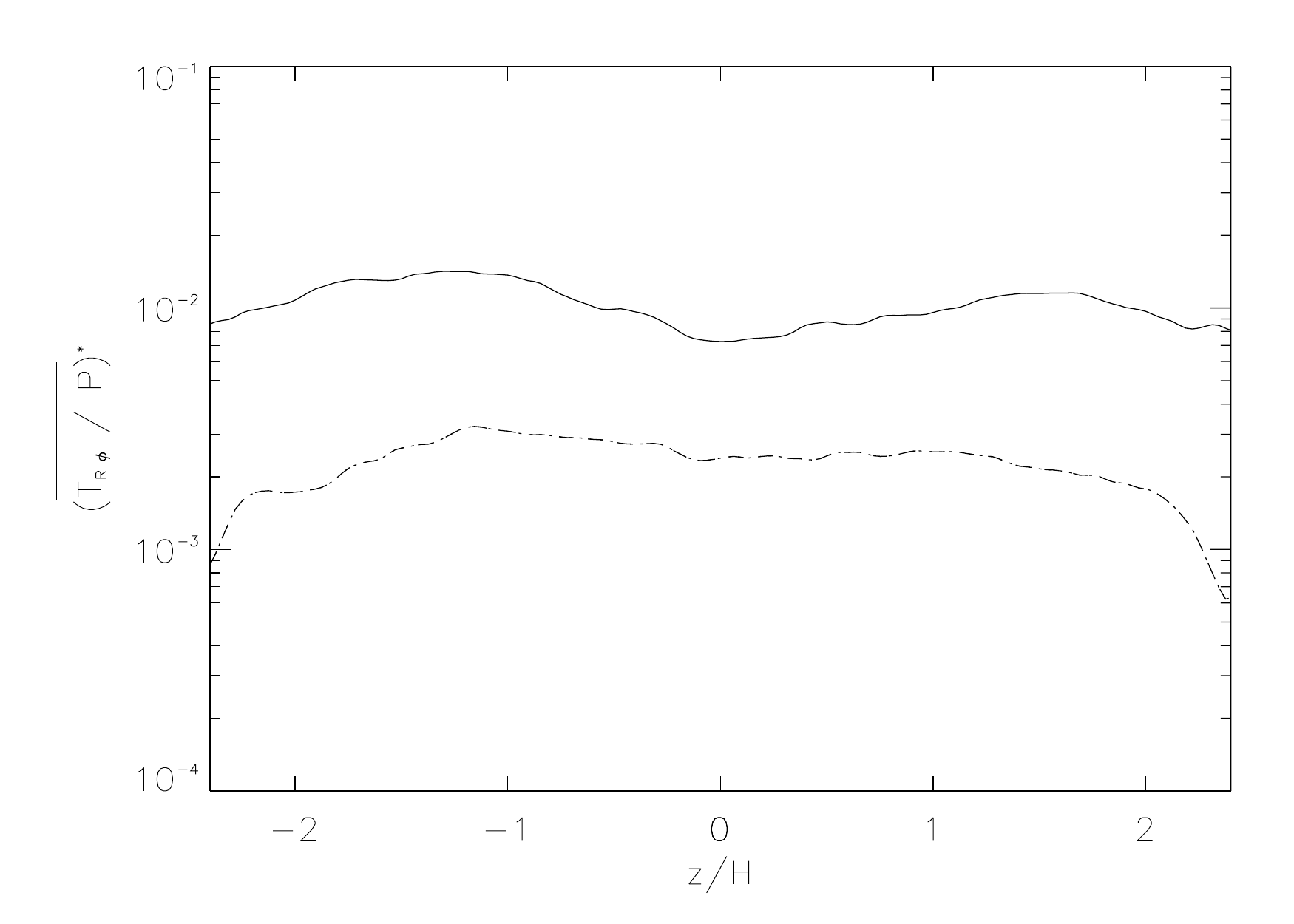}
\includegraphics[width=.45\textwidth]{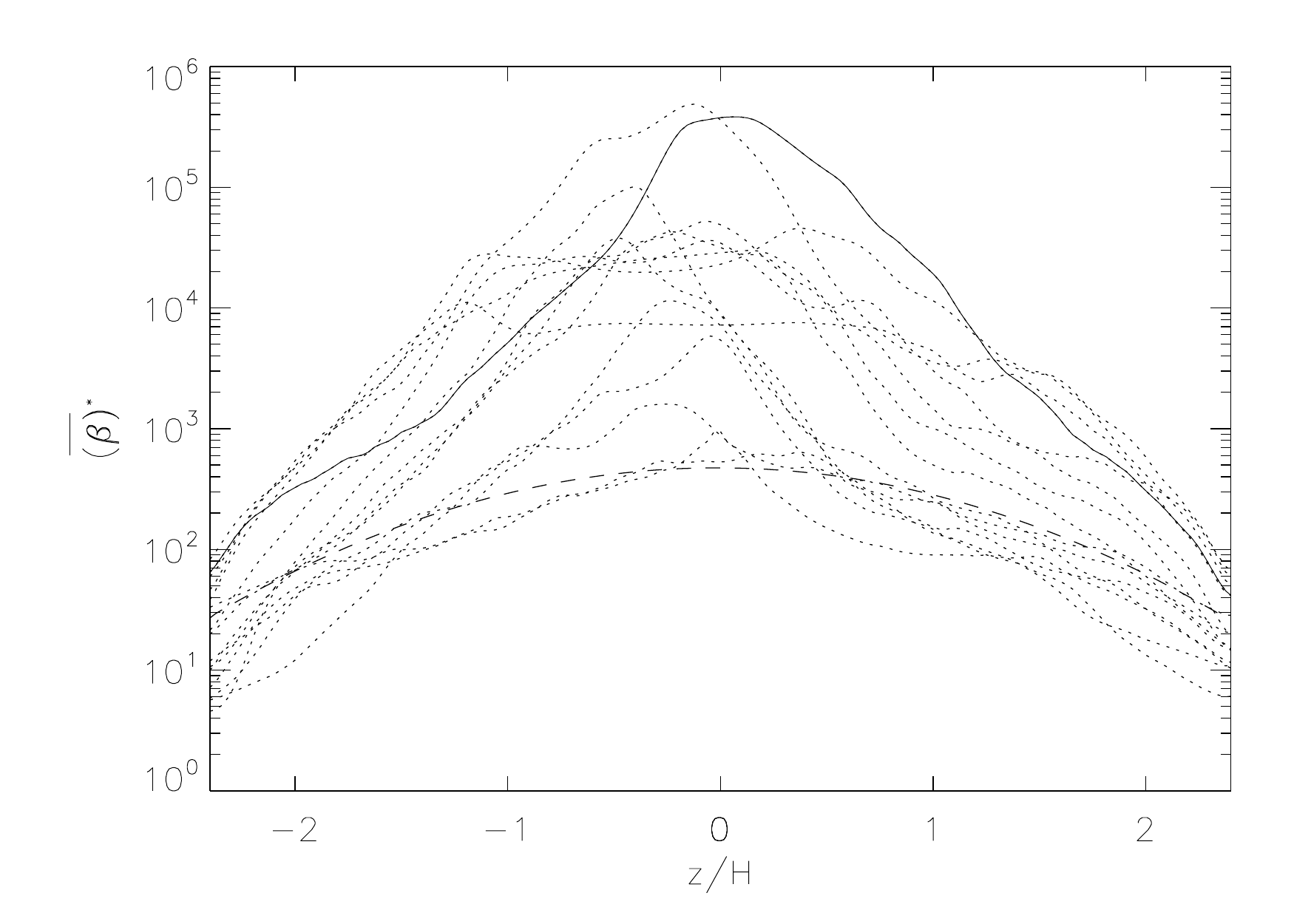}
\includegraphics[width=.45\textwidth]{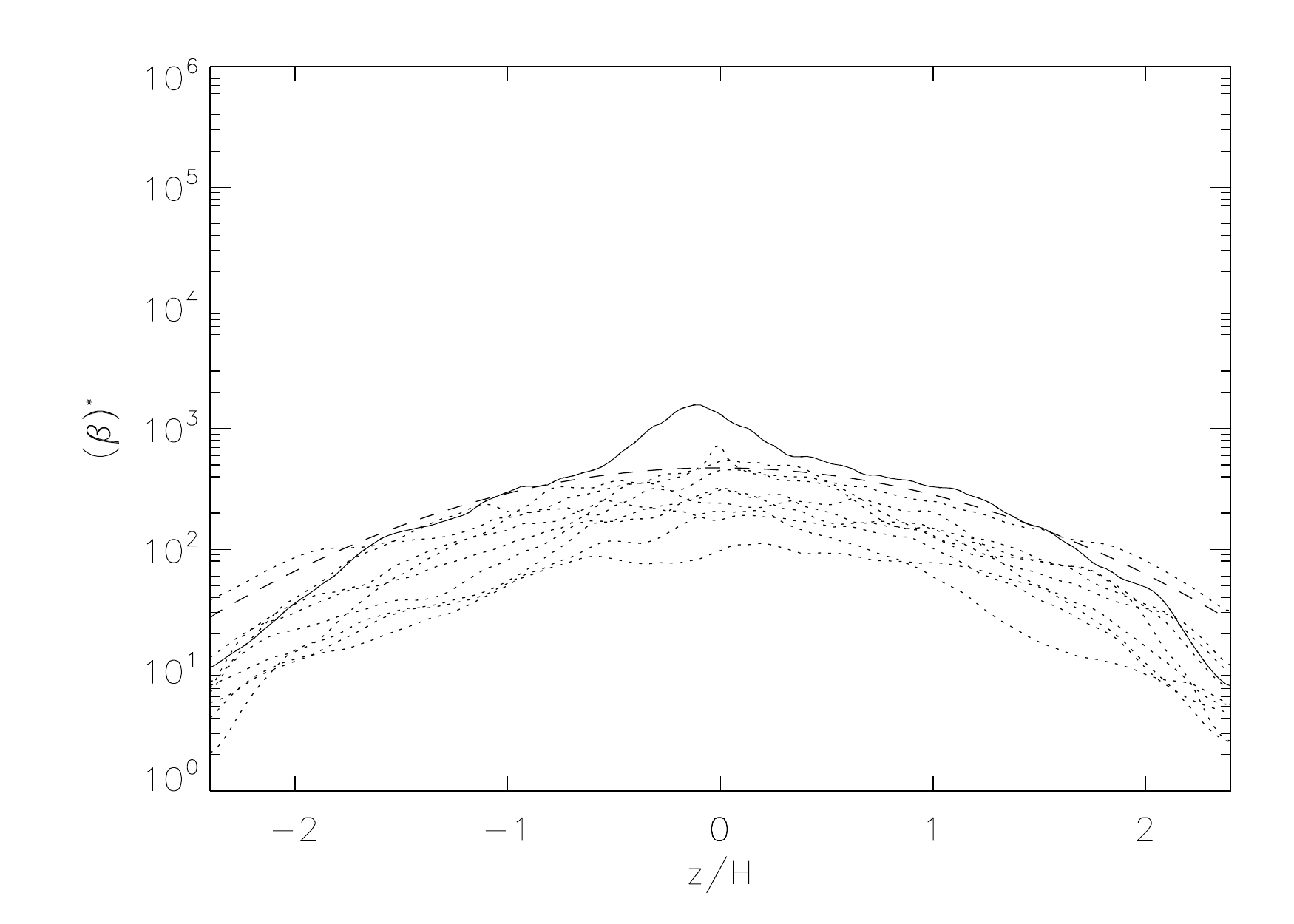}
\caption{\texttt{model1}/\texttt{model2} - Time averaged vertical profiles of the horizontally 
averaged density, normalised Maxwell and Reynolds stresses $T_{R\Phi}^{}/P_0^{}$, 
and plasma parameter $\beta$. The profiles on the left refer to the results obtained with 
the model assuming a fixed resistivity (\texttt{model1}). Results obtained with  \texttt{model2} 
are shown in the right panels. Both shearing box models are calculated for $R =  10 \ \rm AU$ 
and  $x_{\rm Mg}^{} = 5 \times 10^{-11}$. $\alpha_{\rm Max}^{}$ (solid line) and $\alpha_{\rm Rey}^{}$ 
(dashed - dotted line) refer to a time average taken over [100,200] orbits (for \texttt{model1}) and 
[0,100] orbits for \texttt{model2}. For all the other quantities, the time averages are taken over 10 
orbit intervals, starting from $t = 0$ (dashed line). For \texttt{model1} the solid line denotes averages 
takenbetween $t=[140,150]$ orbits, and for \texttt{model2} the solid lines represented averages 
taken between $t=[90,100]$ orbits. The dotted lines refer to time averages taken over 
$t = [0,10], \ [10,20], \ \cdots, [80,90] \, {\rm for \ \texttt{model2}}, \ \cdots, [130,140]$ for \texttt{model1}.}
\label{figure:4}
\end{figure*}
%
\begin{figure*}[ht]
\centering
\includegraphics[width=.45\textwidth]{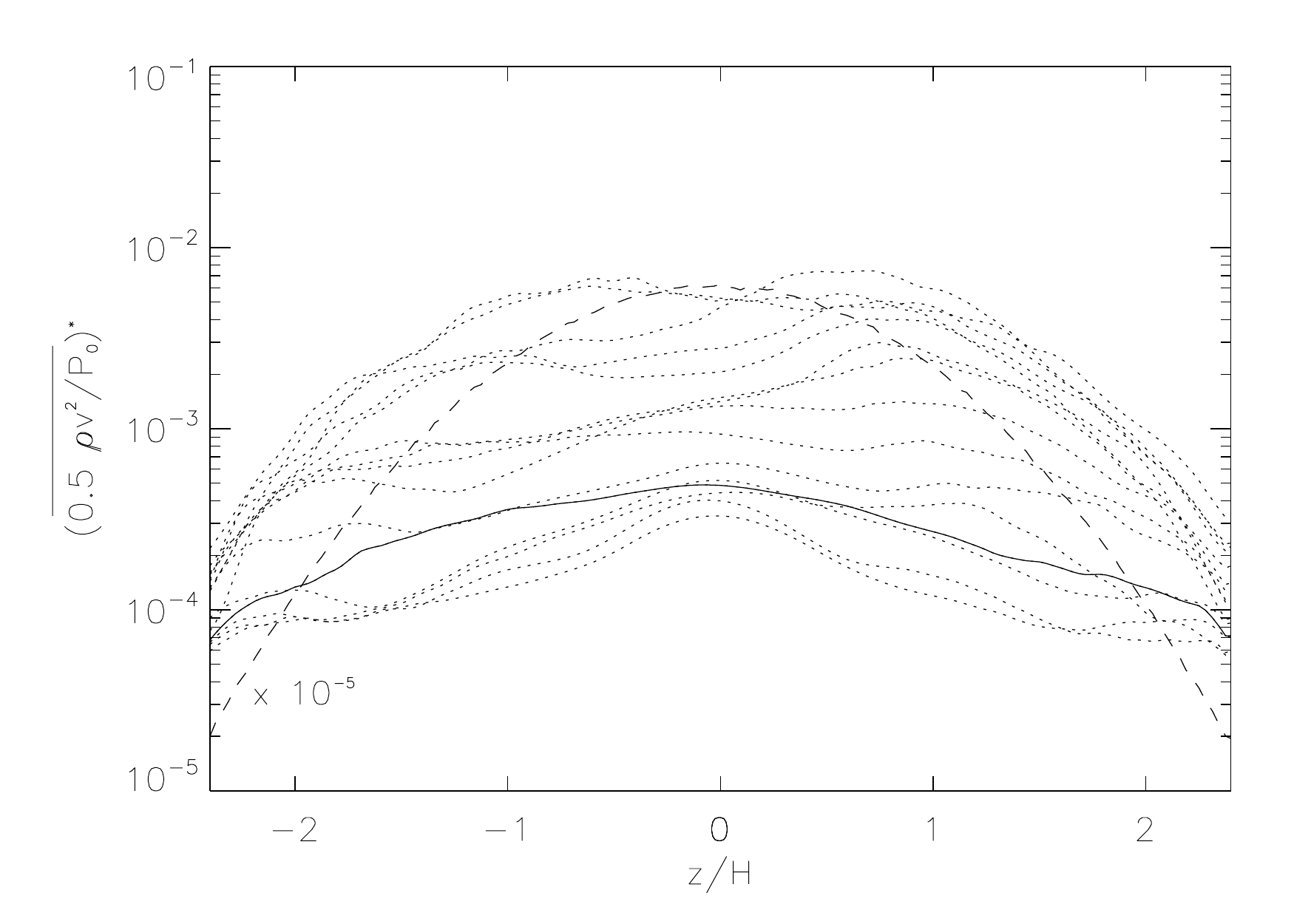}
\includegraphics[width=.45\textwidth]{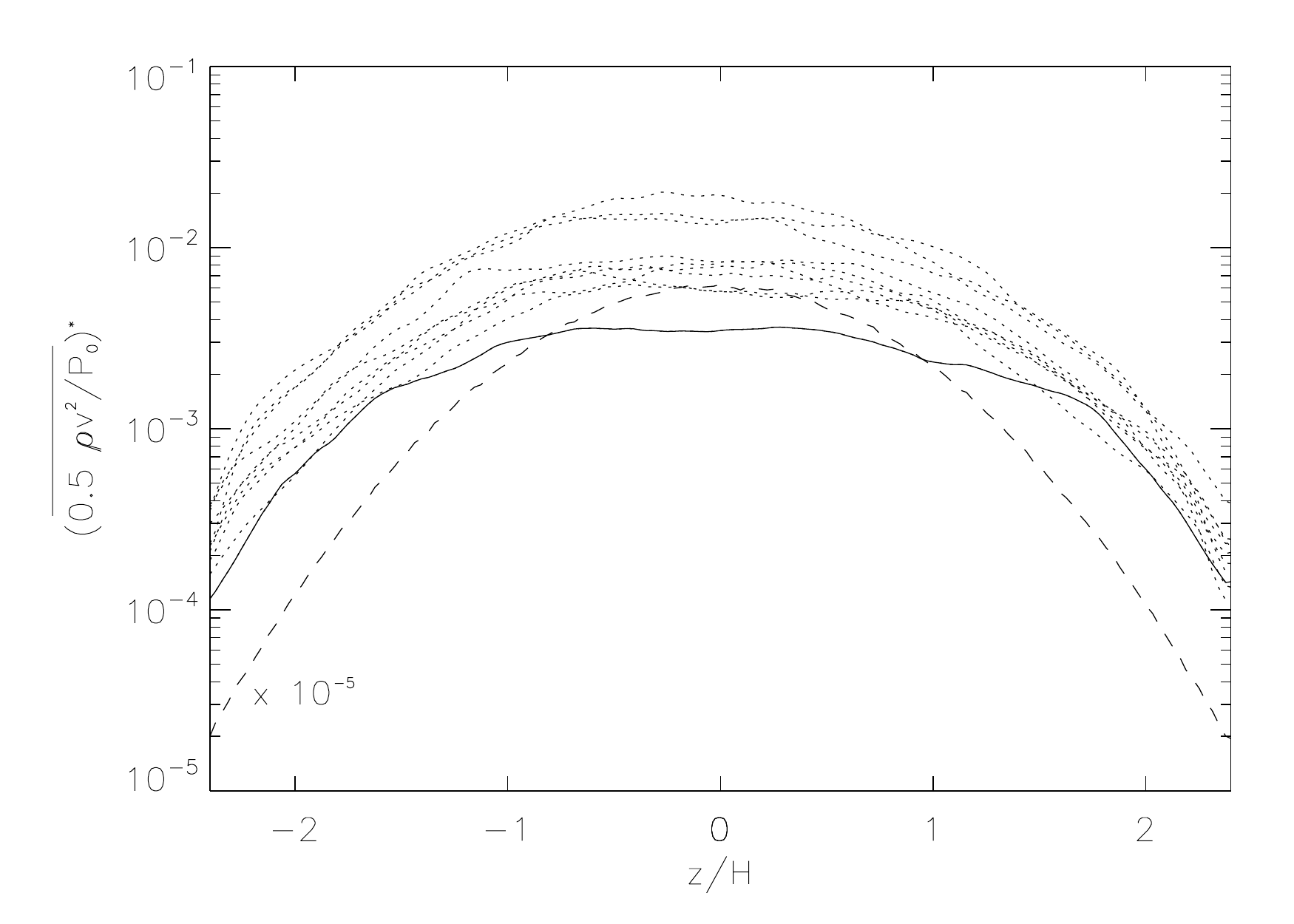}
\includegraphics[width=.45\textwidth]{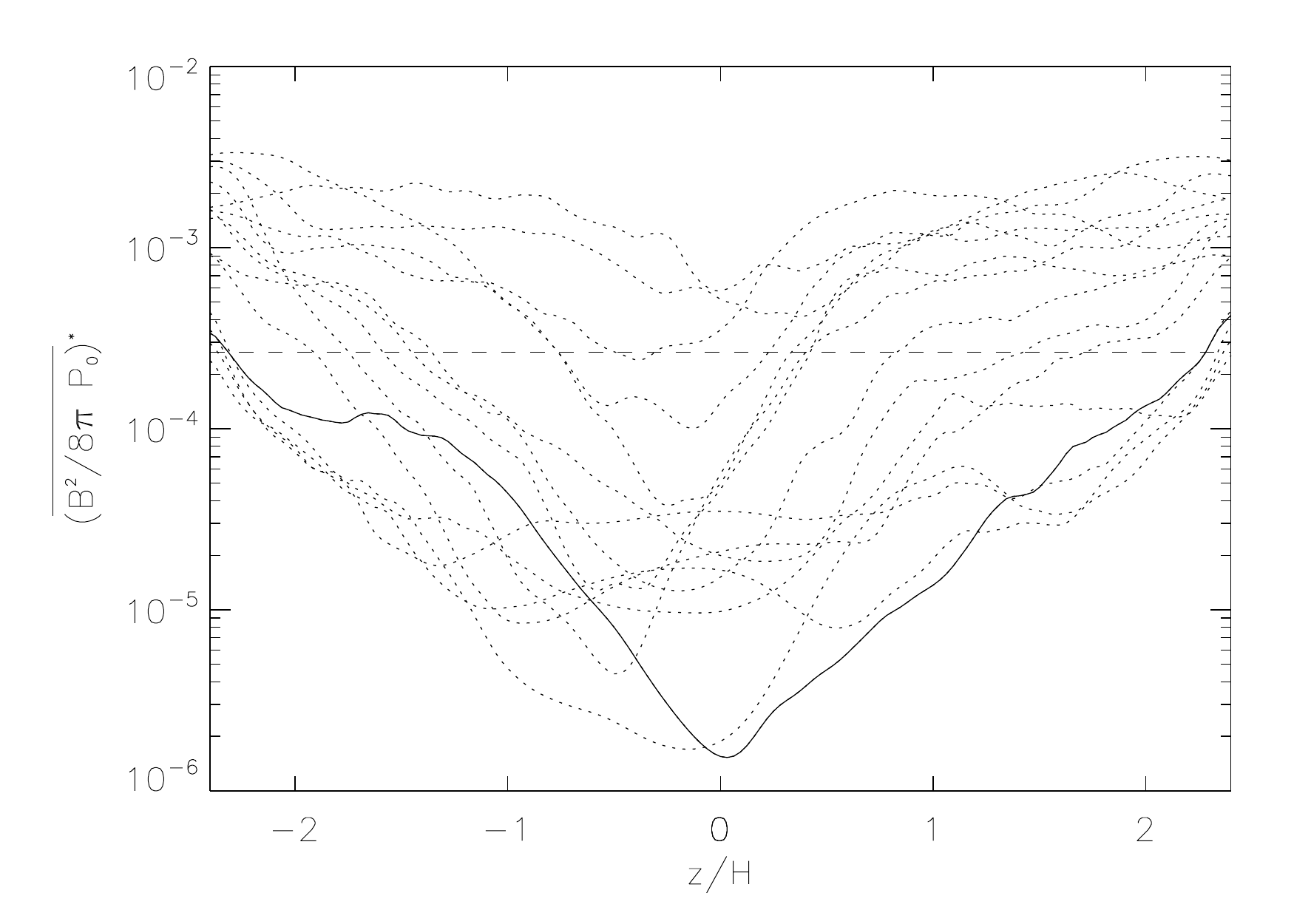}
\includegraphics[width=.45\textwidth]{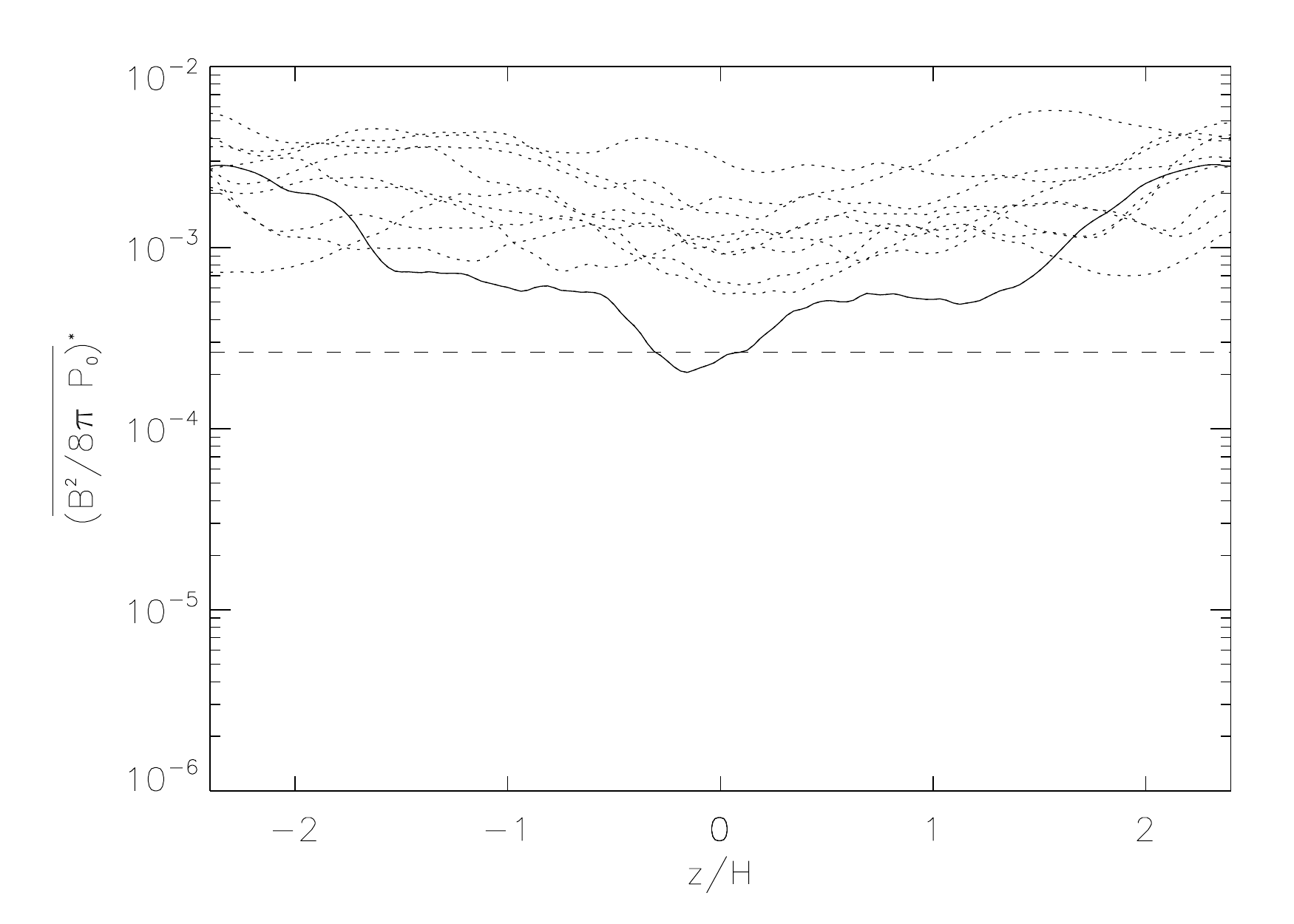}
\caption{\texttt{model1}/\texttt{model2} - Time averaged vertical profiles of the horizontally 
averaged kinetic energy, $0.5 \varrho v^2 /P_0$, and the magnetic energy $B^2 / (8\pi P_0)$. 
The profiles on the left panel show  results obtained assuming a fixed resistivity profile
(\texttt{model1}), while the results obtained with \texttt{model2} are shown in the right panels. 
Both shearing box models are calculated for $R =  10 \ \rm AU$ and  $x_{\rm Mg}^{} = 5 \times 10^{-11}$. 
The time average is taken over 10 orbit intervals, starting from 
$t = 0$ (dashed line) toward the interval $t=[90,100]$ (solid line, \texttt{model2})
and $t=[140,150]$ (solid line, \texttt{model1}), respectively. 
The dotted lines refer to time averages taken over $t = [0,10], \ [10,20], \ \cdots, 
[80,90] ([130,140])$ orbits.}
\label{figure:5}
\end{figure*}
%
\begin{figure*}[ht]
\centering
\includegraphics[width=.45\textwidth]{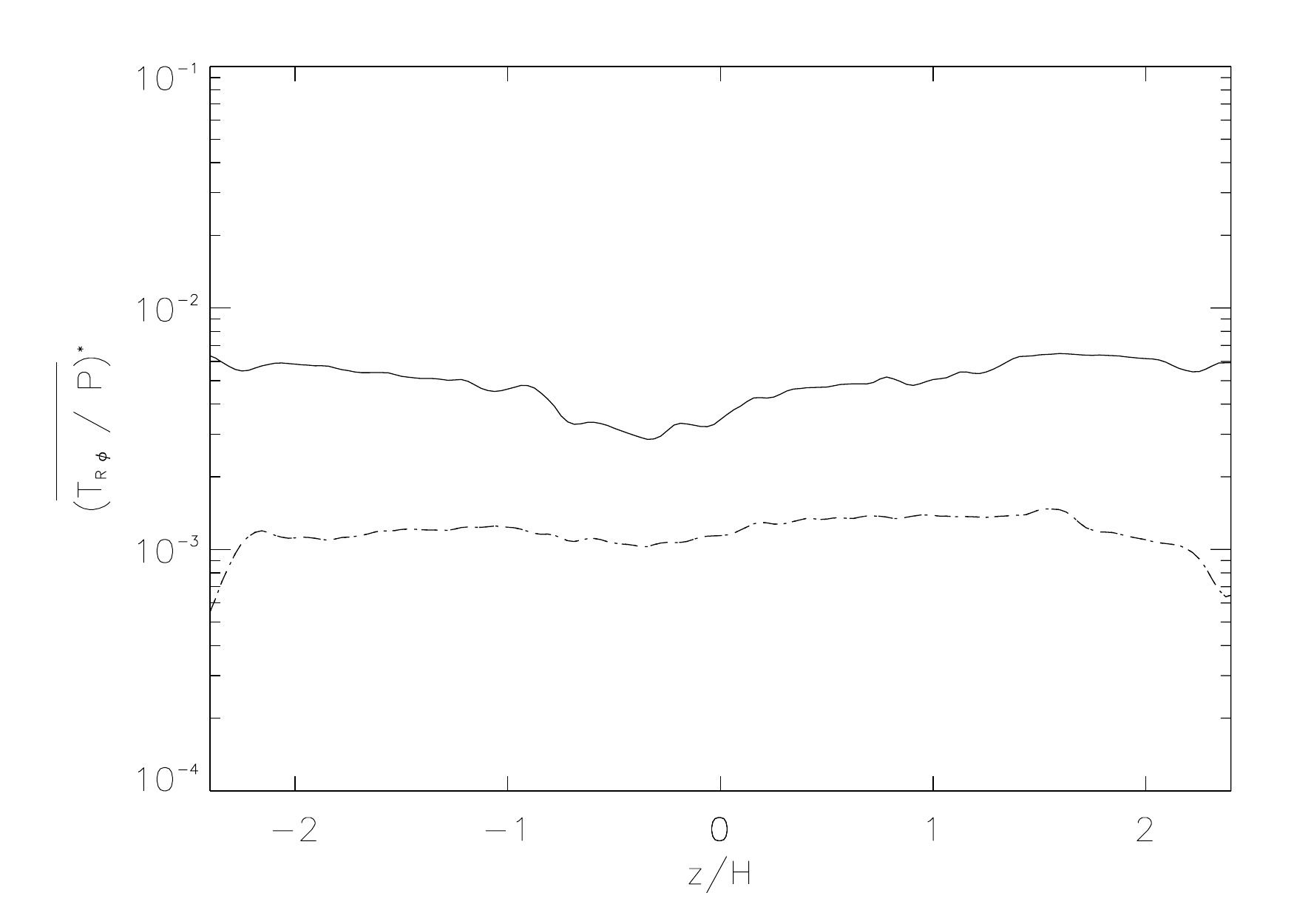}
\includegraphics[width=.45\textwidth]{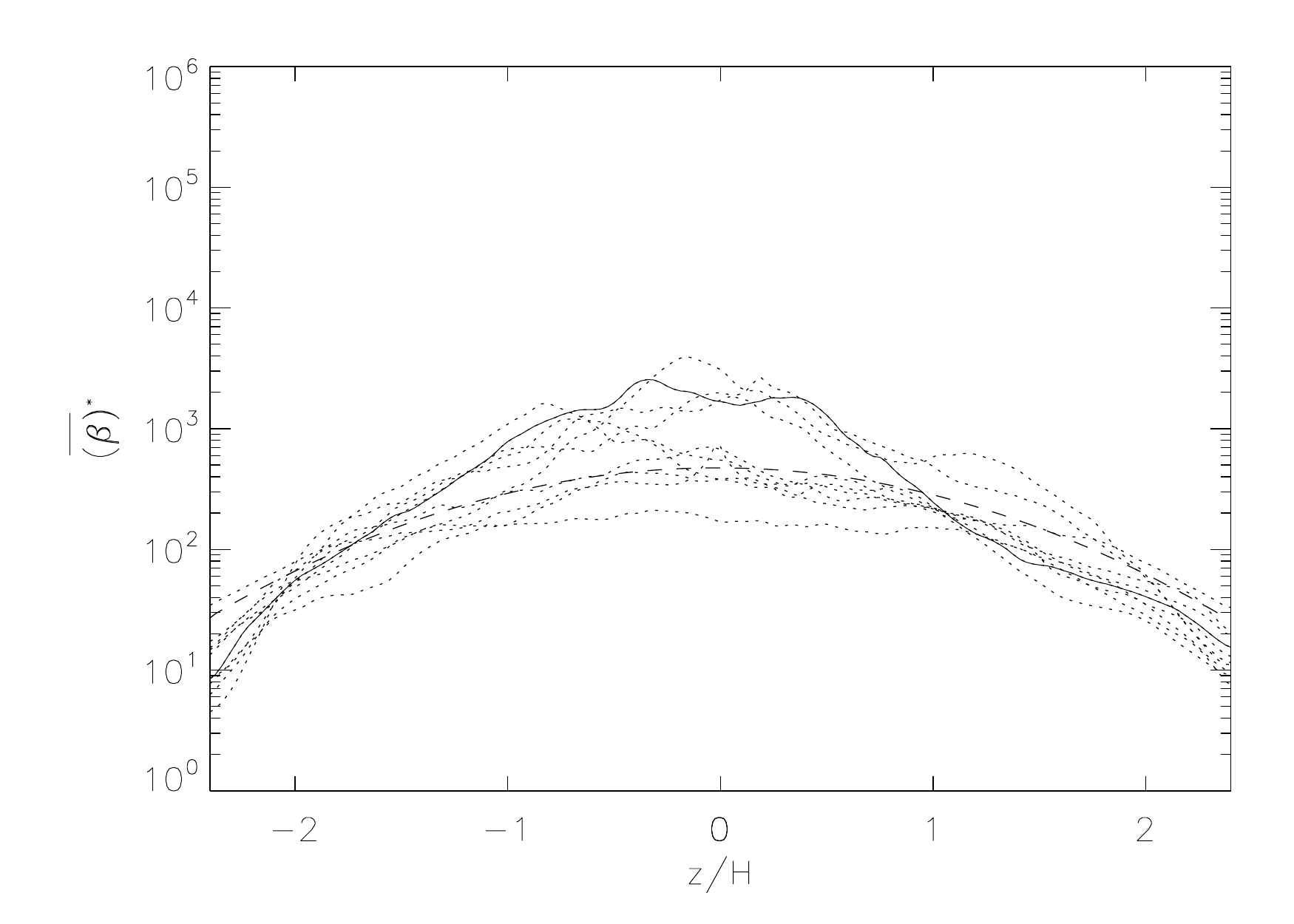}
\includegraphics[width=.45\textwidth]{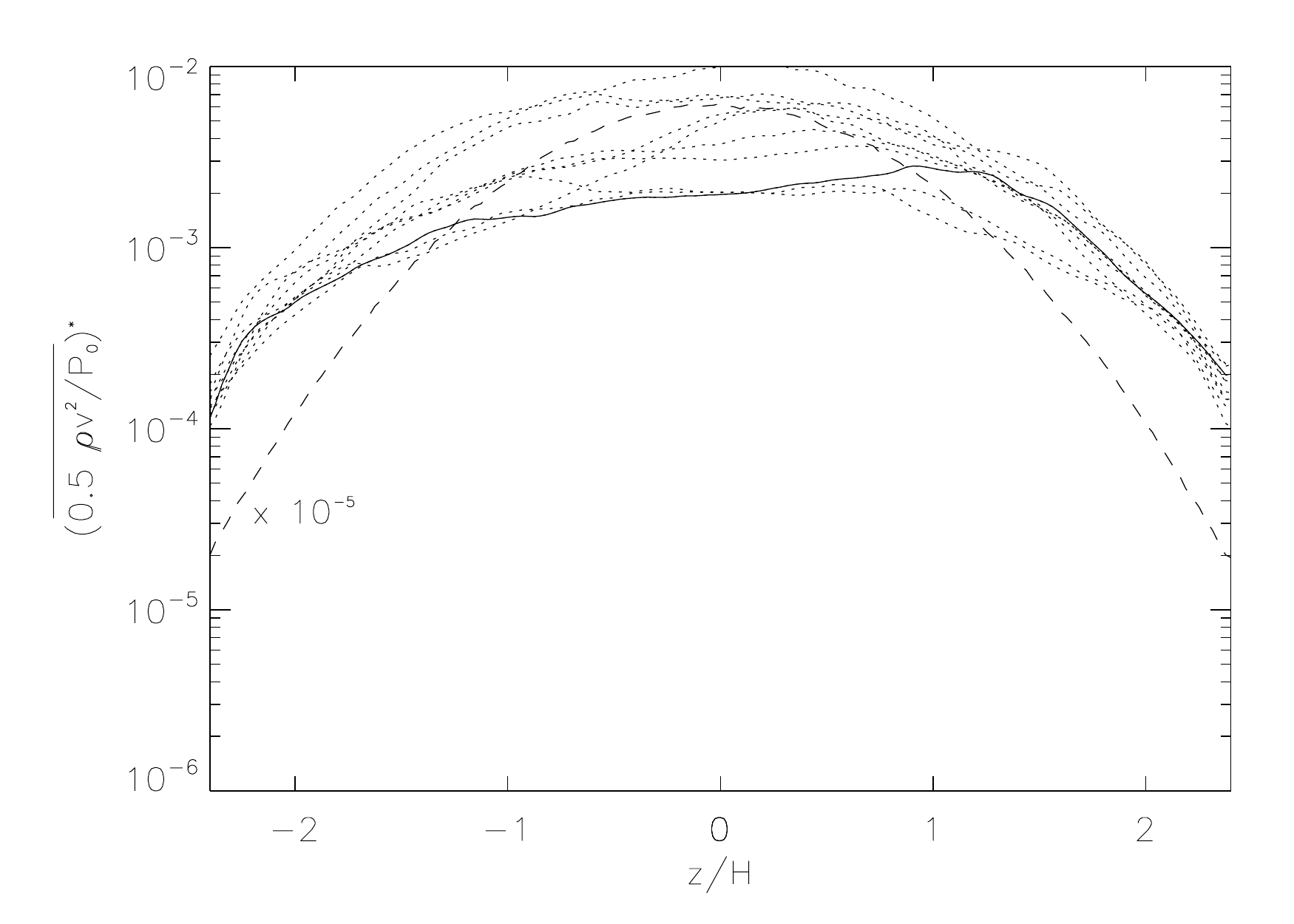}
\includegraphics[width=.45\textwidth]{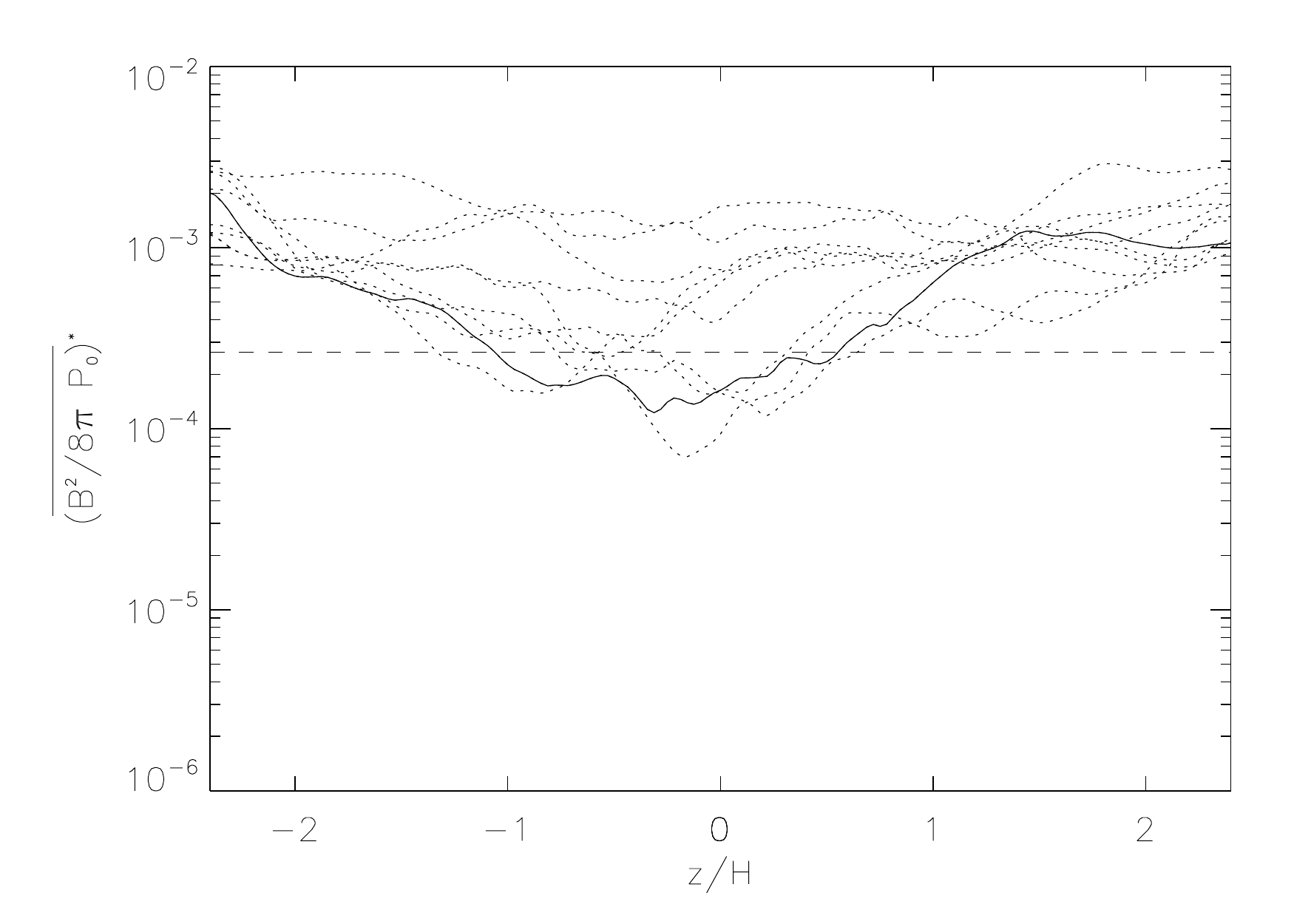}
\caption{\texttt{model3}: - Time averaged vertical profiles of the horizontally averaged value of 
normalised Maxwell and Reynolds stresses $T_{R\Phi}^{}/P_0^{}$, the plasma parameter 
$\beta$, the kinetic energy $0.5 \varrho v^2 /P_0$, and the magnetic energy $B^2 / (8\pi P_0)$.  
As in the previous figures, the shearing box model is calculated for $R =  10 \ \rm AU$ and  
$x_{\rm Mg}^{} = 5 \times 10^{-11}$. $\alpha_{\rm Max}^{}$ (solid line) and $\alpha_{\rm Rey}^{}$ 
(dashed - dotted line) refer to a time average taken over [0,100] orbits. For all the other quantities, 
the time average is taken over 10 orbit intervals, starting from $t = 0$ (dashed line) towards $[90,100]$ 
(solid line). The dotted lines refer to time averages taken over the intervals 
$t = [0,10], \ [10,20], \ \cdots, [80,90]$ orbits.}
\label{figure:6}
\end{figure*}
In this subsection we present a suite of models which demonstrate that 
turbulent mixing and continuing chemical evolution of the gas are able 
to enliven a dead zone. All three models presented in this subsection 
correspond to a radial location in the disc $R=10$ AU, and have a gas--phase 
magnesium abundance $x_{\rm Mg}^{} = 5 \times 10^{-11}$. Because of 
the identical initial conditions used for these three models, we can identify 
the specific effects of the chemistry on the evolution of the MRI .\\[1em]
\noindent
{\bf \underline{\texttt{model1}}}\\[.5em]
\noindent
The resistivity in this model is calculated from the equilibrium electron 
abundance predicted by the chemical model presented in Sect.~\ref{basic-eqns}, 
and is held constant throughout the simulation. The variation of magnetic 
Reynolds number with height is shown in Fig.~\ref{figure:2}, which shows 
that the resistivity profile is intermediate between that applied by Fleming \& 
Stone (2003) in their ``small dead zone'' model, and the model F56 presented 
by Turner et al. (2007). Specifically $\rm Re_m$ takes the following values: 
$1.4 \times 10^3$ at the disc midplane; $6.0 \times 10^3$ at $z/H=1$;
$1.4 \times 10^4$ at $z/H=2$; $1.3 \times 10^5$ at $z/H=3$.

In agreeement with our expectations, this simulation resulted in
a disc with well-defined dead and active zones, with the boundary
between these occuring at $z/H \approx 1$ where $\rm Re_m \simeq 6 \times 10^3$
and $\rm Lu \simeq 1$. This dead zone is larger than that obtained by 
Fleming \& Stone (2003) in their model whose resistivity profile is shown 
in Fig.~\ref{figure:2}, for which the dead zone was confined to $z/H \le 0.4$.

The time and volume averaged sum of the Maxwell and Reynolds 
stresses was found to be $\overline{\alpha_{}^{\ast\ast}} = 4.89 \times 10^{-3}$
when the time average was taken over the interval [20,100] orbits. When 
taken over an interval [100,200] orbits, the stresses were found to decrease 
to $\overline{\alpha_{}^{\ast\ast}} = 8.49 \times 10^{-4}$. This occurs
because the stresses are higher during the development of the non linear 
stage of evolution early on in the simulation. We note that a fully active disc 
is expected to have a value $\overline{\alpha_{}^{\ast\ast}} \ga 10^{-2}$ (see later).

The vertical profiles for the horizontally averaged density $\varrho$, the $\alpha$ 
values associated with the Maxwell and Reynolds stresses, the plasma parameter 
$\beta$, the kinetic and magnetic energy are shown in the left hand panels of 
Figs.~\ref{figure:4} and~\ref{figure:5}. The time averages are taken over 10 orbit 
intervals, starting from $t =0$ (dashed line) towards $[140,150]$ (solid line). The 
dotted lines refer to profiles averaged over $t = [0,10], [10,20], \ \cdots [130,140]$. 
For each of the profiles, the same qualitative behaviour reported in Fleming \& Stone 
(2003) is observed: \\
(i) The vertical density profile remains unchanged throughout the nonlinear evolution 
of the MRI.\\
(ii) A significant decline in the magnetic field energy towards the disc midplane is observed 
as compared to surface regions; $\beta$ at $z/H = 0$ is between 2 and 3 orders of magnitude 
greater than in the active zones.\\
(iii) Dominance of the Reynolds stress over the Maxwell stress is observed in 
dead zones, due to the penetration of sound waves emitted in the overlying active 
zones, while the Maxwell stress is the dominant mechanism by which transport in 
active zones occurs. The transition occurs at $z/H \simeq 1$ where the cross-over 
between active and dead zones occurs.\\[.5em]
{\bf \underline{\texttt{model2}}}\\[.5em]
In this model the initial free electron abundance and resistivity profile is calculated 
from the equilibrium solution to Eqns.~(\ref{chem1}) -- (\ref{chem4}). The full set of 
multifluid equations, and the chemical network, are evolved together so that the local 
resistivity can change through turbulent transport of ions and chemical evolution 
(recombination/ionisation) of the gas. For this model we estimate that the turbulent 
mixing time corresponding to $\alpha \simeq 0.01$ is shorter than the recombination
time, such that chemical mixing should enliven the dead zone. This is indeed what 
we find, as the simulation results in a turbulent flow which fills the entire volume of 
the disc, with $\overline{\alpha_{}^{\ast\ast}} = 1.31 \times 10^{-2}$, where the time 
average was performed in the interval [20,100] orbits. We plot the vertical profiles of 
the various physical quantities that we have already described for \texttt{model1} in 
Figs.~\ref{figure:4} and \ref{figure:5}. Comparing the figures for \texttt{model1} and 
\texttt{model2} we can make the following observations: \\
(i) The mean vertical density profile remains approximately constant in both models. \\
(ii) Whereas the Reynolds stress dominates near the midplane in \texttt{model1} and 
the Maxwell stress dominates in the disc surface layers, we see that the Maxwell stress 
is dominant throughout in \texttt{model2}. \\
(iii) The plasma $\beta$ parameter is found to become very high ($\beta \simeq 10^5$)
in the midplane of the disc in \texttt{model1} as the magnetic field strength there 
becomes very low, whereas it remains in the range $10 \le \beta \le 10^3$ for $|z/H| \le  2$ 
for \texttt{model2}, indicative of a fully active disc. \\
(iv) The kinetic energy throughout the disc, but especially near the midplane, is much 
higher in \texttt{model2} than in \texttt{model1} as the turbulent velocity field is driven 
by the MRI. \\
(v) The magnetic energy near the disc midplane for \texttt{model2} is more than two 
orders of magnitude greater than for \texttt{model1} due to the continuing dynamo action 
associated with the MRI.\\
We conclude that \texttt{model2} shows unambiguously that the dead zone in the disc 
can be removed by turbulent mixing and continuing chemical evolution under 
circumstances where the local mixing time scale is smaller than the recombination time, 
in basic agreement with the prediction of the reaction--diffusion model presented in 
paper II.

\noindent
{\bf \underline{\texttt{model3}}}\\[.5em]
At $t = 0$ the initial resistivity profile is set up in the same way as described for 
\texttt{model1} and \texttt{model2}. For $t > 0$, the local resistivity is updated after 
every MHD time step because of the transport of free electrons and ions only. 
Due to the inhibition of recombination and ionisation in this model, free electrons 
diffuse and cause the resistivity to become homogeneous on long time scales. In 
the presence of sufficient numbers of free electrons in the initial ionisation state of 
the disc, we expect that mixing will lead to a fully active disc, and indeed this is 
what we find. Instead of the two--layer structure obtained using \texttt{model1} above, 
with dead and active zones, the MHD turbulence now fills the full vertical extent of 
the disc. For $t > 40$ orbits, we observe a quasi steady state characterised by small 
fluctuations around the mean value $\overline{\alpha_{}^{\ast\ast}} = 5.75 \times 10^{-3}$, 
where the time average was taken in the interval [20,100] orbits. The vertical profiles 
of various quantities are shown in Fig.~\ref{figure:6}. Compared with the results obtained 
for \texttt{model1}, we see that the volume averaged turbulent stresses in each case 
are very similar, even though \texttt{model1} had a dead zone. The reason for this is 
that the resistivity in \texttt{model3} is higher near the disc surface because of mixing, 
and so reduces the strength of the turbulence there. The subsequent enlivening of the 
midplane in \texttt{model3} does not lead to a substantial increase in the overall stress 
because the now--uniform resistivity is sufficient to damp the strength of the turbulence 
compared to its state in an ideal MHD calculation. 

Comparing \texttt{model3} with \texttt{model2} we see that
higher stresses and more vigorous turbulence are generated by \texttt{model2}.
This is because \texttt{model3}
generates a disc without a dead zone, but in which the
resisivity is higher than for \texttt{model2}, such that the
strength of the resulting turbulence is suppressed somewhat.
The results for \texttt{model3} show that mixing the initial 
free electron population
throughout the disc can cause the dead zone to disappear, but
that allowing the chemical evolution of the disc to continue
during the turbulent mixing leads to a more active disc. This
is because the continuing ionisation of species near the disc
surface, followed by mixing toward the midplane, produces
a higher ionisation fraction overall.
\begin{figure*}[ht]
\centering
\includegraphics[width=.78\textwidth]{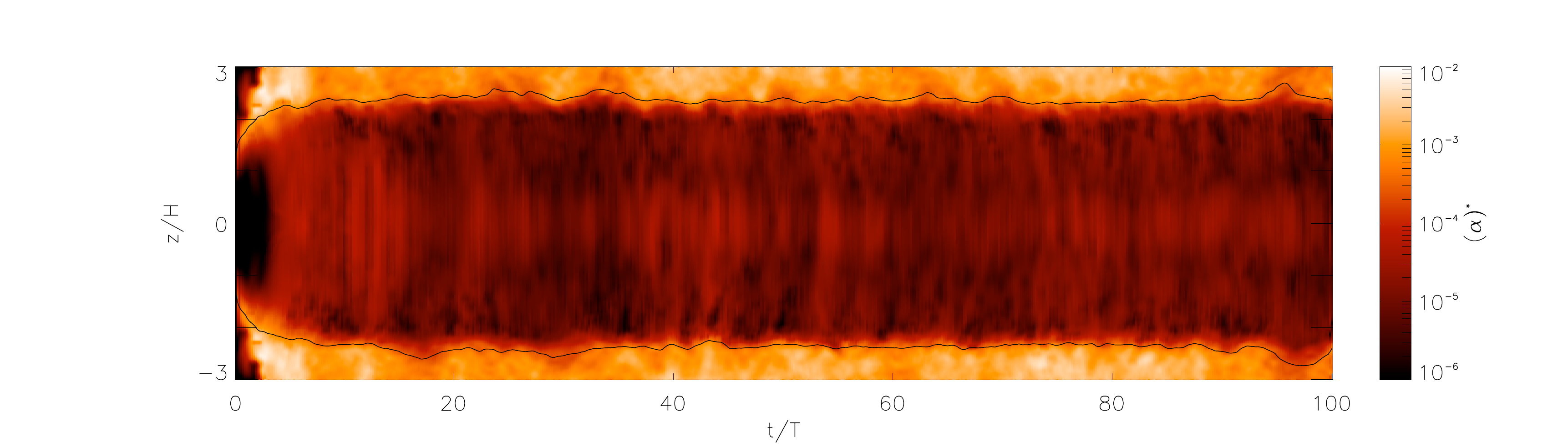}
\includegraphics[width=.78\textwidth]{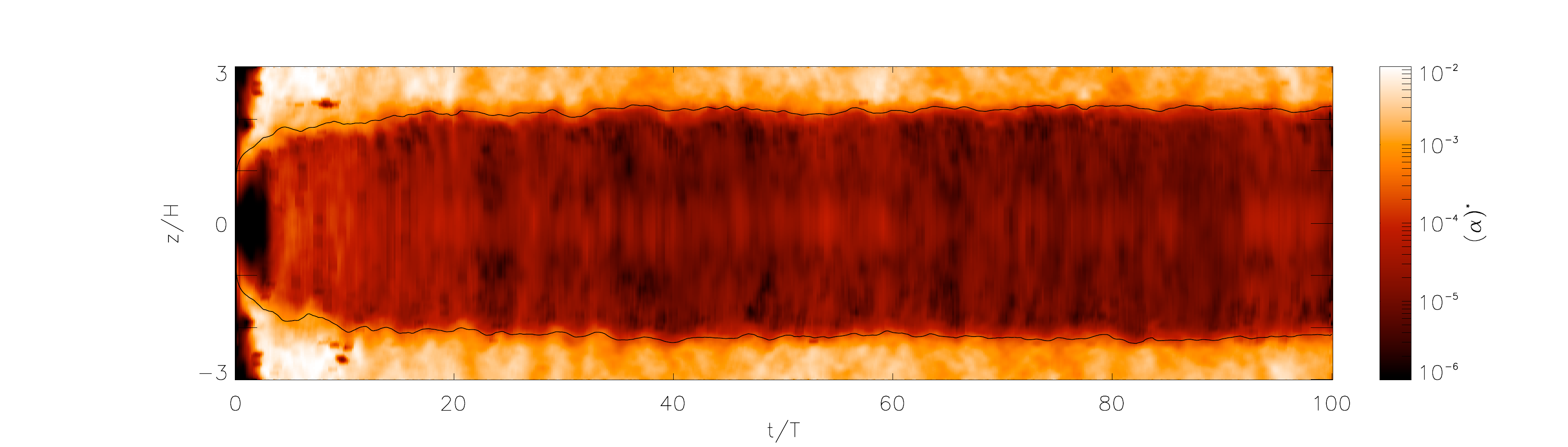}
\includegraphics[width=.78\textwidth]{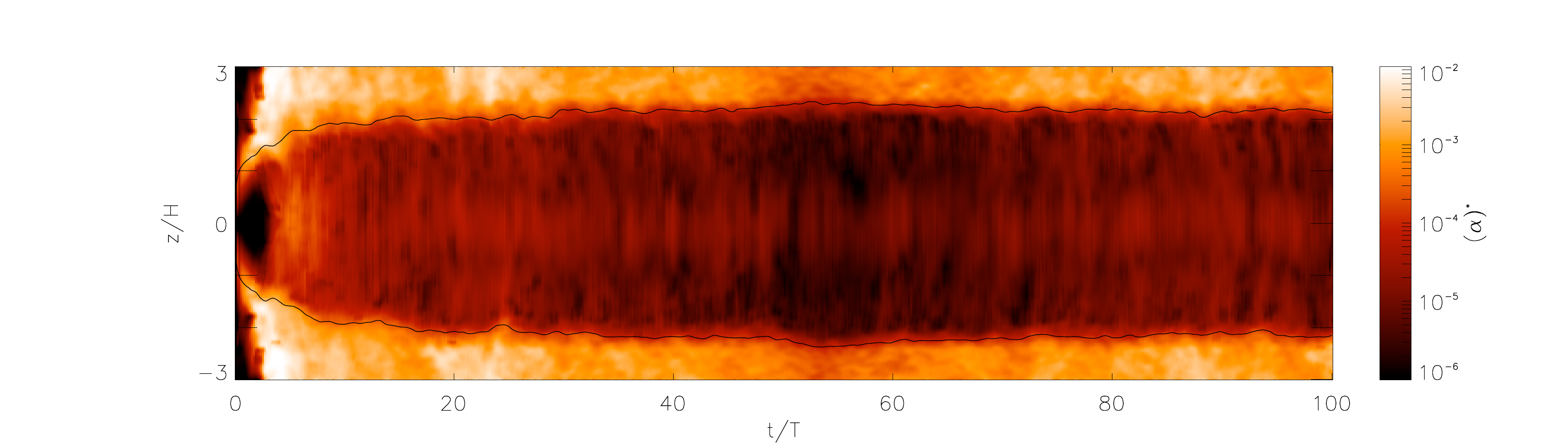}
\includegraphics[width=.78\textwidth]{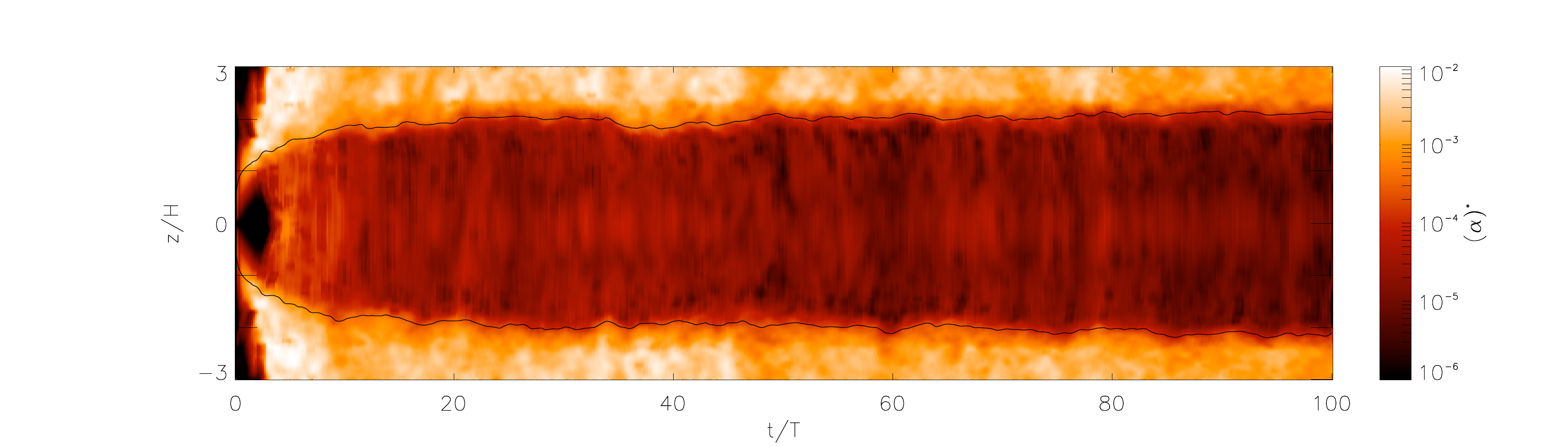}
\includegraphics[width=.78\textwidth]{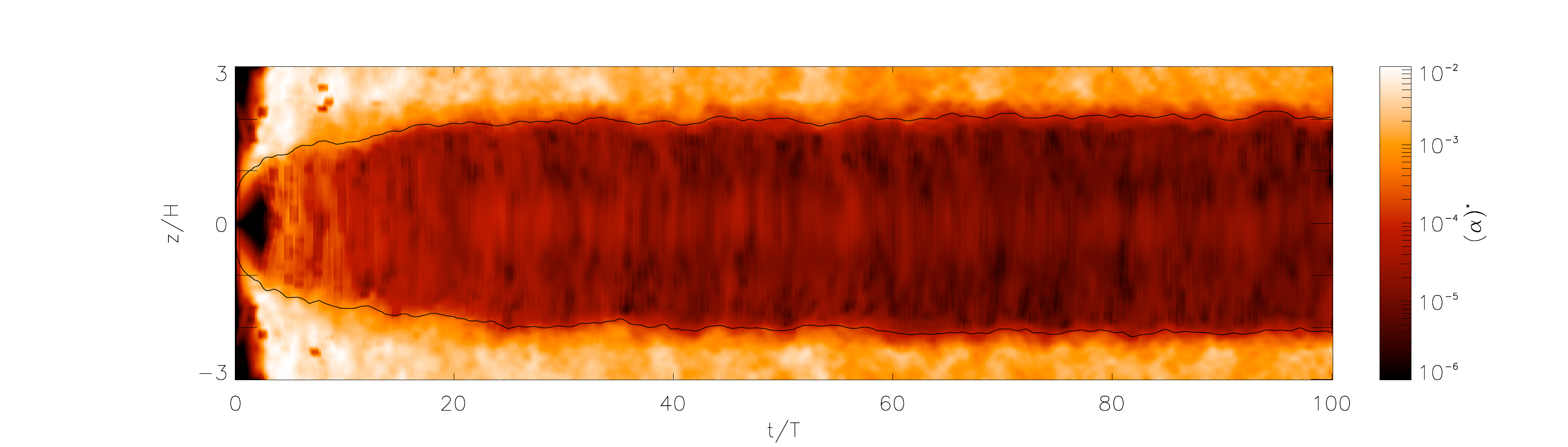}
\caption{\texttt{model2}: - Space-time plots of the horizontally averaged value of $\alpha$ 
calculated at different radial positions $R$ and a metal elemental abundance of
$x_{\rm Mg}^{} = 0$. The cylindrical radius $R$ varies from 1 AU (top panel) to 10 AU 
(bottom panel) passing through $R \rm =3$, 5, and \ 7 \ AU. The solid line drawn at the 
boundary between the live and dead zones indicates the position where the horizontally 
averaged Lundquist number $\rm Lu=1$.}
\label{figure:7}
\end{figure*}
%
\begin{figure*}[ht]
\centering
\includegraphics[width=.78\textwidth]{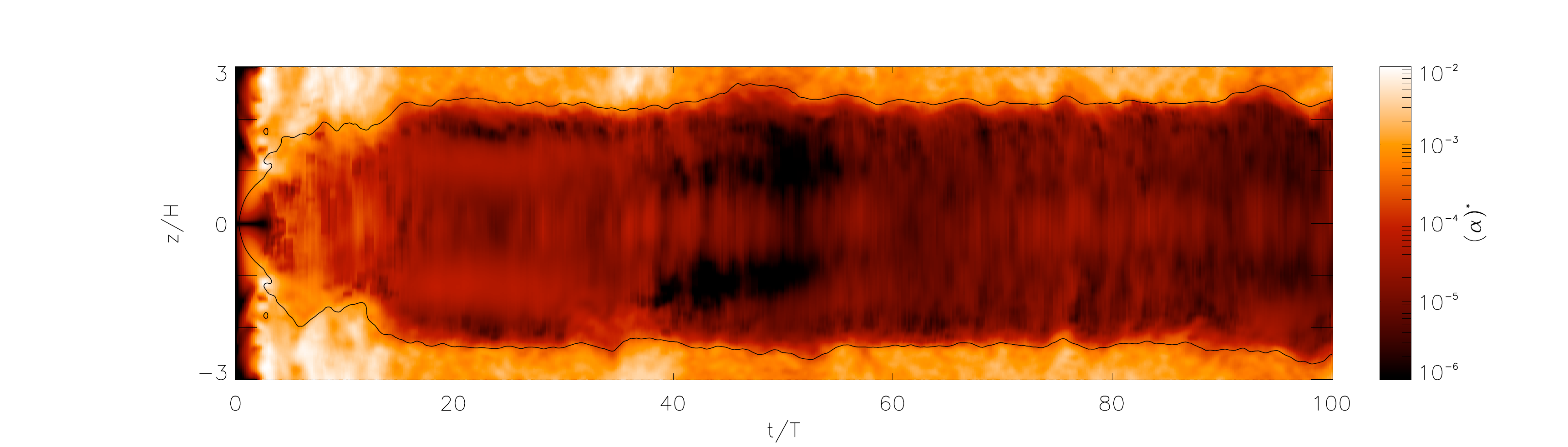}
\includegraphics[width=.78\textwidth]{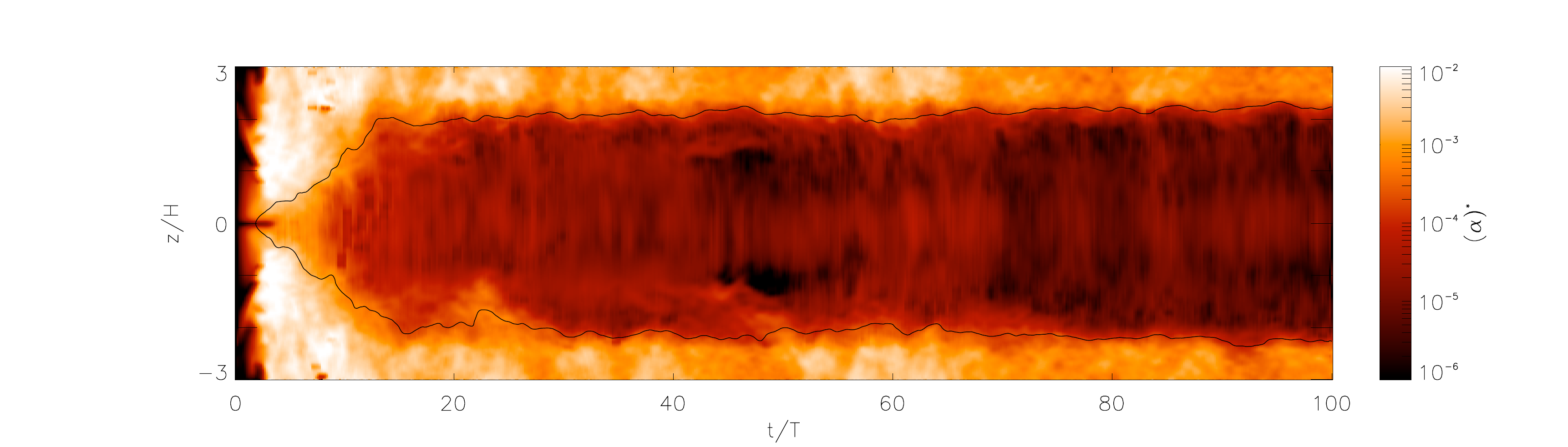}
\includegraphics[width=.78\textwidth]{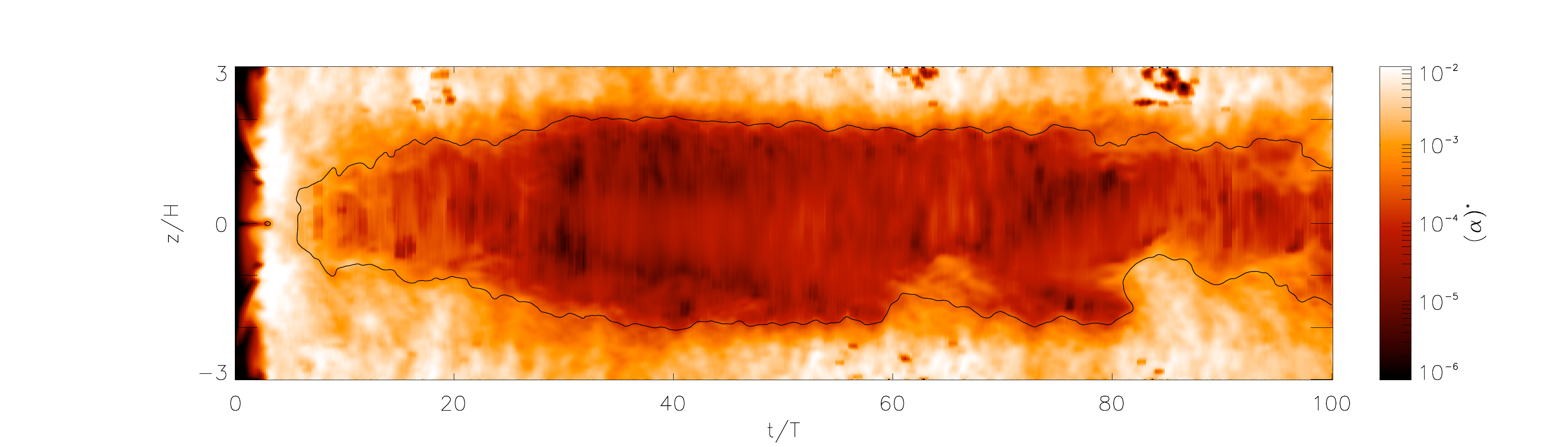}
\includegraphics[width=.78\textwidth]{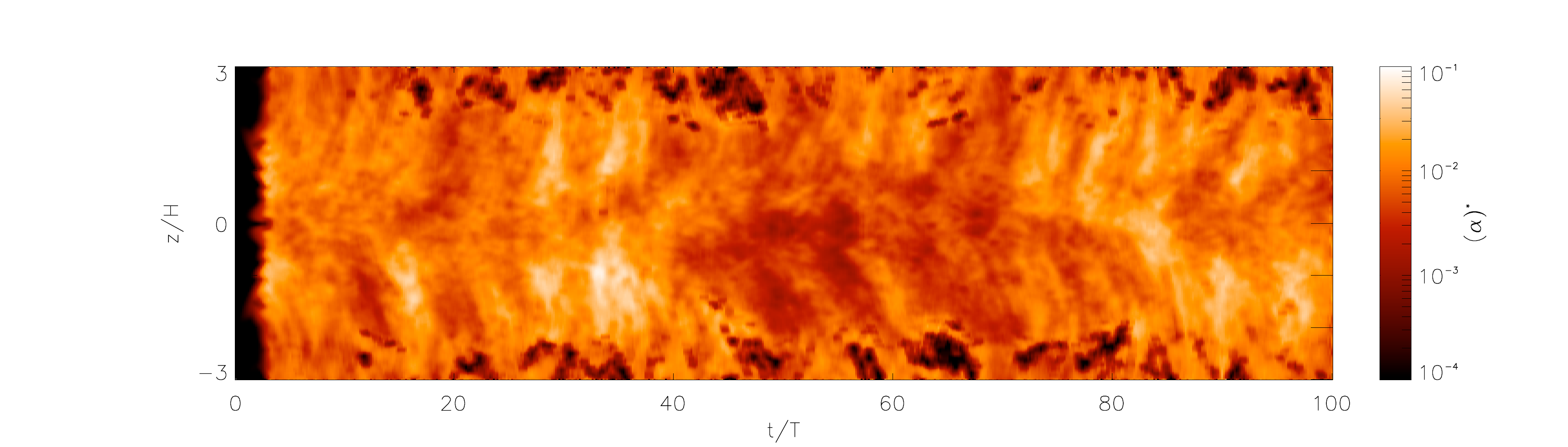}
\includegraphics[width=.78\textwidth]{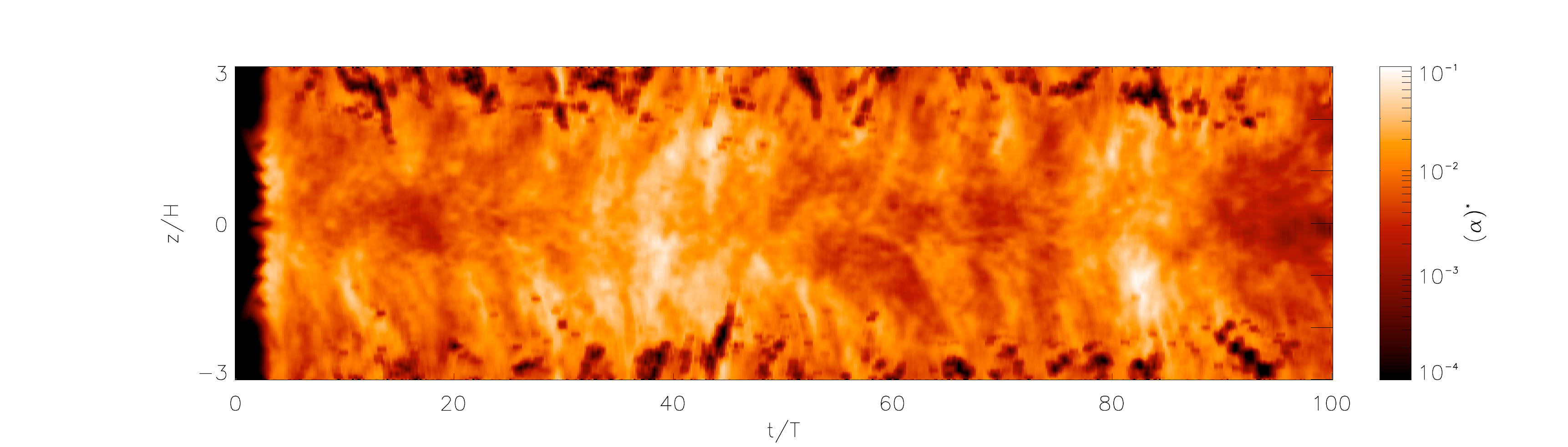}
\caption{\texttt{model2}: - Space-time plots of the horizontally averaged value of $\alpha$ 
calculated at different radial positions $R$ and a metal elemental abundance of
$x_{\rm Mg}^{} = 5\times 10^{-11}$. The cylindrical radius $R$ varies from 1 AU (top panel) 
to 10 AU (bottom panel) passing through $R \rm =3$, 5, and \ 7 \ AU. The solid line drawn 
at the boundary between the live and dead zones indicates the position where the horizontally 
averaged Lundquist number $\rm Lu=1$.}
\label{figure:8}
\end{figure*}
%
\subsection{Results as function of $x_{\rm Mg}$ and orbital radius}
\label{Mg_and_rad}
A primary motivation for this paper was the indication in paper II that dead zones 
could be enlivened by a combination of turbulent mixing and sufficient abundance 
of gas--phase magnesium atoms. In that paper we presented calculations of the 
ionisation fraction in standard $\alpha$-discs using reaction--diffusion models. The 
main results were that turbulent mixing could only change the structure of a dead 
zone if:
(i) the abundance of magnesium was sufficient (so as to increase the recombination time);
(ii) one was considering locations further out in the disc where the lower temperatures 
and densities increase the recombination time relative to the local turbulent transport 
time.

The purpose of the simulations presented in the following subsections are to examine
how turbulent mixing affects dead zone structure as a function of magnesium 
abundance and radial position in the disc, as a test of the predictions contained
in paper II. We also perform a detailed comparison between some of our MHD 
simulations and the predictions of the reaction--diffusion model. These simulations 
solve the full set of multifluid equations in combination with the chemical model 
described in Sect.~\ref{basic-eqns}. We first present shearing box simulations of discs 
at various locations between 1 and 10 AU with gas--phase magnesium abundance 
equal to zero, before considering a similar set of models with gas--phase magnesium 
abundance $x_{\rm Mg}=5 \times 10^{-11}$, which is about $10^{-6}$ of the solar value.

\subsubsection{Models with $x_{\rm Mg}=0$}
We begin our discussion by first examining the dead zone structure after saturation 
obtained for \texttt{model2} with $x_{\rm Mg}=0$. We examine the disc evolution at
radii $R=$ 1, 3, 5, 7 and 10 AU. In basic agreement with the results obtained in paper 
II, the dead zone structures obtained when magnesium is absent are very similar for 
all radii considered. In Fig.~\ref{figure:7} we present space-time plots of the horizontally 
averaged value of $\alpha$, and it is clear that the disc sustains a two--zone structure 
for all radii and all time, which consists of a large dead zone which extends from the 
midplane to $|z/H | \simeq 2$ where the magnetic Reynolds number 
$\rm Re_m \ga 4000$ and the Lundquist number $\rm Lu \ga 1$ (shown by the solid 
lines in Fig.~\ref{figure:7}). Across the region bounded by $2 \le |z/H| \le 2.3$ we find 
that the horizontally averaged $\alpha$ varies by more than two orders of magnitude, 
and this is maintained for the duration of the simulation (100 orbits).

\subsubsection{Models with $x_{\rm Mg}=5 \times 10^{-11}$}
We now consider the dead zone structures obtained for a magnesium abundance of 
$x_{\rm Mg}^{} = 5 \times 10^{-11}$  at radii $R=$ 1, 3, 5, 7, and 10 AU. Space-time 
plots of the hoizontally averaged value of $\alpha$ are shown in Fig.~\ref{figure:8}. 
At $R = 1 \ \rm AU$ and $3 \ \rm AU$, the location of the boundary separating the 
dead from the active layer matches very well the structure obtained for 
$x_{\rm Mg}^{} = 0$, since the height at which $\rm Re_m$ begins to exceed 4000 is 
$|z/H| \simeq 2$ (which is also the height at which the Lundquist number $\rm Lu > 1$,
shown by the solid lines in Fig.~\ref{figure:8}). Significant changes in the dead zone 
structure start to become evident at $R \ge 5 \ \rm AU$. At this radius the recombination 
time is similar to the turbulent mixing time, allowing the resistivity to be reduced there 
through the transport of free electrons into the dead zone. Over longer time scales we 
see that the dead zone that is established early on in the simulation starts to diminish, 
and between 80--100 orbits there is evidence that the dead zone size has decreased 
significantly. Nonetheless, at 100 orbits we find that there remains a region in the vicinity 
of the midplane that retains an average values of $\alpha \simeq 10^{-4}$.

By contrast, the space-time plots of the horizontally averaged value of $\alpha$ at 
$R = 7$ and $10 \ \rm AU$ show that the two--layer structure has completely 
disappeared, because mixing reduces the resistivity, and hence increases the 
magnetic Reynolds number to values $\ga 4000$. Turbulence now fills the entire 
computational domain which is confirmed by the time and volume averaged values 
of $\overline{\alpha_{}^{\ast\ast}}$:  $1.07 \times 10^{-2}$ for $R = 7 \ \rm AU$ and 
$1.31 \times 10^{-2}$ for $R = 10 \ \rm AU$. 

Our results indicate that dead zones can be reduced or removed altogether through 
turbulent mixing. The criteria for sustaining MHD turbulence have already been 
discussed in the context of reaction--diffusion models in paper II, and also apply to 
the shearing box models considered here. These criteria are:\\
({\it i}) 
There are sufficient metal atoms available in the gas phase so that recombination 
between metal ions and electrons becomes the dominant process by which the local 
ionisation fraction is determined. \\
({\it ii}) The turbulent mixing time scale is shorter than the dominant recombination 
time on which free electrons are removed.\\[1em]

\noindent
We now compare in detail the results of some of our MHD simulations with
those obtained using the reaction--diffusion model presented in paper II and 
discussed in Sect.~\ref{reaction-diffusion}.
\noindent
\subsection{Comparing shearing box simulations and reaction--diffusion models}
\label{r-d-results}
Encouraged by the good qualitative agreement obtained between the MHD 
simulations and the results presented in paper II, we now examine in detail 
the level of agreement between the simulations and the reaction--diffusion 
model. As mentioned in Sect.~\ref{reaction-diffusion}, we assume that the 
diffusion coefficient, $\cal D$, which governs the rate at which chemical 
species mix vertically in the disc, is equal to an effective kinematic viscosity 
generated by the turbulence $\nu_m = \alpha_m c_s^2/\Omega$, where 
$\alpha_m$ refers to a dimensionless parameter that measures the rate of 
vertical mixing (not to be confused with the value of $\alpha$ associated 
with the radial transport of angular momentum). When solving the 
reaction--diffusion equations we use a range of $\alpha_m$ values to 
obtain different solutions, which we then compare with the results of the 
MHD simulations. Previous work on the vertical mixing of dust grains and 
molecules (Carballido et al. 2005; Johansen \& Klahr 2005; Turner et al. 2006; 
Fromang \& Papaloizou 2006) suggests that the ratio of the rate of angular 
momentum transport to the rate of transport of chemical species by the 
turbulence should lie in the range $\nu/\nu_m \simeq 1$--3, and we examine 
how well our best--fit value of $\nu_m$ agrees with this expectation. 
Furthermore, the results of Turner et al. (2006) and Fromang \& Papaloizou (2006) 
show that the diffusion coefficient is not constant with height above the midplane, 
but increases with height because the turbulent velocities increase in proportion 
to the Alfv\'en speed. Although we consider only a constant value of ${\cal D}$ for
each reaction--diffusion model that we run, we examine the quality of the best--fit 
that we obtain, and quantify this by stating the error obtained in predicting the 
magnetic Reynolds number (which is a proxy for the free--electron abundance)
at the disc midplane and surface. In general we find that using a uniform diffusion 
coefficient leads to a slight overestimate of the mixing rate near the midplane, 
and a slight underestimate near the surface.

Overall we find good agreement between the reaction--diffusion model and 
the MHD simulation when the value of $\alpha_m$ adopted in the former is 
between a factor of 1--2 times lowerer than the time and volume averaged 
value of $\alpha$ obtained in the latter. The $\alpha$ values corresponding 
to each MHD simulation are listed in table~\ref{table:3}. The best--fit values 
of $\alpha_m$ are listed in table~\ref{table:4}, along with the Schmidt number 
which measures $\alpha/ \alpha_m$.
\begin{figure*}[ht]
\centering
\includegraphics[width=.5\textwidth]{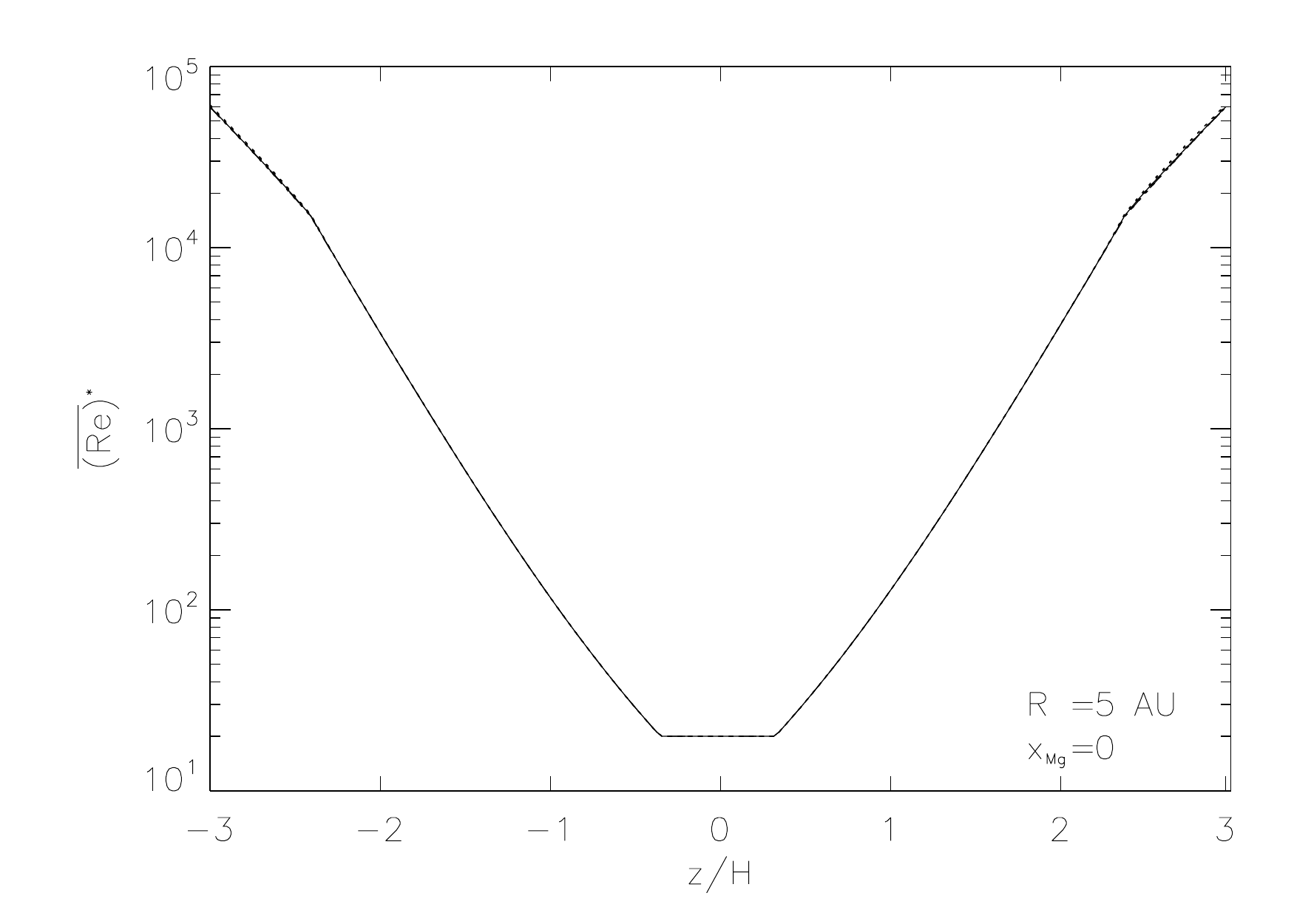} 
\includegraphics[width=.3\textwidth]{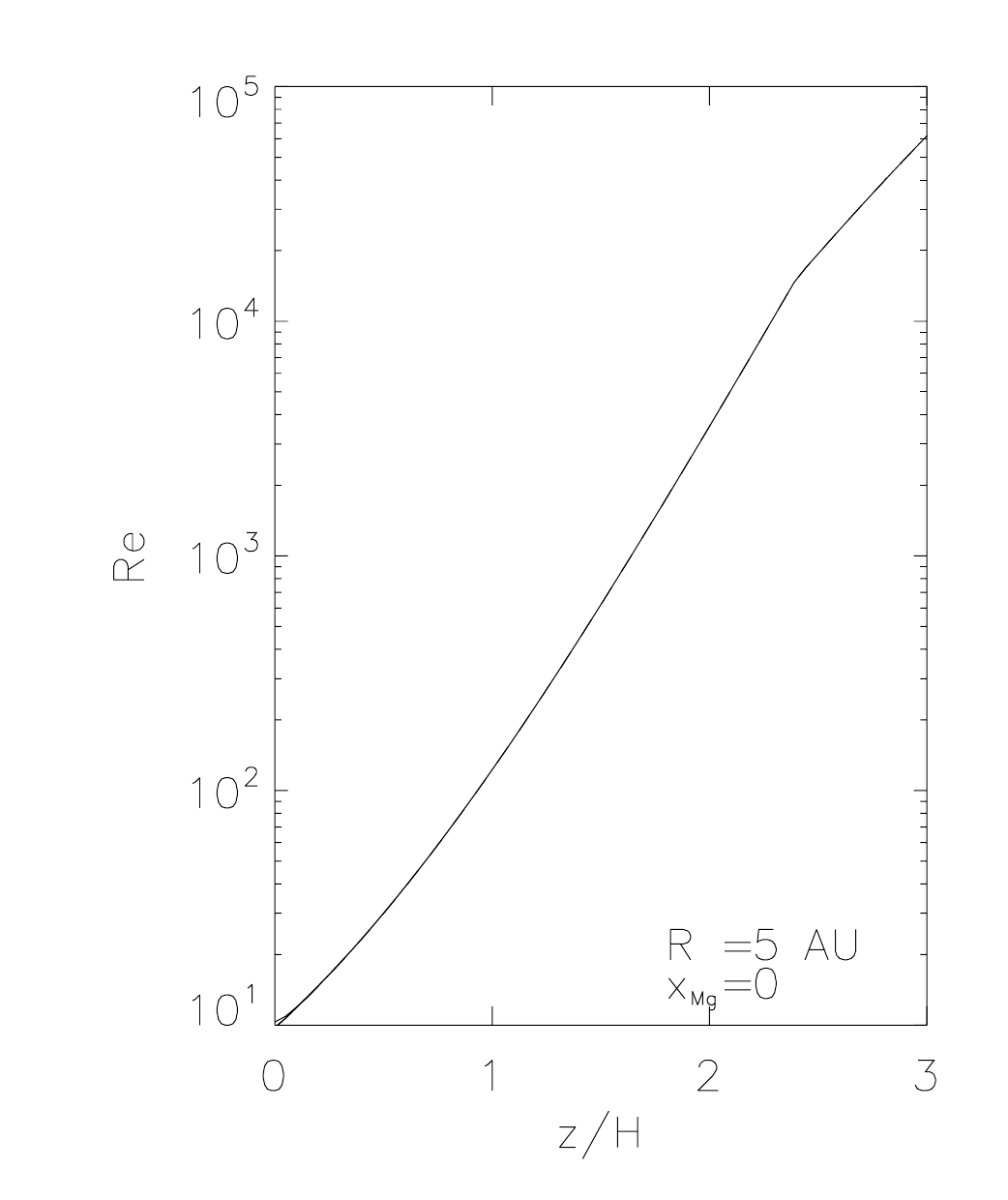}\\
\includegraphics[width=.5\textwidth]{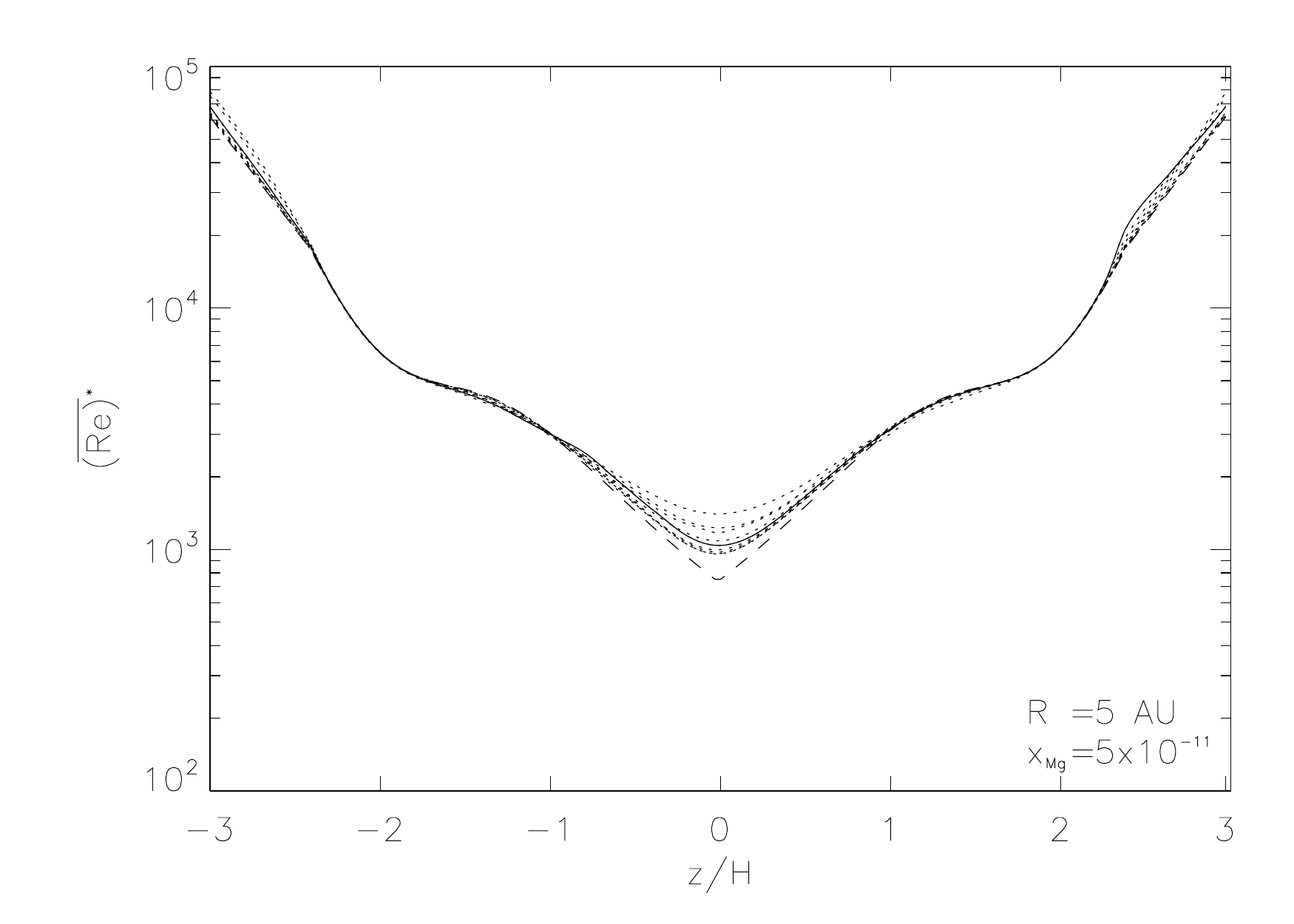}
\includegraphics[width=.3\textwidth]{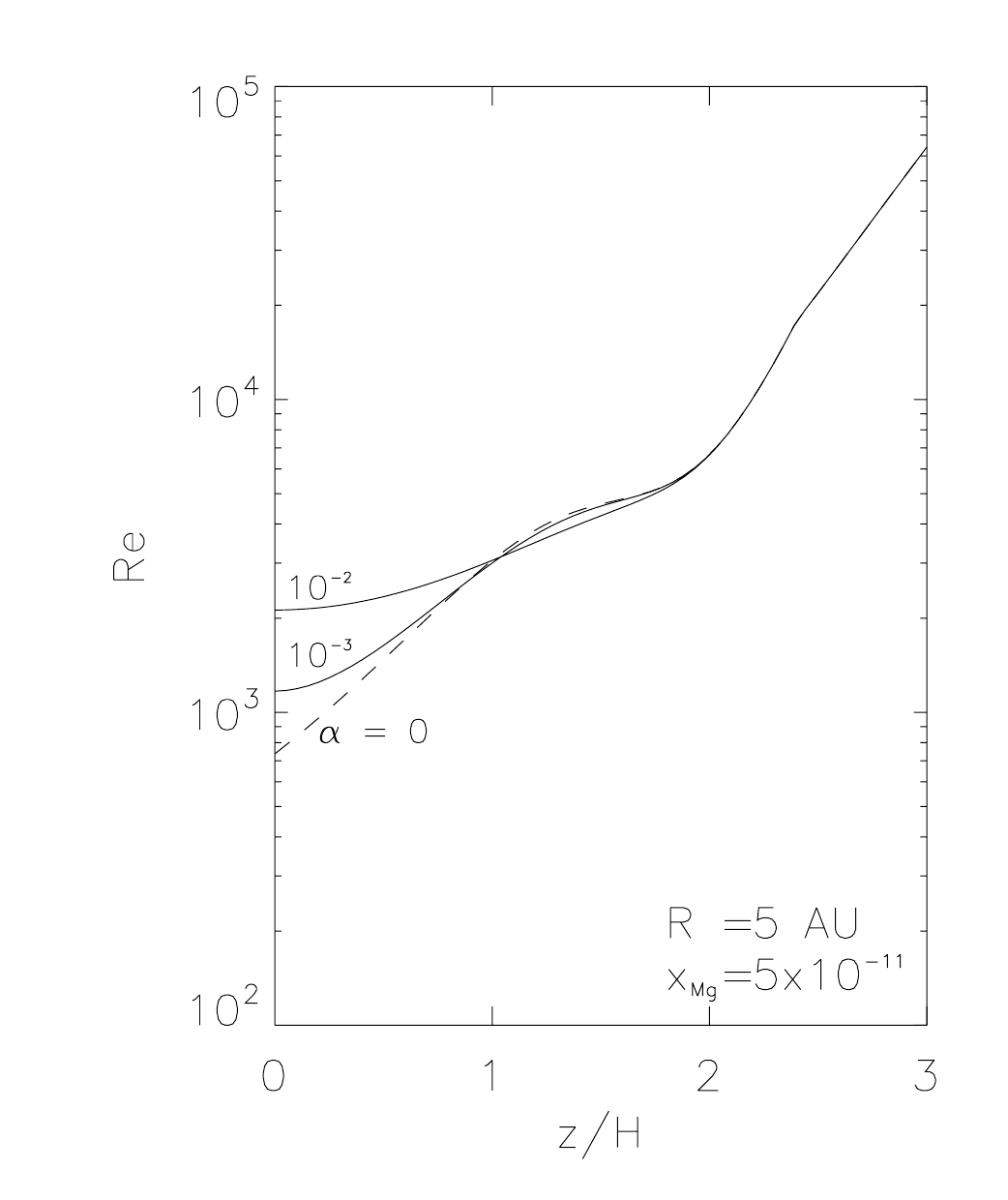}
\caption{\texttt{model2}: -  Comparison between vertical profiles of the magnetic 
Reynolds number obtained for $R \rm = 5 \ AU$ and $x_{\rm Mg}^{} = 0 $ 
and $x_{\rm Mg}^{} = 5 \times 10^{-11}$, respectively. Left panels show 
shearing box simulation results, and right panels reaction--diffusion model 
results. In the left panels, the time averaged vertical profiles of the magnetic 
Reynolds number $\overline{\rm Re^{\ast}}$ are shown by taking the time 
average over 10 orbit intervals, starting from $t = 0$ (dashed line) towards 
$[90,100]$ (solid line). The other profiles (dotted lines) refer to the intervals
$t = [0,10], \ [10,20], \ \cdots [80,90]$ orbits. The resistivity profiles of the 
magnetic Reynolds number $\rm Re_m$ shown in the right panels refer to 
the equilibrium profiles obtained with the reaction--diffusion model assuming 
different values of $\alpha_m$. In particular for $x_{\rm Mg} = 0$, the value 
of the magnetic Reynolds number is not affected by the actual value of 
$\alpha_m$, leading to identical profiles for $\alpha_m = 0$, 
$\alpha_m = 10^{-3}$, and $\alpha_m = 10^{-2}$. Note that both shearing 
box simulations and the reaction--diffusion models are initiated with the steady 
state profile obtained for $\alpha_m = 0$.}
\label{figure:9}
\end{figure*}
%

We have compared the results of models located at disc radii
$R=5$, 7 and 10 AU, and with magnesium abundances $x_{\rm Mg}=0$ 
and $x_{\rm Mg}=5 \times 10^{-11}$. In each case we examine
how the vertical profile of the magnetic Reynolds number $\rm Re_m$ 
evolves with time, and in the case of the reaction--diffusion model
we calculate the equilibrium value of $\rm Re_m$ from the underlying
ionisation fraction. In each case we initiate the calculation
assuming that the initial chemical abundance profile is equal
to the equilibrium state in the absence of mixing.
For each disc radius, we plot the profile of
the evolving $\rm Re_m$ below. In the left panels we present 
the results from the MHD simulations,
and in the right panels the equilibrium profile from
the corresponding reaction--diffusion model assuming different values of
$\alpha_m$. 
The values of $\rm Re_m$ plotted for shearing box simulations are horizontal
and time averages, where each time average was performed over 10 orbit
intervals. We plot the initial value of $\rm Re_m$ using the dashed line,
and subsequent values are plotted using dotted lines starting at
$t=[0,10]$ and moving up to $t=[80,90]$. The final values at $t=[90,100]$
are shown using the solid line.
\begin{figure*}[ht]
\centering
\includegraphics[width=.5\textwidth]{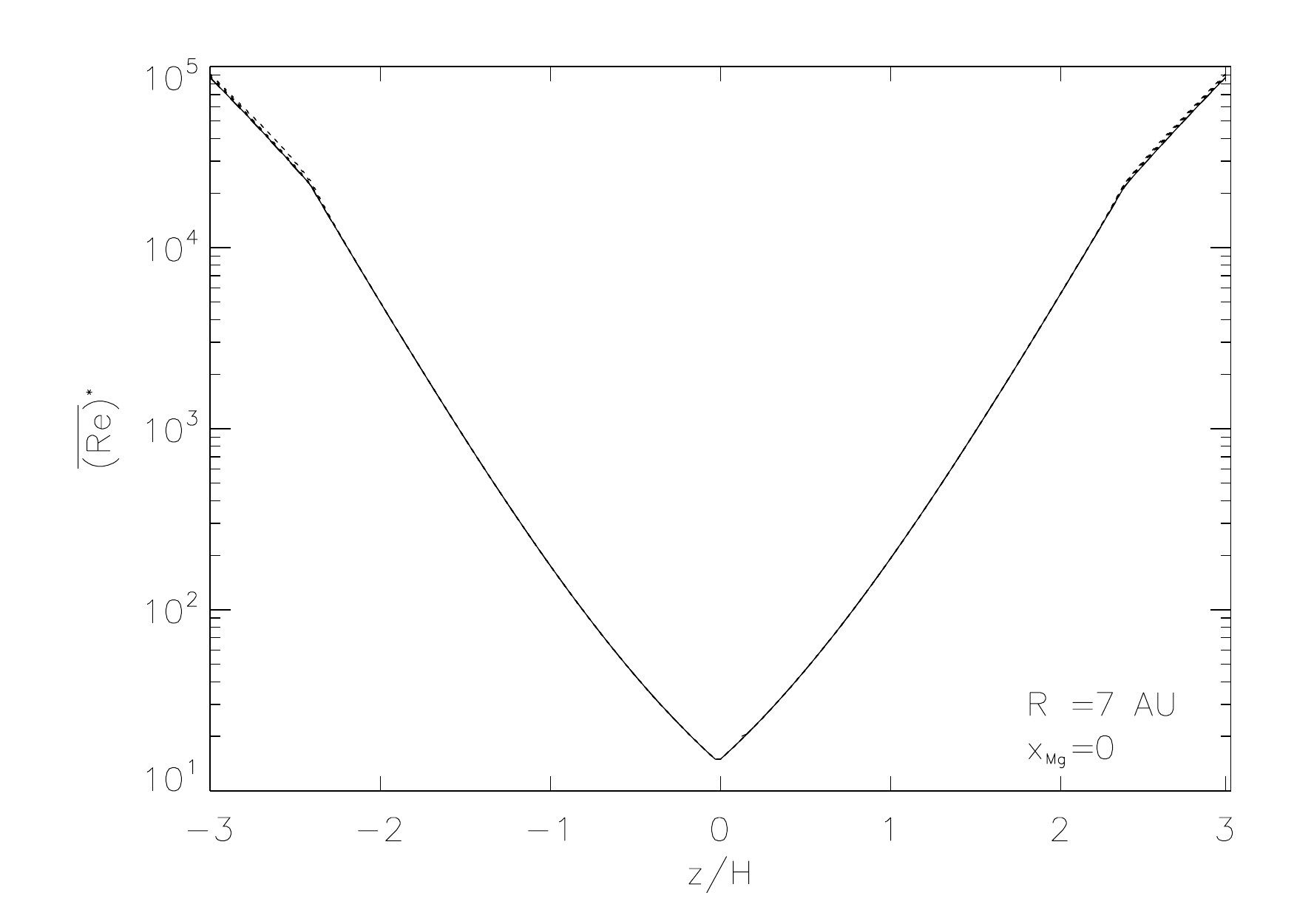} 
\includegraphics[width=.3\textwidth]{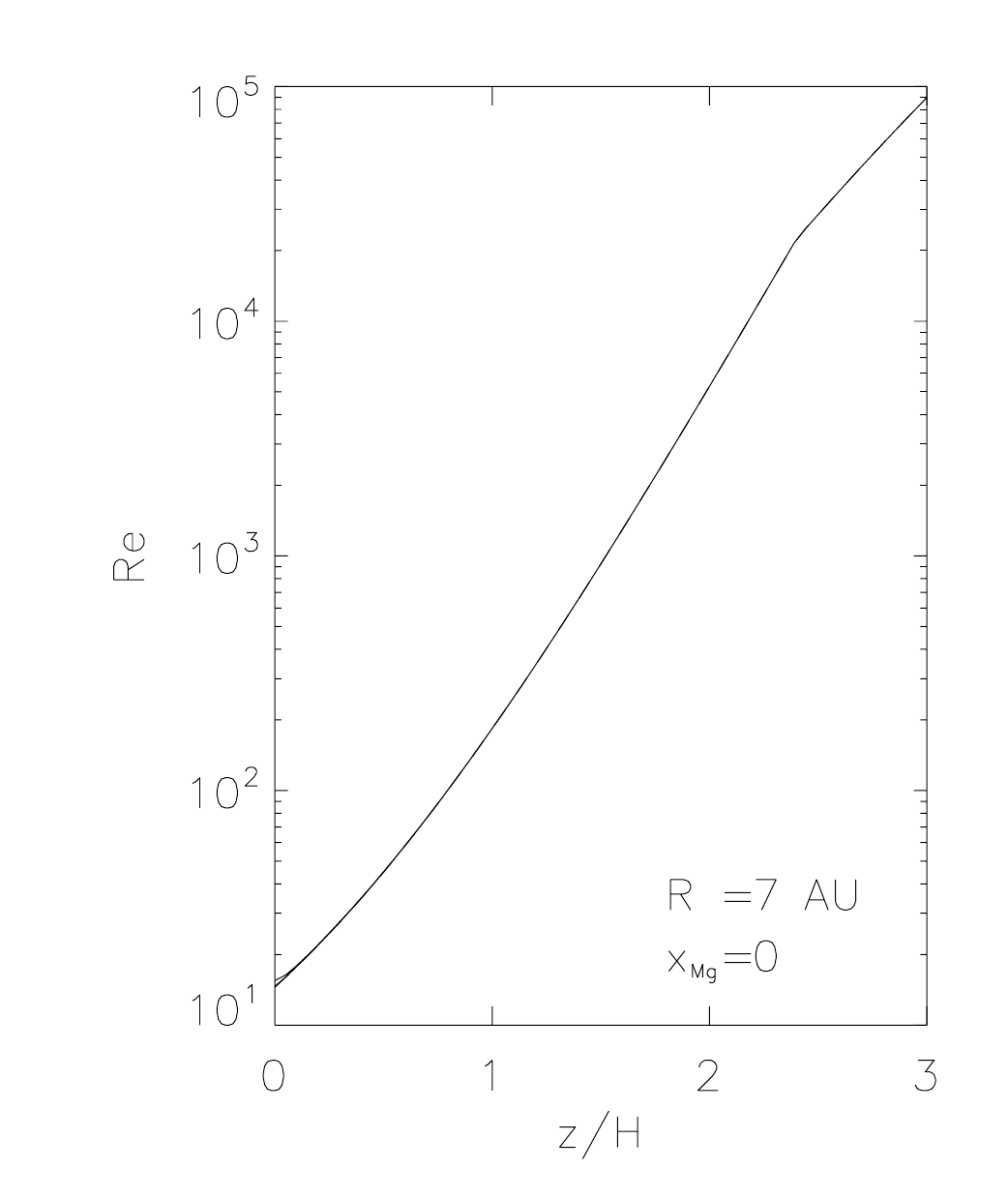}\\
\includegraphics[width=.5\textwidth]{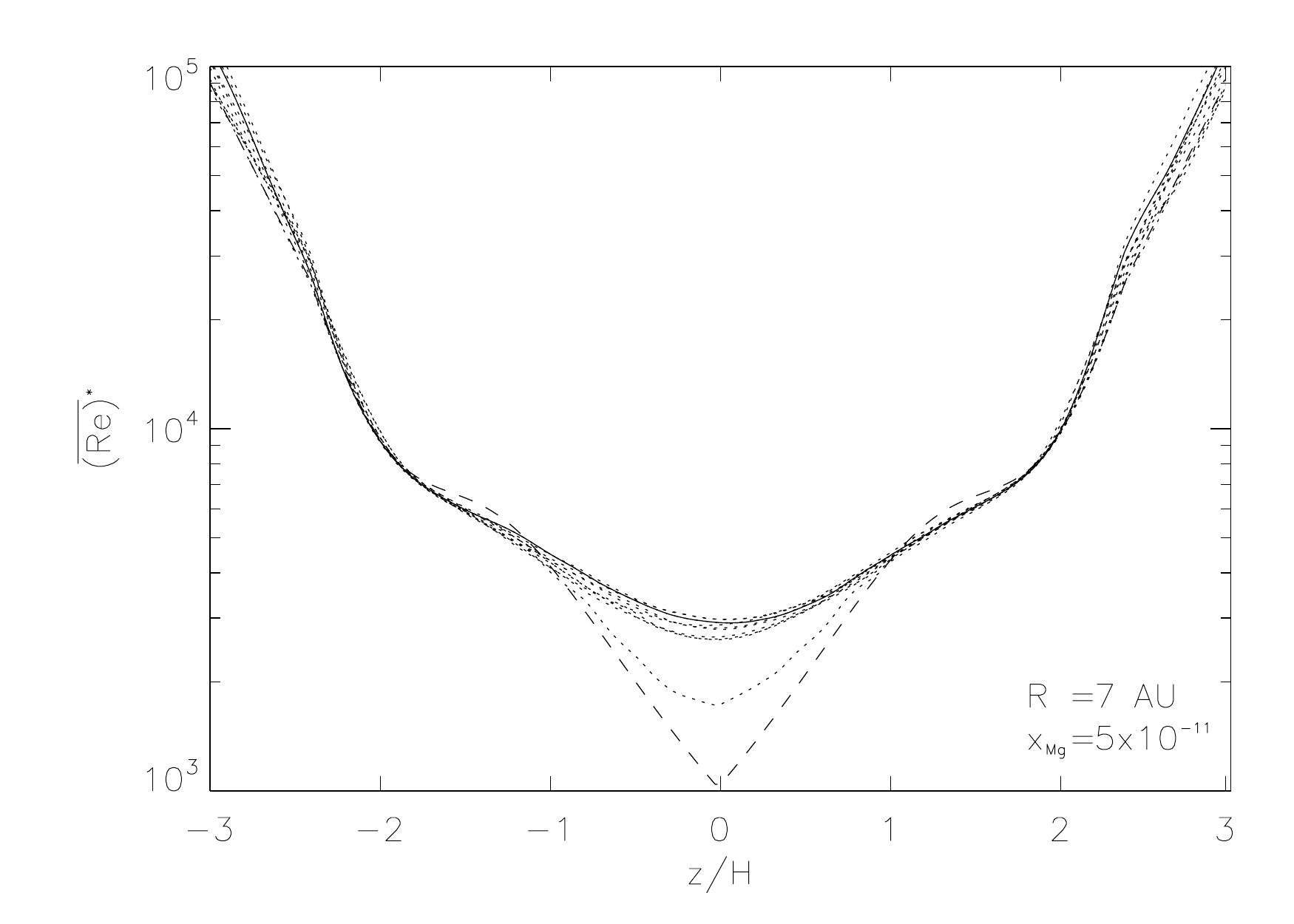} 
\includegraphics[width=.3\textwidth]{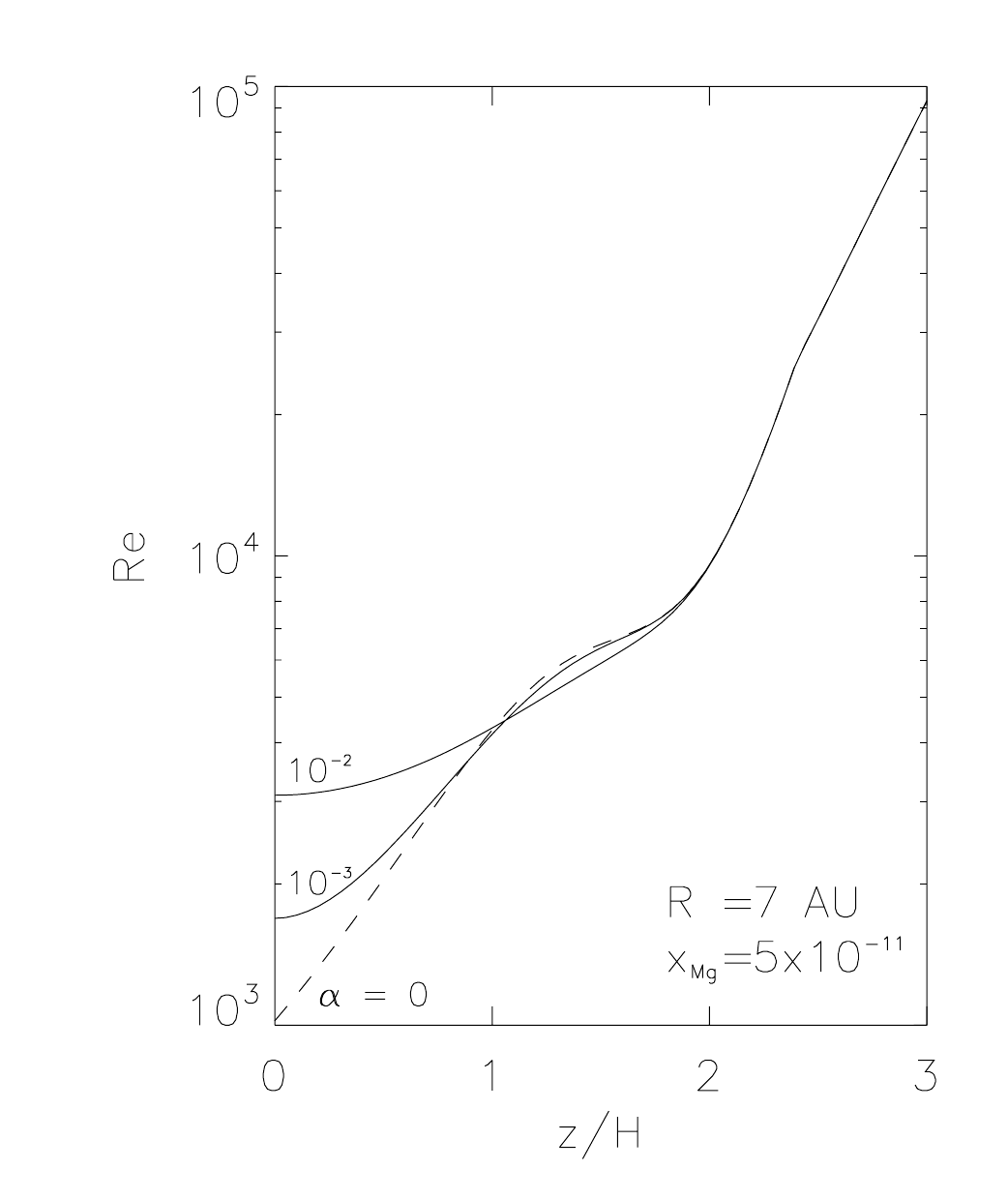}
\caption{\texttt{model2}: -  Comparison between vertical profiles of the magnetic
Reynolds number obtained for $R \rm = 7 \ AU$ and $x_{\rm Mg}^{} = 0 $
and $x_{\rm Mg}^{} = 5 \times 10^{-11}$, respectively. Left panels show shearing 
box simulation results, and right panels reaction--diffusion model results. In the left 
panels, the time averaged vertical profiles of the magnetic Reynolds number
$\overline{\rm Re^{\ast}}$ are shown by taking the time average over 10 orbit 
intervals, starting from $t = 0$ (dashed line) towards $[90,100]$ (solid line). The 
other profiles (dotted lines) refer to the intervals $t = [0,10], \ [10,20], \ \cdots [80,90]$ 
orbits. The resistivity profiles of the magnetic Reynolds number $\rm Re_m$ shown 
in the right panels refer to the equilibrium profiles obtained with the reaction--diffusion
model assuming different values of $\alpha_m$. In particular for $x_{\rm Mg} = 0$,
the value of the magnetic Reynolds number is not affected by the actual value of
$\alpha_m$, leading to identical profiles for $\alpha_m = 0$, $\alpha_m = 10^{-3}$, 
and $\alpha_m = 10^{-2}$. Note that both shearing box simulations and the reaction--diffusion 
models are initiated with the steady state profile obtained for $\alpha_m = 0$.}
\label{figure:10}
\end{figure*}
\begin{table}[h]
\centering
\caption{Time and volume averaged values of
$\overline{\alpha^{\ast\ast}_{}}$
obtained for the models described in the paper. The first column gives the 
model label and the cylindrical radius considered. The values of 
$\overline{\alpha^{\ast\ast}_{}}$ are listed in the $2^{\rm nd}$ and 
$3^{\rm rd}$ column assuming a elemental metal abundance $x_{\rm Mg}$ of 0 
and $5 \times 10^{-11}$, respectively. Apart from the value denoted with 
the symbol \dag, the time average was taken over [20,100] orbits, while
()$^{\dag}$ refers to the time average taken over [100,200] orbits.}
\begin{tabular}{lcc} \hline \hline \\[-.5em]
 & $ \overline{\alpha^{\ast\ast}_{}}(x_{\rm Mg}^{}=0)$
&$\overline{\alpha^{\ast\ast}_{}}(x_{\rm Mg}^{}=5\cdot 10^{-11})$ \\[.3em]
\hline
\texttt{model1} &  & \\
~ at $10 \ \rm AU$ & & $4.89 \cdot 10^{-3}$ \\
& & $(8.49 \cdot 10^{-4})^{\dag}$ \\ \hline
\texttt{model3} & & \\
~ at $10 \ \rm AU$ & & $5.75 \cdot 10^{-3}$ \\ \hline
\texttt{model2} & & \\
 at \ $1 \ \rm AU$ & $2.67 \cdot 10^{-4}$ & $2.51 \cdot 10^{-4}$\\
~ at \ $3 \ \rm AU$ & $3.01 \cdot 10^{-4}$ & $3.46 \cdot 10^{-4}$\\
~ at \ $5 \ \rm AU$ & $2.84 \cdot 10^{-4}$ & $1.11 \cdot 10^{-3}$\\
~ at \ $7 \ \rm AU$ & $5.44 \cdot 10^{-4}$ & $1.07 \cdot 10^{-2}$\\
~ at $10 \ \rm AU$ & $5.10 \cdot 10^{-4}$ &  $1.31 \cdot 10^{-2}$\\
\hline \hline \label{table:3}
\end{tabular}
\end{table}

\begin{table}[h]
\centering
\caption{Values of the diffusion coefficients applied to the
reaction-diffusion
model which best matches the corresponding MHD results for \texttt{model2} with
$x_{\rm Mg} = 5 \cdot 10^{-11}$. The Schmidt number $S_c$ is listed in
the 3rd column refering to time and volume averaged values of
$\overline{\alpha^{\ast\ast}}$ between [20,100] orbits.}
\begin{tabular}{rcc} \hline \hline \\[-.5em]
R [AU] & $\alpha_m$ & $S_C$ \\[.3em] \hline
 5   & $5.89 \cdot 10^{-4}$ & 1.88 \\
 7   & $8.67 \cdot 10^{-3}$ & 1.23 \\
10   & $9.28 \cdot 10^{-3}$ & 1.41 \\
\hline \hline \label{table:4}
\end{tabular}
\end{table}

\begin{figure*}[ht]
\centering
\includegraphics[width=.5\textwidth]{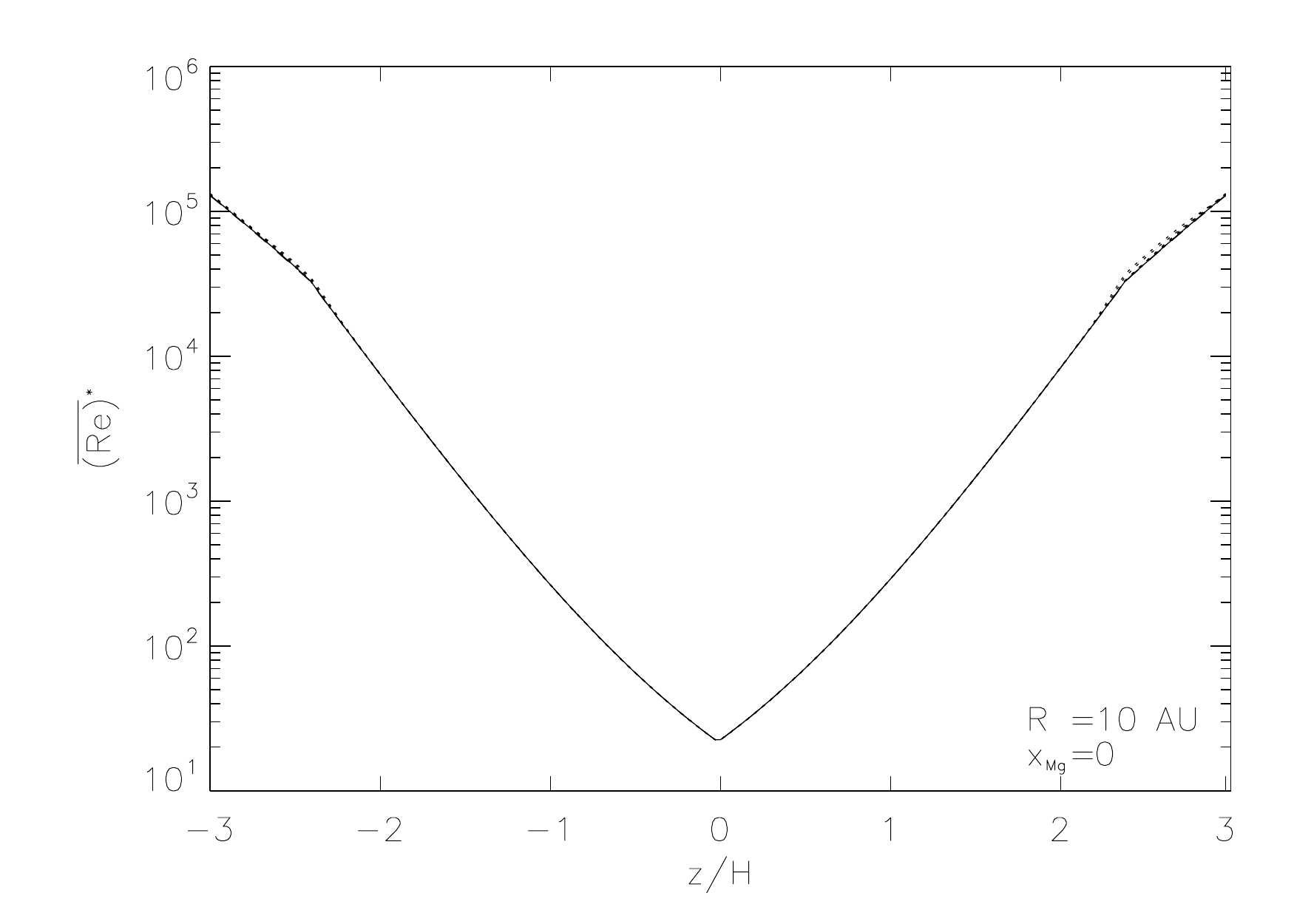} 
\includegraphics[width=.3\textwidth]{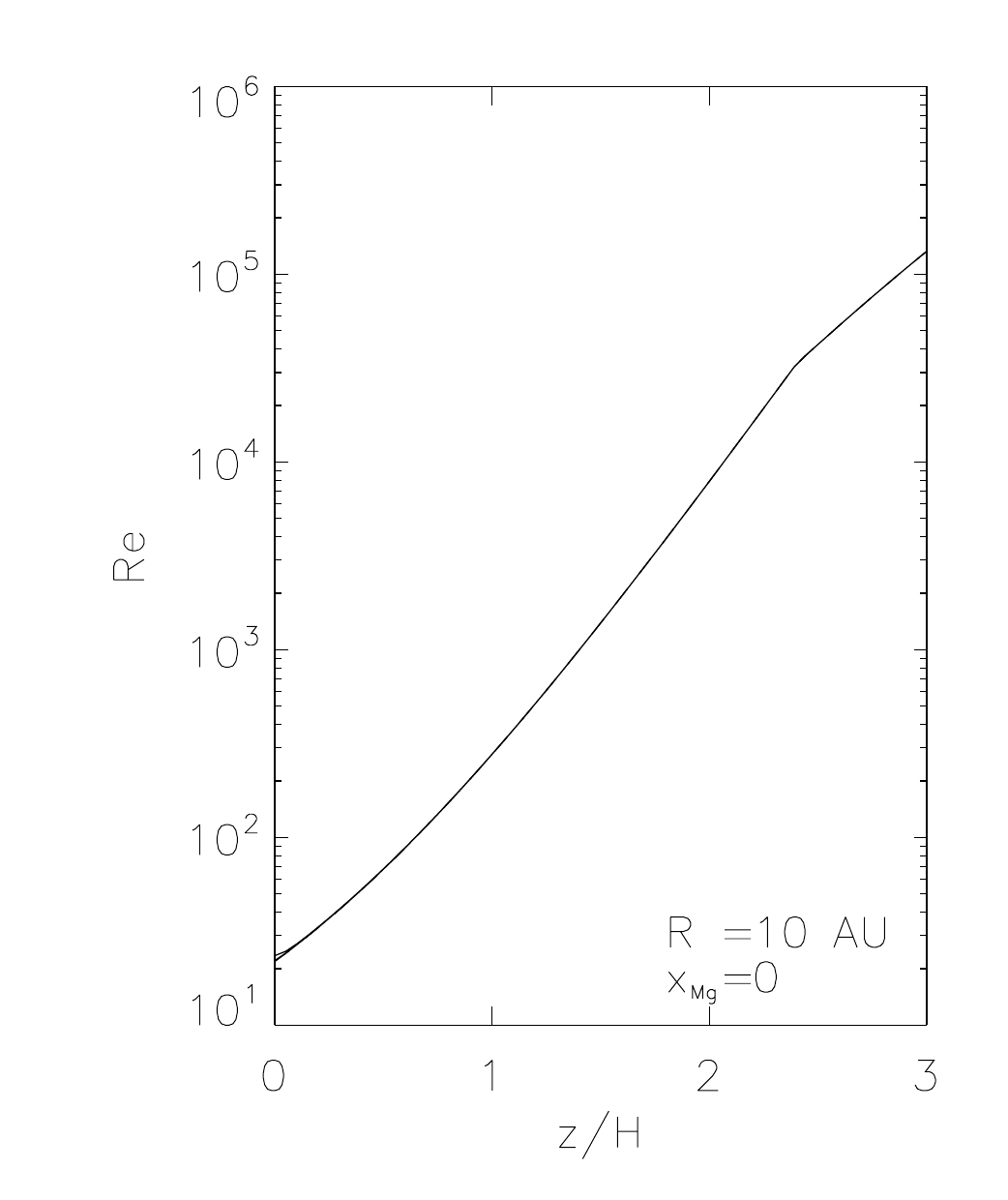}\\
\includegraphics[width=.5\textwidth]{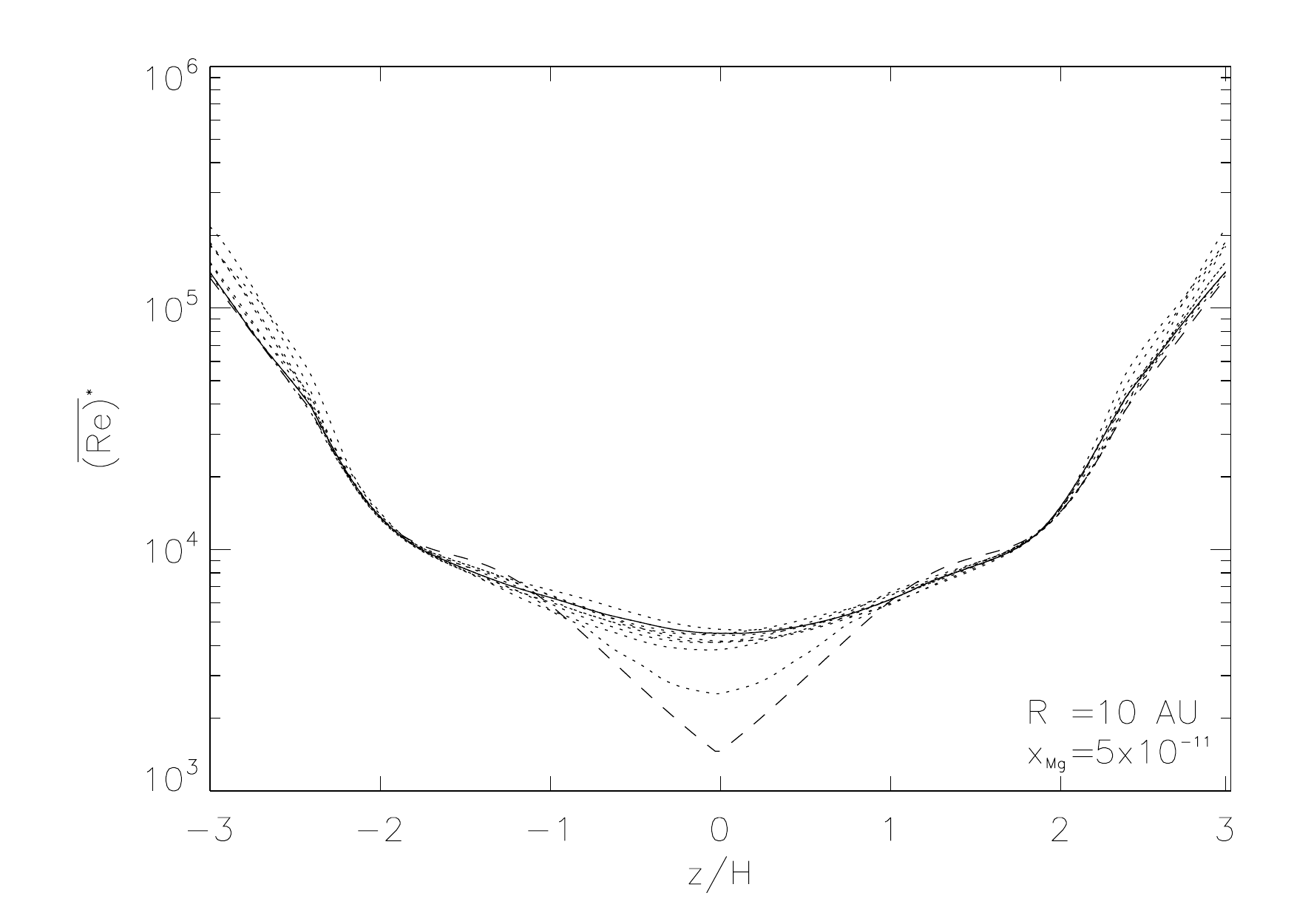}
\includegraphics[width=.3\textwidth]{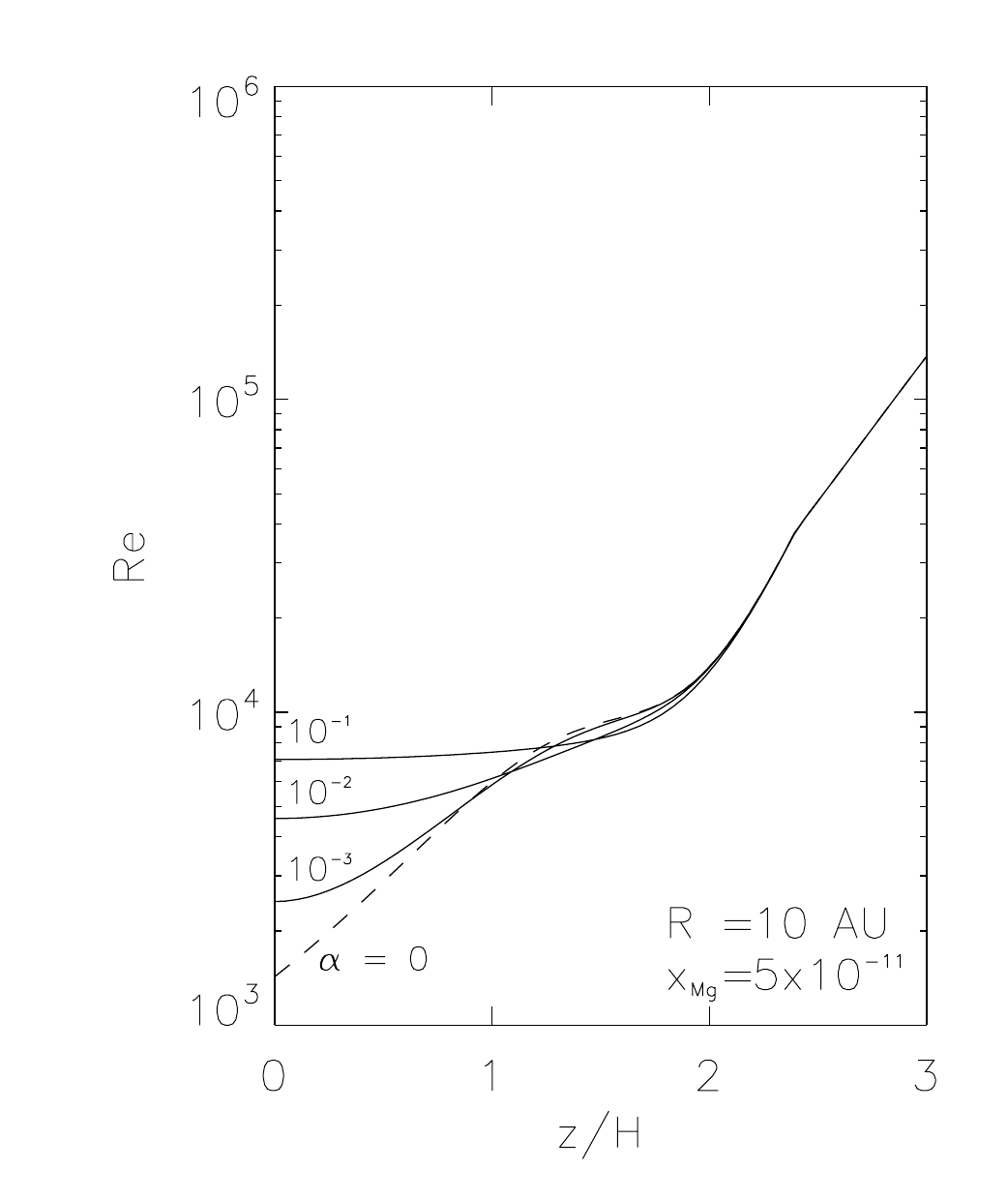}
\caption{\texttt{model2}: -  Comparison between vertical profiles of the magnetic
Reynolds number obtained for $R \rm = 10 \ AU$ and $x_{\rm Mg}^{} = 0 $
and $x_{\rm Mg}^{} = 5 \times 10^{-11}$, respectively. Left panels show shearing 
box simulation results, and right panels reaction--diffusion model results.
In the left panels, the time averaged vertical profiles of the magnetic Reynolds 
number $\overline{\rm Re^{\ast}}$ are shown by taking the time average over 10 
orbit intervals, starting from $t = 0$ (dashed line) towards $[90,100]$ (solid line). 
The other profiles (dotted lines) refer to the intervals $t = [0,10], \ [10,20], \ \cdots [80,90]$ 
orbits. The resistivity profiles of the magnetic Reynolds number $\rm Re_m$ shown in 
the right panels refer to the equilibrium profiles obtained with the reaction--diffusion
model assuming different values of $\alpha_m$. In particular for $x_{\rm Mg} = 0$,
the value of the magnetic Reynolds number is not affected by the actual value of
$\alpha_m$, leading to identical profiles for $\alpha_m = 0$, $\alpha_m = 10^{-3}$, 
and $\alpha_m = 10^{-2}$. Note that both shearing box simulations and the 
reaction--diffusion models are initiated with the steady state profile obtained for 
$\alpha_m = 0$.}
\label{figure:11}
\end{figure*}

\subsubsection{Results at 5 AU}
\label{5AU}
We first discuss the results for $R = 5 \ \rm AU$ which are
shown in Fig.~\ref{figure:9}. Note that in the upper left panel of
Fig.~\ref{figure:9}, the magnetic Reynolds number shows a well defined
minimum, and this arises because this MHD simulation was performed
with a ceiling being adopted for the resistivity in the induction equation,
corresponding to a minimum value of $\rm Re_m=20$. This was done 
to ensure that the time step constraint arising from the diffusive term in
the induction equation was not too severe. The minimum value of 
$\rm Re_m$ that we calculated from the electron fraction during this 
simulation was $\rm Re_m = 10.007$.

The upper left panel of Fig.~\ref{figure:9} simply shows that 
the model at 5 AU with $x_{\rm Mg}=0$ maintains a significant
dead zone with $|z/H| \le 2$ (where $\rm Re_m \simeq 4000$)
throughout, and the magnetic Reynolds number does not change 
from its initial value. The upper right hand panel shows that the 
reaction--diffusion equation agrees with this, as the single line 
plotted is actually three lines overplotted corresponding to $\alpha_m=0$, 
$10^{-3}$ and $10^{-2}$. In the absence of magnesium, the recombination
rate is simply too high to allow mixing to modify the dead zone for
$\alpha_m $ values in this range.

The lower left panel shows the evolution of $\rm Re_m$ with 
$x_{\rm Mg}=5 \times 10^{-11}$. It is clear that the 
$\rm Re_m$ profile in this case is non stationary near the
midplane, even after 100 orbits, and this appears to be
because this particular model is one which maintains
a dead zone throughout the run, but whose parameters
are close to those which would allow mixing to remove the dead zone.
Episodic increases and decreases in turbulent activity modify
the ionisation state near the midplane, causing the $\rm Re_m$ values
to oscillate about a value close to 1000. 

The lower right panel shows the results from the reaction--diffusion
model agree quite well with the mean $\rm Re_m$ profile from the MHD run,
in particular when $\alpha_m= 10^{-3}$. Inspection of table~\ref{table:3}
shows that mean value of $\alpha$ obtained from the MHD run was
$1.52 \times 10^{-3}$. Table~\ref{table:4} shows that the best fit 
reaction--diffusion model has a value of $\alpha_m=5.89 \times 10^{-4}$ 
(such that the Schmidt number equals 1.88). The error in the predicted value
of $\rm Re_m$ at the midplane was 3 \%, while the error at the disc surface was
$\simeq 15 \%$, showing good overall agreement even when a uniform
diffusion coefficient is adopted in the reaction--diffusion models.
 
\subsubsection{Results at 7 AU}
\label{7AU}
The vertical profiles of $\rm Re_m$ obtained at $R=7$ AU are
shown in Fig.~\ref{figure:10}. The upper left and right panels
show results from the MHD simulation and reaction--diffusion
calculation, respectively for magnesium abundance $x_{\rm Mg} = 0$.
Once again we see that mixing has no effect on the magnetic Reynolds number
profile in the absence of magnesium, and a substantial dead zone is
maintained throughout both calculations, which show excellent agreement.

The lower left and right panels show models for which 
$x_{\rm Mg}=5 \times 10^{-11}$. Here we see that there is very
significant change in the magnetic Reynolds number profile as
turbulent mixing ensues. In the MHD simulation we see that
the minimum value of $\rm Re_m$ changes from 1000 to $\simeq 4000$,
which is high enough for the disc midplane to become active.
The lower right panel shows good agreement with the MHD simulation
when $\alpha =10^{-2}$. We see from table~\ref{table:3}
that the average value of $\alpha$ from the MHD run is 
$1.07 \times 10^{-2}$. Table~\ref{table:4} shows
that the best fit reaction--diffusion model
has a value of $\alpha_m=8.67 \times 10^{-3}$ (such
that the Schmidt number equals 1.23). The error in the predicted value
of $\rm Re_m$ at the midplane was 2 \%, while the error at the disc surface was
$\simeq 15 \%$. 
\begin{figure*}[ht]
\centering
\includegraphics[width=.5\textwidth]{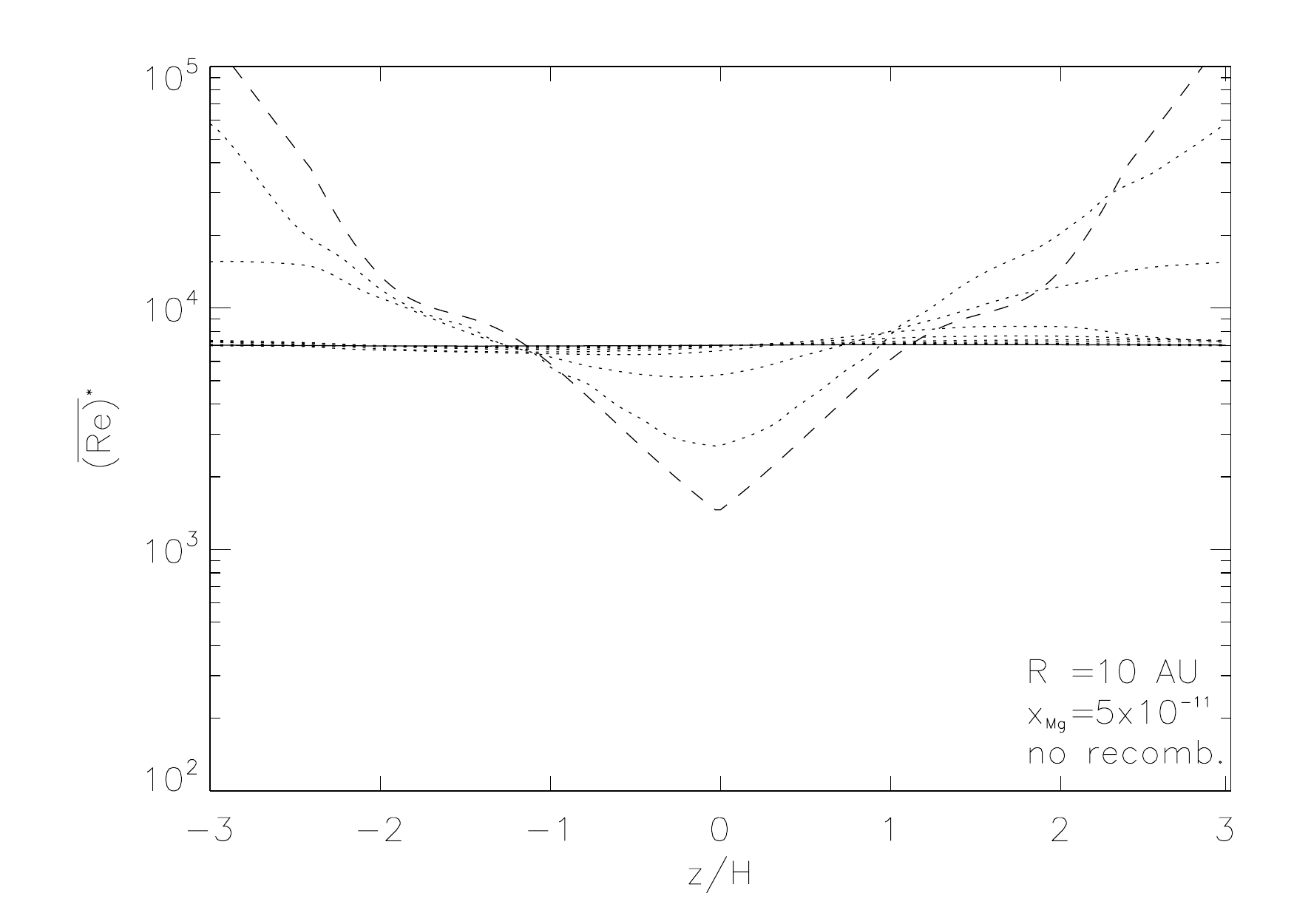} 
\includegraphics[width=.3\textwidth]{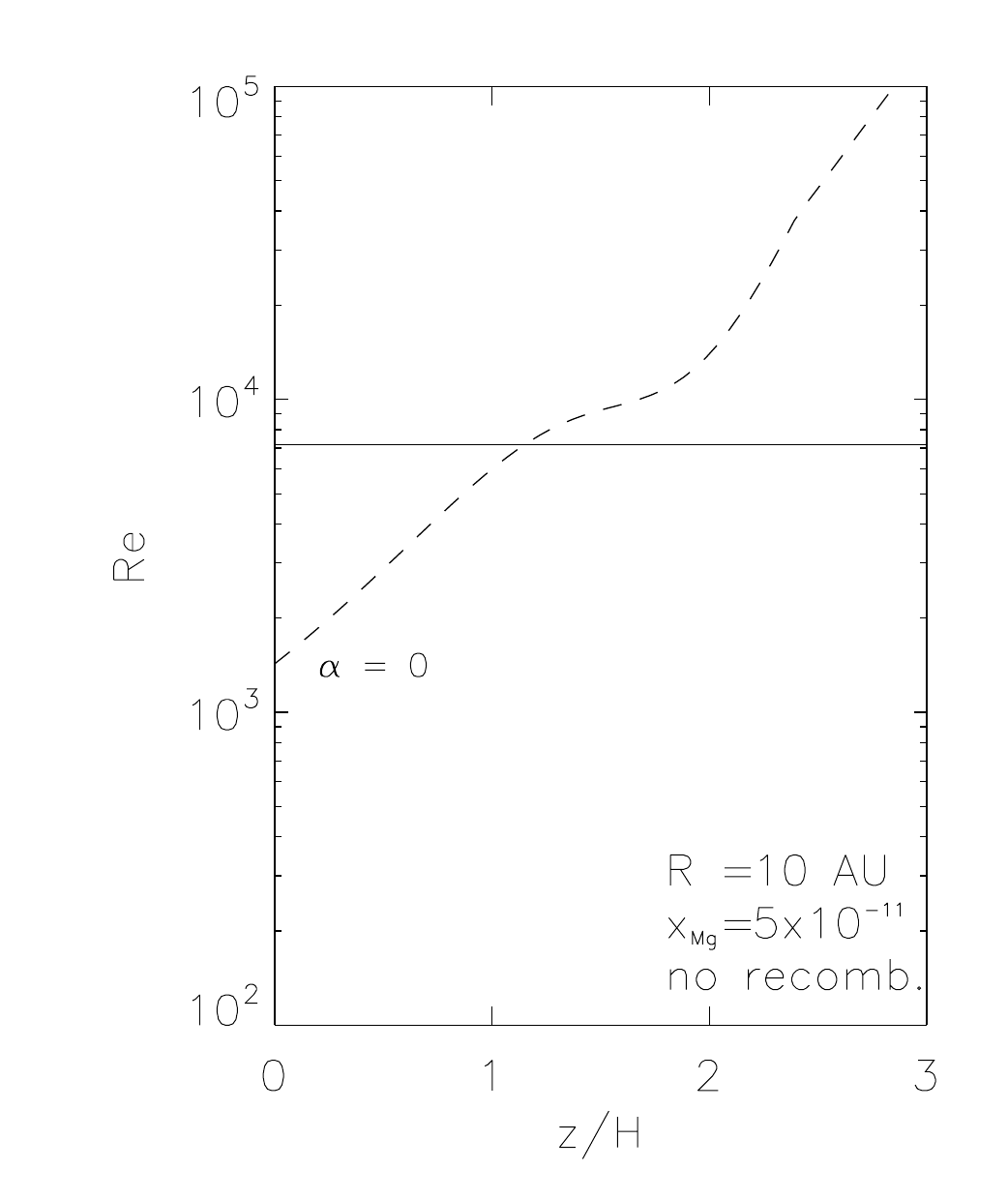}
\caption{\texttt{model3}: -  Comparison between vertical profiles 
of the magnetic Reynolds number obtained for $R \rm = 10 \ AU$ and 
$x_{\rm Mg}^{} = 5 \times 10^{-11}$ obtained by
the shearing box simulation (left panel) and the 
reaction--diffusion model (right panel). 
The results refer to the simulation where
recombination and ionisation are switched off 
(\texttt{model3}).
In the left panel, the time averaged vertical profiles 
of the magnetic Reynolds number $\overline{\rm Re_m^{\ast}}$ are shown 
by taking time averages
over 10 orbit intervals, starting from $t = 0$ (dashed line) towards 
$[90,100]$ (solid line). The other 
profiles (dotted lines) refer to $t = [0,10], \ [10,20], \ \cdots [80,90]$. 
The resistivity profiles of the magnetic Reynolds number $\rm Re_m$ shown 
in the right panel refer to the equilibrium profiles obtained 
with the reaction--diffusion model assuming different values of 
$\alpha_m$. In fact, the equilibrium 
profile obtained for \texttt{model3} is not affected by the actual 
value of $\alpha_m > 0$. Note that both shearing box simulation and the 
reaction--diffusion model are initiated with the equilibrium
 profile obtained for $\alpha_m = 0$.}
\label{figure:12}
\end{figure*}

\subsubsection{Results at 10 AU}
\label{10AU}
The profiles obtained at $R = 10 \ \rm AU$ are shown in 
Fig.~\ref{figure:11}. The upper panels are again in good
agreement when $x_{\rm Mg}=0$, showing that mixing has
no effect on the dead zone structure.
The lower panels show that the $\rm Re_m$ profile
is changed significantly by mixing when 
$x_{\rm Mg}=5 \times 10^{-11}$, such that the dead zone
is enlivened completely. In the MHD simulation the minimum value of $\rm Re_m$
changes from $\simeq 1000$ to $\simeq 6000$, allowing the
dead zone to become MRI--active and the disc to be turbulent
throughout its height. Good agreement in the
$\rm Re_m$ profile is obtained using the reaction--diffusion
model when $\alpha_m \simeq 10^{-2}$, which as expected is slightly
lower than the value
$\alpha = 1.31 \times 10^{-2}$ listed in table~\ref{table:3}
as arising from the MHD simulation. Table~\ref{table:4} shows
that the actual best fit reaction--diffusion model
has a value of $\alpha_m=9.28 \times 10^{-3}$ (such
that the Schmidt number equals 1.41). The error in the predicted value
of $\rm Re_m$ at the midplane was $<1$ \%, 
while the error at the disc surface was
$\simeq 15 \%$, which again illustrates the fact that reasonable
agreement can be obtained when using a uniform diffusion coefficient.
\subsubsection{Reaction-diffusion results for \texttt{model3}}
\label{RD-model3}
We finally present a comparison between the MHD simulation
for \texttt{model3} and a corresponding reaction--diffusion
model. To recap: \texttt{model3} allows the free electrons and
ions to diffuse, but does not include recombination or on--going
ionisation. The MHD simulation and reaction-diffusion model
were initiated with the equilibrium chemical abundance for
the case $x_{\rm Mg}= 5 \times 10^{-11}$ at $R=10$ AU.
The expectation is that turbulent mixing will cause the
$\rm Re_m$ profiles to change from their initial values
to become uniform. Inspection of Fig.~\ref{figure:12} confirms
that our models agree with this expectation.
\noindent
\section{Conclusions}
\label{conclusions}
We have presented the results from a series of shearing box multifluid MHD simulations
aimed at examining the evolution and structure of dead zones in protoplanetary discs, 
in the presence
of turbulent transport of ions, on--going chemical evolution of the gas, and ionisation
due to X--rays emitted by the central star.
We have adopted a number of simplifying assumptions, including the
absence of small grains whose presence in even modest numbers would
lead to rapid removal of free electrons (Sano et al. 2000; Ilgner \& Nelson 2006a).
As such, our results are likely to be most applicable to protostellar discs at a fairly
late stage of evolution after grains have accumulated to form larger bodies.

A primary objective of this work was to use MHD simulations to re--examine the results 
of Ilgner \& Nelson (2006b),
who used a simple reaction--diffusion model to calculate the effects of turbulent
mixing on dead zone structure. The simple model predicted that turbulent transport
can be effective at enlivening a dead zone provided that: (i) the abundance of gas--phase
magnesium is sufficient; (ii) one considers regions of the disc beyond radii $\ga 5$ AU
where turbulent mixing occurs on a shorter time scale than recombination.
The main conclusions of this paper are that full multifluid MHD simulations
are in good agreement with these predictions. Models simulated at radii 
between 1 -- 10 AU, and with no magnesium 
in the gas--phase, showed a two--layer structure consisting of 
an actively accreting zone near the disc surface, and a magnetically
inactive region near the midplane. The addition of gas--phase magnesium
with fractional abundance $x_{\rm Mg} = 5 \times 10^{-11}$ led to significant
dead zones persisting for radii $R \le 5$ AU, but models at 7 and 10 AU
resulted in fully active discs without dead zones. The implications for
protoplanetary discs is that at late times, when most of the small submicron sized
dust grains have grown to much larger sizes, the dead zone beyond 5 AU may be
enlivened because of turbulent transport of ions toward the midplane.
Regions interior to 5 AU will, however, retain their dead zones.

A further conclusion of our work is that detailed comparison
between the simple reaction--diffusion model and the MHD simulations
leads to very good agreement in the vertical profiles of resistivity and
magnetic Reynolds number when an appropriate diffusion coefficient 
is chosen. Typically we find that the best--fit vertical diffusion coefficient
corresponds to a ratio in the range 1--2 between the rate at which 
angular momentum is transported radially and the rate at which diffusion 
of chemical species occurs vertically. This result is consistent with those
presented by Carballido et al. (2005), Johansen \& Klahr (2005), Turner 
et al. (2006) and Fromang \& Papaloizou (2006) who showed that turbulent 
diffusion of dust (and chemical tracers) occurs on a slightly slower time scale
than the transport of angular momentum, since it is driven through correlations 
in the perturbed flow velocities only.

It has traditionally been assumed that MRI-driven MHD turbulence in discs
can be sustained against the damping effects provided by ohmic resistivity
if magnetic field diffusion over the characteristic wavelength of the
instability occurs on a time scale longer than the growth time.
Indeed Turner et al. (2007) showed that such a condition provides
a good indicator of where the transition between active and dead zones
in a disc will occur. They showed that the transition zone occurs
where the Lunquist number ${\rm Lu} \equiv v_A^2/(\eta \Omega) \simeq 1$, 
and our simulation results are in good agreement with this. 
Recent work by Fromang \& Papaloizou (2007) and Fromang et al. (2007), however,
has shown that in the case of non stratified shearing box simulations, turbulence
is only sustained in discs where the magnetic Prandtl number $P_m \equiv \nu/\eta > 1$
(where $\nu$ is the physical (molecular) viscosity),
even when ${\rm Lu} > 1$ in the initial state. 
The interpretation is that MRI-driven turbulence 
cascades the magnetic field down to the smallest scales available, by virtue
of the turbulent velocity field twisting the field up. If the characteristic
scale on which velocity fluctuations are damped is smaller than the resistive scale,
then a zero net flux field will be dissipated and turbulence will die.
Interestingly, the magnetic Prandtl number in protostellar discs is expected to be
$P_m < 1$, since resistivity is high and viscosity is low. Fromang \& Papaloizou (2007)
also show that the intrinsic numerical magnetic Prandtl number of the ZEUS code is $>1$,
at least for simulations with resolutions feasible on current computers. 
This suggests that the results in this paper, and those in other papers that
have looked at dead zones, are modelling discs which only fullfil one of the
necessary criteria for MHD turbulence to be sustained in a physical way, namely that
${\rm Lu} > 1$. The condition for $P_m >1$ is satisfied  because of the nature
of numerical dissipation in the code. We note that the effects
observed by Fromang \& Papaloizou (2007) and Fromang et al. (2007) occur
for the particular case of non stratified shearing box simulations, and
that a mechanism for maintaining active MRI turbulence may be the generation
of large scale magnetic field through magnetic buoyancy effects and 
field stretching in vertically
stratified discs, such as those we consider in this paper. Nonetheless, 
it is clearly necessary to examine these issues by including the
appropriate viscous as well as resistive transport coefficients in simulations,
and we will do this in a future publication.

There are two additional issues that we have not addressed in this paper.
The first is that scattering of X--rays toward the disc midplane
may increase the ionisation rate in the disc by up to an order of
magnitude (Igea \& Glassgold 1999), and this can have an obvious effect
on the structure of the dead zone. Although we have not undertaken
an extensive analysis of the effect of this, we have run a model at 1 AU
with $x_{\rm Mg} = 5 \times 10^{-11}$ with the X--ray luminosity increased
by two orders of magnitude. We find only a small change in the dead zone
structure in this case. We would expect in general that increases
in the X--ray luminosity due to scattering will move the radial boundary of the
dead zone inward slightly, but will not completely remove the dead zone. 
A final issue that we have not addressed in this paper is that of X--ray flares.
Observations of T Tauri stars by CHANDRA have shown that they emit regular
outbursts of X--rays which may increase the X--ray luminosity by a few orders
of magnitude, and also harden the X--ray spectrum (Favata et al 2005; Wolk et al. 2005). 
This issue was examined by Ilgner \& Nelson (2006c), who showed that
the flaring could significantly modify dead zones in protoplanetary discs.
We will address this issue in a future paper using multifluid MHD simulations 
with chemistry, so that both the effects of X--ray flaring and chemical mixing
on dead zone structure can be examined.

\begin{acknowledgements}
We would like to thank S\'ebastien Fromang, who very kindly made his version
of the ZEUS code available to us. The referee, Neal Turner, provided numerous 
comments which improved this paper. The simulations presented here were 
performed on the QMUL High Performance Computing Facility purchased under 
the SRIF initiative.
\end{acknowledgements}


\end{document}